\documentclass[aps,prb,showpacs,superscriptaddress,twocolumn,10pt,floatfix]{revtex4-1}
\usepackage{graphicx}
\usepackage{natbib}
\usepackage{amsmath}
\usepackage{epsfig}
\usepackage{physics}
\usepackage{multirow}
\usepackage{siunitx}
\usepackage{natbib}
\usepackage{indentfirst}
\usepackage{siunitx}
\usepackage{mathtools}
\usepackage{amssymb}
\usepackage{bbold}
\usepackage{xcolor}
\usepackage{bm}
\usepackage{eufrak}
\usepackage{epstopdf}
\usepackage{fancyhdr}

\begin{document}

\title{Skyrmion zoo in graphene at charge neutrality in a strong magnetic field}

\author{Jonathan Atteia}
\email{jonathan.atteia@u-psud.fr}
\affiliation{Laboratoire de Physique des Solides, Universit\'e Paris Saclay, CNRS UMR 8502, F-91405 Orsay Cedex, France}

\author{Yunlong Lian}
\email{lianyl@ihep.ac.cn}
\affiliation{Spallation Neutron Source Science Center, Institute of High Energy Physics, Chinese Academy of Sciences, Dongguan 523803, China}

\author{Mark Oliver Goerbig}
\email{mark-oliver.goerbig@universite-paris-saclay.fr}
\affiliation{Laboratoire de Physique des Solides, Universit\'e Paris Saclay, CNRS UMR 8502, F-91405 Orsay Cedex, France}

\date{\today}
\begin{abstract}
As a consequence of the approximate spin-valley symmetry in graphene, the ground state of electrons in graphene at charge neutrality is a particular SU(4) quantum-Hall ferromagnet to minimize their exchange energy. If only the Coulomb interaction is taken into account, this ferromagnet can appeal either to the spin degree of freedom or equivalently to the valley pseudo-spin degree of freedom. This freedom in choice is then limited by subleading energy scales that explicitly break the SU(4) symmetry, the simplest of which is given by the Zeeman effect that orients the spin in the direction of the magnetic field. In addition, there are also valley symmetry breaking terms that can arise from short-range interactions or electron-phonon couplings. Here, we build upon the phase diagram, which has been obtained by Kharitonov [Phys. Rev. B \textbf{85}, 155439 (2012)], in order to identify the different skyrmions that are compatible with these types of quantum-Hall ferromagnets. Similarly to the ferromagnets, the skyrmions at charge neutrality are described by the $\text{Gr}(2,4)$ Grassmannian at the center, which allows us to construct the skyrmion spinors. The different skyrmion types are then obtained by minimizing their energy within a variational approach, with respect to the remaining free parameters that are not fixed by the requirement that the skyrmion at large distances from their center must be compatible with the ferromagnetic background. We show that the different skyrmion types have a clear signature in the local, sublattice-resolved, spin magnetization, which is in principle accessible in scanning-tunneling microscopy and spectroscopy. 
\end{abstract}

\maketitle

\section{Introduction}

Graphene, a one-atom thick layer of carbon atoms is the prototype of a large class of 2D materials such as transition metal dichalcogenides\cite{Manzeli2017}, van der Waals heterostructures\cite{Geim2013} or twisted bilayers\cite{Lu2019} which present striking properties such as superconductivity, correlated or topological phases. A salient feature of graphene is its linear electronic dispersion relation which is analogous to massless Dirac fermions\cite{Novoselov2005}. These fermions come in two flavors corresponding to the two degenerate valleys located at the corners of the first Brillouin zone. Upon application of a perpendicular magnetic field, flat Landau levels (LLs) are formed and one observes the relativistic quantum Hall effect\cite{Novoselov2005,Zhang2005,Novoselov2007} characteristic of Dirac fermions, where the Hall resistance is quantized in half-integer units of $4e^2/h$. The factor 4 originates from the spin and valley degeneracy. From the non-interacting electron point of view and in the absence of a Zeeman effect, the system has thus an SU(4) symmetry associated with the fourfold spin and pseudo-spin (valley index) degeneracy. 

Upon increasing the magnetic field or synthesizing higher quality samples, additional quantum Hall plateaus in the conductance are observed at values $\nu e^2/h$ where $\nu=n_{el}/n_B$ is the filling factor of the LL, in terms of the electronic $n_{el}$ and the flux densities $n_B=1/2\pi l_B^2$, respectively, where $l_B=\sqrt{\hbar/eB}$ is the magnetic length\cite{Zhang2006,Jiang2007,Young2012}. This indicates that each Landau level is indeed composed of four sub-Landau levels (sub-LL) due to the above-mentioned fourfold spin-valley degeneracy, which is gradually lifted at higher magnetic fields. At the filling factors $\nu=0$ and $\pm 1$, this degeneracy lifting is due to electron-electron (Coulomb) interactions that represent a substantially larger energy scale than e.g. the Zeeman effect \cite{Goerbig2006,Goerbig2011,Nomura2006}. These interactions favor quantum-Hall ferromagnetic (QHFM) states that can be understood in the following manner. In order to minimize the Coulomb interaction between the electrons, a maximally \textit{anti-symmetric} orbital wave function is favored such that the electrons can maximally avoid each other. As a consequence of the fermionic nature of the electrons and the requirement of a totally anti-symmetric wave function, the anti-symmetry in the orbital part must be accompanied by a \textit{symmetric} wave function for the internal degrees of freedom, i.e. the spin and the valley pseudo-spin in the case of graphene. The resulting ferromagnetism is thus of a particular SU(4) type that does not only lead to a macroscopic spin magnetization but also to a valley magnetization\cite{Alicea2006,Sheng2007,Nomura2009,Jung2009,Abanin2006,Kharitonov2012}. The fourfold broken-symmetry states also have an edge state signature which has been observed in the $n=0$ LL graphene using atomic force microscopy\cite{Kim2020}.

A particularly intriguing feature of QHFM is the nature of its quasiparticles. Indeed, the addition of an electron (with an opposite spin or pseudo-spin) to a QHFM state in the lowest LLs perturbs locally the magnetization such that the electron is dressed by a spin-pseudospin texture that is known as a skyrmion. Most saliently, such a skyrmion is a topological object, the topological charge of which is directly proportional to the electric charge. In a general context, skyrmions exist in quantum Hall systems\cite{Sondhi1993,Moon1995,Fertig1994,Fertig1997,Arovas1999,Yang2006}, in chiral magnets\cite{Rossler2006,Schulz2012,Nagaosa2013,Freimuth2013} and in magnetic thin films\cite{Bogdanov2001,Sampaio2013,Moreau-Luchaire2016}. They have been observed using scanning tunnelling microscopy and spectroscopy (STM/STS) in the latter two systems but not in the former yet. In fact, their existence in two-dimensional electron gases (2DEG) of semiconductor heterostructures in the quantum Hall regime was discovered experimentally via different indirect techniques such as nuclear magnetic resonance\cite{Barrett1995,Gervais2005,Mitrovic2007}, heat capacity measurements\cite{Bayot1996,Melinte1999} or thermally activated transport\cite{Schmeller1995,Goldberg1996}. However, these techniques only showed indirect evidence for the presence of skyrmions because the 2DEG in semiconductor heterostructures is not directly accessible by near-field spectroscopy, such as STS. This problem can in principle be circumvented in graphene where the 2DEG is directly situated at the surface unless the graphene sheet is encapsulated by thick insulating crystal slabs. Furthermore, the skyrmions in graphene inherit the underlying SU(4) spin-valley symmetry from the QHFM ground states\cite{Yang2006,Doucot2008,Lian2016,Lian2017,Jolicoeur2019}. Additionally, in the $n=0$ LL of graphene, the valley index is locked to the sublattice index, rendering therefore possible a direct imaging of the valley polarization of the phase by measuring the sublattice polarization \cite{Lian2016,Lian2017}.

In the present paper, we extend the work by Lian \textit{et al}\cite{Lian2016,Lian2017} for the phase diagram of the skyrmion in the $n=0$ LL at filling $\nu=\pm1$ to the case $\nu=0$ corresponding to charge neutrality. In the SU(4) case at half-filling, there are two electrons per Landau orbit, where a Landau orbit corresponds to the area occupied by an electron in his cyclotron motion. Hence, there are two electrons located at the same position in the $n=0$ LL implying that they must thus be orthogonal in SU(4) space. The situation is thus more complex than that at $\nu=\pm 1$, where a complete spin polarization can come along with a full pseudo-spin polarization, e.g. if the spin-up branch of the $K$-valley sublevel is completely filled. This is no longer possible at $\nu=0$. Indeed, if one polarizes the spin of all electrons in the $n=0$ LL, one needs imperatively to fill both valley branches such that the pseudo-spin magnetization vanishes. Similarly, if the valley pseudo-spin is completely polarized, the spin magnetization vanishes. Therefore, choices in the magnetization need to be made, and these choices are related to explicit spin or valley symmetry-breaking terms. In addition to the natural Zeeman term, we consider also pseudo-spin anisotropic terms $u_\perp$ and $u_z$ that may find their origin in short-range electron-electron or also in electron-phonon interactions. Similarly to the work by Kharitonov\cite{Kharitonov2012}, we do not fix the values of these parameters, which are all on the same order of magnitude and likely sample- or substrate-dependent. Instead, we use them as control parameters that allow us not only to span the QHFM phase diagram but also to identify the lowest-energy skyrmion types that are compatible with the QHFM background. From a technical point of view, we use a classical non-linear sigma model (NLSM) obtained from a gradient expansion of the energy of the spin and pseudo-spin magnetizations. In the graphene SU(4) case at filling factor $\nu=\pm1$, the NLSM appeals to a four-components CP$^3$ spinor field that describes the skyrmion (anti-skyrmion) in the completely empty (filled) LL. At filling $\nu=0$, a CP$^3$ description is no longer possible because two LL subbranches are filled, and we need instead use the Grassmannian $\text{Gr}(2,4)$ which is a $2\times 4$ matrix field decribing the spin and pseudo-spin of the two indistinguishable electrons.

To check our Grassmannian parametrization of the spinors, we first recover the phase diagram of QHFM ground state originally discovered by Kharitonov\cite{Kharitonov2012}, as a function of the anisotropic parameters and the Zeeman energy. The ground state is composed of four phases, two of which are fully or partially spin polarized [the ferromagnetic (F) and canted anti-ferromagnetic (CAF) phases] and the other two are pseudo-spin polarized [the charge density wave (CDW) and Kekul\'e disortion (KD) phases] and spin unpolarized. We find an additional symmetry restoration at the F-CDW transition where the spin and pseudo-spin spinors are interchanged. Most saliently, the CAF phase is characterized by a spin-valley entanglement, which is similar to that found at $\nu=\pm1$ \cite{Lian2016,Lian2017} and where the spin and valley magnetizations are at least partially locked. This allows one to minimize the valley anisotropic energy while paying a little price in Zeeman energy. If the latter tends to zero, the CAF phase becomes a true anti-ferromagnet with spin-up electrons situated on one sublattice and spin-down electrons on the other one. Notice, however, that this anti-ferromagnetic phase is simply a manifestation of an SU(4) ferromagnetic phase : in fact, this phase can be obtained from a \textit{true} ferromagnet with all spins on both sublattices pointing in the same direction by a global SU(4) rotation. From an experimental point of view, the true ground state is yet under debate. While it is mostly accepted that the ground state at $\nu=0$ is insulating, such as the CDW, KD and CAF phases with a vanishing Hall conductance\cite{Zhang2006,Zhang2010,Checkelsky2008,Young2012,Amet2013}, recent experiments\cite{Veyrat2020} on a substrate with a large dielectric constant and thus a large screening show helical edge states that are in line with the F phase\cite{Abanin2006}. Furthermore, first experimental indications of a CAF phase\cite{Young2014} are now challenged by STM experiments that are in line with a KD phase\cite{Li2019}. We note additionally that the QHFM ground state is susceptible to be modified near an edge sample\cite{Knothe2015}.

In addition to the QHFM phases, we consider skyrmions of unit charge that are described by the Grassmannian $\text{Gr}(2,4)$ field.  At infinity, this field must recover the two sub-LLs spinors of the QHFM background, while at the center, one of the spinors corresponds to one of the two empty sub-LLs. The skyrmion texture is thus constructed by the interpolation of one of the spinors describing an empty sub-LL at the center to one of filled sub-LLs that describe the QHFM at distances far away from the skyrmion center. The only constraints are the orthogonality of the two spinors in the interpolation and the compatibility with the QHFM background, such that the skyrmions are still characterized by a finite number (six) of parameters that we use in a variational approach, in which we minimize the skyrmion energy with respect to these parameters. We thus obtain a true skyrmion zoo within our phase diagram, and we characterize these skyrmions by their sublattice-resolved spin magnetization at the center as compared to that in the QHFM background. These patterns may be a guide in the identification of SU(4) skyrmions in spin-resolved STM experiments. 

The paper is organized as follows. In Sec. \ref{sec:ground_state} we concentrate on the QHFM ground states within a Grassmannian description, which has the advantage of being generalizable to quantum-Hall systems with even larger components than 4. We discuss the parametrization of these states and describe how the measurable spin and valley pseudo-spin magnetizations are affected by entanglement. In contrast to the case at $\nu=\pm 1$, we find that entanglement is described in terms of two angles instead of a single one. We then present the phase diagram and describe the four different phases with their spin magnetization and electronic density on the A and B sublattices. In Sec. \ref{sec:skyrmions}, we construct the Grassmannian fields that describe charge-one skyrmions as  the solutions of the NLSM. We discuss their energy and size as a function of the anisotropic and Coulomb energies. Section \ref{sec:phase_diag_skyrm} is devoted to the phase diagram for the skyrmions in the different QHFM backgrounds. We visualize the skyrmion spinors with the help of their spin magnetization and, if relevant, their electronic density on the A and B sublattices. Our results for the energy and the size of the skyrmions are presented in sec. \ref{sec:energy_size}, where we also describe in detail the symmetry restoration at the different transition lines.

\section{Ground state}

\label{sec:ground_state}

In this section, we review the basics of the SU(4) quantum Hall ferromagnetism in graphene in order to recover Kharitonov's\cite{Kharitonov2012} results for the ground state of graphene in the quantum Hall regime at $\nu=0$. Our parametrization of the spinors of the electrons allows us to recover Kharitonov's four difference phases, namely ferromagnetic, Kekul\'e-distortion, charge density wave and canted anti-ferromagnetic. Our resulting two-particle states are identical to Kharitonov's state for the first three phases, howevever, for the case of the canted anti-ferromagnetism, we obtain more general spinors.

Under a strong perpendicular magnetic field applied to a graphene sheet, Landau levels (LL) are formed with (non-interacting) energies $E_{\lambda n}=\lambda\hbar\omega_c\sqrt{n}$ where $\lambda=\pm$ is the band index, $n$ is the LL index, $\omega_c=\sqrt{2}v/l_B$ is the cyclotron energy, and $v$ is the Fermi velocity of graphene. The magnetic length scales as $l_B\sim 1/\sqrt{B}$ with the magnetic field $B$ perpendicular to the graphene plane. These LL have a very high orbital degeneracy characterized by the guiding center $m$ of the orbital wavefunction. They also have an additional fourfold degeneracy due to the spin and valley degeneracy (when neglecting Zeeman splitting and possible valley splittings due to lattice interactions).  The characteristic energy scale of the non-interacting spectrum is given by the LL separation (for the $n=0$ and $n=1$ LL) by $E_1-E_0=\hbar\omega_c\approx1330\sqrt{B[T]}\si{K
}$.

When an integer number of sub-LL's are filled, one can use the Hartree-Fock theory to describe electron interactions. The characteristic energy scale of the Coulomb interaction at the magnetic length scale of graphene on a hexagonal Boron-Nitride (hBN) substrate taking into the screening is\cite{Goerbig2011} $E_C=e^2/\varepsilon\varepsilon_\infty l_B=625\sqrt{B[T]}\si{K}/\varepsilon$, where $\varepsilon$ is the dielectric constant of the environment the graphene sheet is embedded into, and $\varepsilon_\infty$ takes into account interband screening. We can see that the Coulomb energy scale is small compared to the LL spacing and one can thus project the wavefunction on the lowest Landau level (LLL). 

In the theory of quantum Hall ferromagnetism, the Coulomb interaction favors a maximally antisymmetric orbital wavefunction due to the exchange interaction, which in turn leads to a symmetric wavefunction in the valley and spin indices. Neglecting symmetry breaking terms such as Zeeman coupling or intervalley scattering for example, whose energy scales are negligible compared to the Coulomb interaction, the system has an approximate SU(4) symmetry. For $\nu=1$, one obtains thus a quantum Hall ferromagnet where all electrons have the same spin and valley index orientation at each Landau orbit. For $\nu=0$, however, there are two electrons per Landau orbit and the spin and pseudo-spin wavefunction of these two electron can therefore not be fully symmetric. There is thus a compromise to find in the spin-valley polarization -- if the spin is fully polarized, both valleys must be occupied such that the pseudo-spin part of the wave function must be anti-symmetric, which is the case of the (spin) ferromagnetic phase. On the other hand, if both pseudo-spin sub-LL are occupied, the spin wave function must be anti-symmetric as in the KD and CDW phases. This needs to be contrasted with the case at $\nu=\pm 1$, where only one sub-LL is fully occupied so that both the spin and the valley pseudo-spin can be completely polarized. Notice that the energy of the Zeeman coupling $\Delta_Z=1.2[B(T)]K$ is on the same order of magnitude as the pseudo-spin symmetry-breaking terms due to lattice interactions or short-range interactions between the electrons unless the latter are suppressed by a dielectric environment with a large dielectric constant such as in a recent experiment on an strontium-titanate substrate \cite{Veyrat2020}. In view of the energy scales, one should therefore search for a minimization of the SU(4)-invariant Coulomb energy, which is precisely at the origin of the SU(4) QHFM with a random spin-valley orientation that is then chosen by the above-mentioned low-energy symmetry breaking terms.

\subsection{Grassmannian}

\label{sec:Grassmannian}

At filling $\nu=0$ of the Landau level $n=0$, there are two electrons per Landau orbit $\mathbf{X}$, where $\mathbf{X}$ corresponds to the guiding center of the cyclotron motion of an electron. The quantum Hall ferromagnet ground state is described by a Slater determinant\cite{Ezawa2005},
\begin{align}
    |G\rangle=\frac{1}{2}\prod_\mathbf{X} g_{\mu\nu}c^\dagger_{\mu,\mathbf{X}} c^\dagger_{\nu,\mathbf{X}}|0\rangle,
    \label{eq:GroundState}
\end{align}
where $c^\dagger_{\mu,\mathbf{X}}$ creates an electron in the Landau orbit $\mathbf{X}$ where $\mu=\{\sigma,\xi\}$ labels the spin ($\sigma$) and valley ($\xi$) index. $g_{\mu\nu}=-g_{\nu\mu}$ is an antisymmetric matrix describing the spontaneously broken symmetry state of the quantum Hall ferromagnet. Because $g_{\mu\nu}$ is a $4\times 4$ antisymmetric matrix, it is described by 6 complex parameters ($g_{12}$,$g_{13}$,$g_{14}$,$g_{23}$,$g_{24}$,$g_{34}$) corresponding to the (complex) 6 dimensional antisymmetric irreducible representation of a two electron state $4\otimes 4=10\oplus6$. Normalizing and eliminating the overall unphysical phase, we are left with 10 real parameters. Moreover, for the state (\ref{eq:GroundState}) to be an eigenstate of the Coulomb Hamiltonian, the matrix $g_{\mu\nu}$ must obey the Pl\"ucker condition\cite{Kharitonov2012,Ezawa2005} 
\begin{align}
    g_{12}g_{34}-g_{13}g_{24}+g_{14}g_{23}=0,
\end{align}
which restricts the number of parameters describing the ground state to eight. It is useful to express the Slater determinant $g_{\mu\nu}$ as 
\begin{align}
    g_{\mu\nu}=f_{1\mu}f_{2\nu}-f_{1\nu}f_{2\mu},
\end{align}
where $f_1$ and $f_2$ are two normalized orthogonal four-component spinors describing the indistinguishable states of the two particles. The order parameter of the ferromagnet is
\begin{align}
    \langle G|c_{\mu,m}^\dagger c_{\nu,m}|G\rangle=f_{1\mu}^*f_{1\nu}+f_{2\mu}^*f_{2\nu}=P_{\nu\mu},
\end{align}
where we have introduced the expression for the order parameter :
\begin{align}
    P=f_1f_1^\dagger+f_2f_2^\dagger
\end{align}
which is a projector that obeys $P^\dagger=P$, $P^2=P$ and $\text{Tr}[P]=f_1^\dagger f_1+f_2^\dagger f_2=\tilde{\nu}=2$, where $\tilde{\nu}=\nu+2$ is the filling factor of the $n=0$ LL relative to the empty LL. The QHFM state is thus characterized completely by its projector $P$ in terms of the four-spinors $f_1$ and $f_2$.

This spontaneously broken symmetry electronic state of the SU(4) invariant Hamiltonian described by the projector $P$ that belongs to the Grassmannian projective space $\text{Gr}(\tilde{\nu},N)$. The Grassmannian can be expressed as the coset space
\begin{align}
    \text{Gr}(2,4)=\frac{\text{SU(4)}}{\text{SU(2)}\times \text{SU(2)}\times \text{U(1)}}
\end{align}
where SU(4) describes the symmetry of the Hamiltonian, and the two SU(2) groups describe the symmetry of the ground state under transformation between filled and empty states, respectively, while U(1) corresponds to the phase difference between filled and empty states. The Grassmannian $\text{Gr}(\tilde{\nu},N)$ has (real) dimension $2\tilde{\nu}(N-\tilde{\nu})=8$ and the ground state is thus described by eight parameters, which agrees with our previous counting.

\begin{figure}[t]
	\includegraphics[width=6cm]{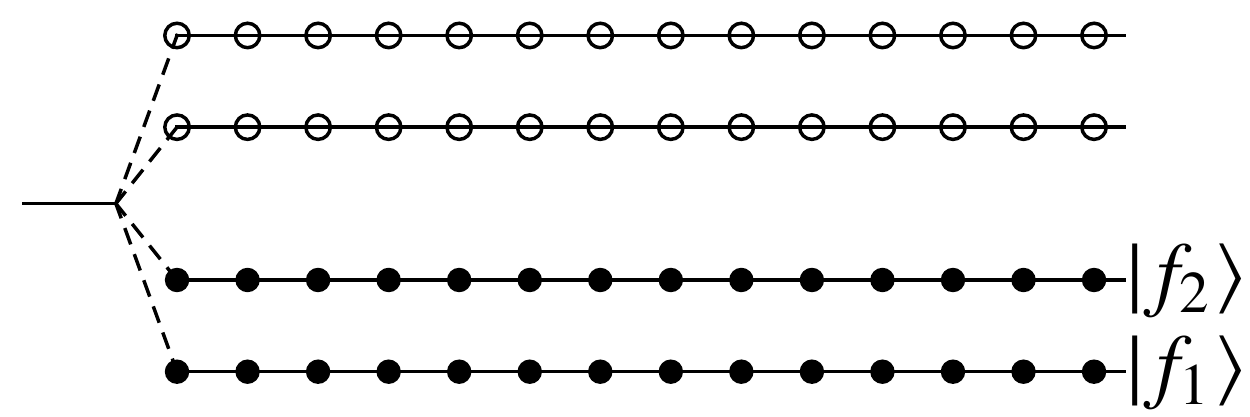}
	\caption{Schematics of the of the four sub-LLs of the QHFM ground state at $\nu=0$. The filled circles correspond to the filled sub-LLs described by the spinor $f_1$ and $f_2$.}
	\label{fig:LL}
\end{figure}

At quarter filling $\tilde{\nu}=1$ where only one sub-LL is filled, the ground state is described by only one spinor $f_1$ such that the projector is simply $P=f_1f_1^\dagger$. At half-filling, $\tilde{\nu}=2$, this corresponds to the filling of two sub-LLs with the spinors $f_1$ and $f_2$ as shown in Fig. \ref{fig:LL}. Notice that $f_1$ and $f_2$ represent schematically unspecified spin-valley sub-LLs or quantum superpositions of the spin and pseudo-spin. The only condition imposed is that $f_1$ and $f_2$ be orthogonal. Notice furthermore that one can generalize this description to any filling $\tilde{\nu}$ and dimension $N$ of the internal degrees of freedom such that the state at $\tilde{\nu}<N$ is described by the projector $P(f_1,...,f_{\tilde{\nu}} )$, with $\text{Tr}[P]=\tilde{\nu}$.

An element of the Grassmannian $\text{Gr}(2,4)$ is a $4\times 2$ matrix spinor
\begin{align}
    f=(f_1,f_2)
\end{align}
which obeys the normalization condition :
\begin{align}
    f^\dagger f=1
\end{align}
while the projector can be expressed as :
\begin{align}
    P=ff^\dagger
\end{align}
We can see that the projector is thus invariant under a SU(2) transformation that mixes the spinors such that :
\begin{align}
    f'=fU
    \label{eq:unitary}
\end{align}
where $U$ is a SU(2) unitary matrix such that $P'=f'f'^\dagger=P$ according to the fact that the spinors $f_1$ and $f_2$ are indistinguishable.

\subsection{Hartree-Fock approximation}

When projecting upon the LLL and neglecting intervalley scattering processes that are suppressed by a factor $a/l_B\sim 0.005\sqrt{B[T]}$\cite{Goerbig2011}, the Coulomb interaction is SU(4) invariant and reads
\begin{align}
    \hat{V}_C=\frac{1}{2}\sum_{\mathbf{q}\neq 0}v(\mathbf{q})\bar{\rho}(\mathbf{q})\bar{\rho}(-\mathbf{q}),
\end{align}
in terms of $v(\mathbf{q})$, the Coulomb potential renormalized by the LLL form factors
\begin{align}
    v(\mathbf{q})=\frac{1}{\mathcal{A}}\frac{2\pi e^2}{\varepsilon|\mathbf{q}|}|\mathcal{F}_0(\mathbf{q})|^2
\end{align}
where $\mathcal{A}$ is the area of the sample and $\mathcal{F}_0(\mathbf{q})$ is the form factor of the LLL (see eg. Ref. [\onlinecite{Goerbig2011}]). $\bar{\rho}(\mathbf{q})$ represents the density operator in momentum space projected into the LLL such that
\begin{align}
\bar{\rho}(\mathbf{q})=\sum_{\mu}\sum_{\mathbf{X},\mathbf{X}'}f_{\mathbf{X},\mathbf{X}'}(\mathbf{q}) c^\dagger_{\mu,\mathbf{X}}c_{\mu,\mathbf{X}}
\end{align}
where $c_{\mu,\mathbf{X}}(c^\dagger_{\mu,\mathbf{X}})$ is the annihilation (creation) operator in the internal state $\mu$ at Landau orbit $\mathbf{X}$ and the terms
\begin{align}
    f_{\mathbf{X},\mathbf{X}'}(\mathbf{q})=\langle \mathbf{X}|e^{-i\mathbf{q}.\mathbf{R}}|\mathbf{X}'\rangle
\end{align}
are the cyclotron orbits form factor with $\mathbf{R}=(X,Y)$ the guiding center of the cyclotron orbit. Applying the Hartree-Fock approximation, we obtain that the SU(4) invariant energy without symmetry breaking terms of the ground state is equal to
\begin{align}
    E_\text{SU(4)}=E_H-E_F
\end{align}
where the Hatree and Fock terms read :
\begin{align}
    E_H&=\frac{1}{2}\sum_\mathbf{q\neq 0}\sum_{XX'}v(\mathbf{q})f_{\mathbf{X}\mathbf{X}}(-\mathbf{q})f_{\mathbf{X}'\mathbf{X}'}(\mathbf{q})\text{Tr}[P]^2=0 \\
    E_F&=\frac{1}{2}\sum_\mathbf{q\neq 0}\sum_{\mathbf{X}\mathbf{X}'}v(\mathbf{q})f_{\mathbf{X}\mathbf{X}'}(-\mathbf{q})f_{\mathbf{X}'\mathbf{X}}(\mathbf{q})\text{Tr}[P^2]=-2N_\phi E_X
\end{align}
The first term vanishes because we have taken into account the positive ionic background at $\mathbf{q}=0$ which cancels with the electronic density of the direct term : $\langle \bar{\rho}(\mathbf{q})\rangle=\sum_{\mathbf{X}}f_{\mathbf{X}\mathbf{X}}(\mathbf{q})\text{Tr}\left[P\right]=\tilde{\nu}\rho_0\delta(\mathbf{q})$ and we have considered a uniform density state. The exchange energy given by
\begin{equation}
E_X=\frac{1}{2(2\pi)^2}\int d^2q\frac{2\pi e^2}{\varepsilon |\mathbf{q}|}|\mathcal{F}_0(\mathbf{q})|^2=\sqrt{\frac{\pi}{8}}E_C
\end{equation}
The Hamiltonian is SU(4) invariant and thus we observe a broken symmetry ground state where the spin and pseudo-spin are aligned in a random direction. Due to this symmetry, even a small perturbation orients the ground state in a particular direction in spin/pseudo-spin space. The chosen ground state is determined by the low-energy symmetry breaking terms which we present in the next section.

\subsection{Symmetry breaking terms}

Inspired by earlier works\cite{Kharitonov2012,Nomura2009,Lian2017} that focus on short-range electron-electron\cite{Alicea2006} and electron-phonon\cite{Kharitonov2012} interactions at the lattice scale, we consider the local anisotropic Hamiltonian 
\begin{align}
    H_A=\int d^2r&\Big \{U_\perp[P_x^2(\mathbf{r})+P_y^2(\mathbf{r})]+U_zP_z^2(\mathbf{r})\nonumber \\
    &-\Delta_ZS_z(\mathbf{r})\Big\},
    \label{eq:SB}
\end{align}
where 
\begin{align}
    \mathbf{S}(\mathbf{r})=\Psi^\dagger(\mathbf{r})(\bm{\sigma}\otimes\tau_0)\Psi(\mathbf{r})\\
    \mathbf{P}(\mathbf{r})=\Psi^\dagger(\mathbf{r})(\sigma_0\otimes\bm{\tau})\Psi(\mathbf{r}) 
\end{align}
are the local spin and pseudo-spin densities, respectively, in terms of the vectors 
$\bm{\sigma}$ and $\bm{\tau}$ of Pauli matrices vectors acting in spin and pseudo-spin spaces respectively while $\sigma_0$ and $\tau_0$ are the identity matrices acting in spin and pseudo-spin spaces respectively. In the following, we will neglect the identity and consider $\bm{\sigma}\equiv\bm{\sigma}\otimes\tau_0$ and $\bm{\tau}\equiv\sigma_0\otimes\bm{\tau}$. The potentials $U_\perp$ and $U_z$ correspond to local interactions that act when two electrons are at the same position, and they act only in valley space thus favoring in-plane or out-of-plane pseudo-spin polarizations. The relative values of $\Delta_Z$ and $U_z$ and $U_\perp$ will thus determine the spin or pseudo-spin polarization of the ground state.

The first term in Eq. (\ref{eq:SB}) represents the electrons' interaction with in-plane phonons which create a Kekul\'e-like distortion\cite{Nomura2009} and is estimated to be of the order of $U_\perp \sim 2.0B[(T)]K$. The term $U_z$ originates from short-range Hubbard type interactions\cite{Alicea2006} and intervalley coupling due to the Coulomb interaction\cite{Goerbig2006}. Electron-phonon interactions with out-of-plane phonons also contribute to $U_z$, which is estimated to be of the order of $\sim 0.5B[(T)]K$. $\Delta_z=g\mu_BB$ corresponds to the Zeeman coupling and is of the order of $\sim 1.2B[(T)]K$. The energies $U_\perp$ and $U_z$ are proportional to the perpendicular magnetic field\cite{Li2019}, while $\Delta_z$ is proportional to the total magnetic field. Notice that these energy scales are all on the same order of magnitude and are likely to be strongly sample-dependent. We thus consider them, here, as tunable parameters that determine the phase diagram of the QHFM ground states as well as that of the skyrmions formed on top of these states. 

Applying the Hartree-Fock approximation, the energy of the anisotropic energy $E_A=\text{Tr}[H_AP]$ can be expressed as\cite{Kharitonov2012} 
\begin{align}
    E_A=N_\phi\left\{u_zt_z(P)+u_\perp\left[(t_x(P)+t_y(P)\right]-\Delta_z M_{S_z}\right\}
    \label{eq:AnisoEn}
\end{align}
with 
\begin{align}
    u_{z,\perp}=\frac{U_{z,\perp}}{2\pi l_B^2}
    \label{eq:aniso_orbital}
\end{align}
and 
\begin{align}
    t_i(P)&=\frac{1}{2}(\text{Tr}[\tau_iP]^2-\text{Tr}[(\tau_iP)^2]) \\
    &=\langle f_1|\tau_i|f_1\rangle\langle f_2|\tau_i|f_2\rangle-|\langle f_1|\tau_i|f_2\rangle|^2 \label{eq:tiP2}\\
    &=M^i_{Pf1}M^i_{Pf2}-|\langle f_1|\tau_i|f_2\rangle|^2,
    \label{eq:tiP}
\end{align}
where we have introduced the pseudo-spin magnetization of the spinors 
\begin{align}
    \mathbf{M_P}_{f_k}&=\langle f_k|\bm{\tau}|f_k\rangle
\end{align}
and also their spin magnetization 
\begin{align}
    \mathbf{M_S}_{f_k}&=\langle f_k|\bm{\sigma}|f_k\rangle.
\end{align}
The total spin and pseudo-spin magnetizations $\mathbf{M_S}$ and $\mathbf{M_P}$ are the sum of the magnetization of each spinor,
\begin{align}
    \mathbf{M_P}&=\text{Tr}[\bm{\tau}P]=\mathbf{M_P}_{f_1}+\mathbf{M_P}_{f_2} \\
    \mathbf{M_S}&=\text{Tr}[\bm{\sigma}P]=\mathbf{M_S}_{f_1}+\mathbf{M_S}_{f_2}.
\end{align}

\subsection{Parametrization of the spinors}

\label{sec:parametrization}

We have seen that the density matrix is described by 8 real parameters. Inspired by Refs. \onlinecite{Lian2017} and \onlinecite{Doucot2008}, we parametrize the spinors $|f_1\rangle$ and $|f_2\rangle$ using a Schmidt decomposition as 
\begin{align}
    |f_1\rangle&=\cos\frac{\alpha_1}{2}|\mathbf{n}\rangle|\mathbf{s}\rangle+e^{i\beta_1}\sin\frac{\alpha_1}{2}|-\mathbf{n}\rangle|-\mathbf{s}\rangle \label{eq:Schmidt1} \\
    |f_2\rangle&=\cos\frac{\alpha_2}{2}|\mathbf{n}\rangle|-\mathbf{s}\rangle+e^{i\beta_2}\sin\frac{\alpha_2}{2}|-\mathbf{n}\rangle|\mathbf{s}\rangle.
    \label{eq:Schmidt2}
\end{align}
where $|\mathbf{n}\rangle|\mathbf{s}\rangle\equiv|\mathbf{n}\rangle\otimes|\mathbf{s}\rangle$ with $|\mathbf{n}\rangle$ and $|\mathbf{s}\rangle$ the SU(2) spinors acting in pseudo-spin and spin space respectively
\begin{align}
|\mathbf{n}\rangle&=\begin{pmatrix}
    \cos\frac{\theta_P}{2} \\ \sin\frac{\theta_P}{2}e^{i\varphi_P}
    \end{pmatrix}, 
    \label{eq:param2} \\
    |\mathbf{s}\rangle&=\begin{pmatrix}
    \cos\frac{\theta_S}{2} \\ \sin\frac{\theta_S}{2}e^{i\varphi_S}
    \end{pmatrix} \label{eq:param1} 
\end{align}
such that $|f_1\rangle$ and $|f_2\rangle$ are four-components spinors in the basis $(K\uparrow,K\downarrow,K'\uparrow,K'\downarrow)$. Let us comment on some features of the decomposition (\ref{eq:Schmidt1}) and (\ref{eq:Schmidt2}). The spinors $|-\mathbf{s}\rangle$ and $|-\mathbf{n}\rangle$ are obtained from $|\mathbf{s}\rangle$ and $|\mathbf{n}\rangle$
by the replacement $\theta\rightarrow\pi-\theta$ and $\varphi\rightarrow\varphi+\pi$ such that we have $\langle \mathbf{s}|- \mathbf{s}\rangle = \langle \mathbf{n}|- \mathbf{n}\rangle=0$. We therefore notice that, for any choice of the parameters $\alpha_1,\alpha_2,\beta_1$, and $\beta_2$ (as well as the angles $\theta_{S/P}$ and $\phi_{S/P}$), the spinors $|f_1\rangle$ and $|f_2\rangle$ are orthogonal.
We thus obtain eight free parameters, in agreement with the counting based on our symmetry analysis in Sec. \ref{sec:Grassmannian}. Notice that the states at a filling factor $\nu=\pm 1$ are described in terms of a single spinor $|f\rangle$ and thus by only six free parameters \cite{Lian2016,Lian2017}. Finally, this parametrization includes the possibility of ``entanglement'' between the spin and the pseudo-spin. In fact, this decomposition of the spinors does not correspond to real entanglement between two particles because here it is the spin and pseudo-spin of the \textit{same} particle which is ``entangled''. However we will refer loosely to the angles $\alpha_1$ and $\alpha_2$ as entanglement angles for simplicity.

We have $\bm{\sigma}.\mathbf{s}|\pm\mathbf{s}\rangle=\pm|\pm\mathbf{s}\rangle$ and $\bm{\tau}.\mathbf{n}|\pm\mathbf{n}\rangle=\pm|\pm\mathbf{n}\rangle$, where
\begin{align}
\mathbf{s},\mathbf{n}=\begin{pmatrix}
\sin\theta_{S,P}\cos\varphi_{S,P}\\
\sin\theta_{S,P}\sin\varphi_{S,P} \\
\cos\theta_{S,P}
\end{pmatrix}
\end{align}
are the unit vectors on the spin and pseudo-spin Bloch spheres, respectively, with $\theta_S,\theta_P\in[0,\pi]$ and $\varphi_S,\varphi_P\in[0,2\pi]$. The angles $\alpha_1,\alpha_2\in[0,\pi]$ and $\beta_1,\beta_2\in[0,2\pi]$ are the angles of the entanglement Bloch spheres of the particles $1$ and $2$. Using this parametrization, the spin and pseudo-spin magnetizations of the spinors $f_1$ and $f_2$ are equal to 
\begin{align}
\mathbf{M_S}_{f_1}=\cos\alpha_1\mathbf{s}, \quad &
    \mathbf{M_S}_{f_2}=-\cos\alpha_2\mathbf{s},\\
    \mathbf{M_P}_{f_1}=\cos\alpha_1\mathbf{n}, \quad & \mathbf{M_P}_{f_2}=\cos\alpha_2\mathbf{n},
\end{align}
such that one finds 
\begin{align}
    \mathbf{M_S}&=(\cos\alpha_1-\cos\alpha_2)\mathbf{s}\\
    \mathbf{M_P}&=(\cos\alpha_1+\cos\alpha_2)\mathbf{n} 
\end{align}
for the total spin and pseudo-spin magnetizations, respectively.
We can see that the entanglement parameters $\alpha_1$ and $\alpha_2$ reduce the total spin and pseudo-spin magnetization. In the absence of entanglement ($\alpha_1,\alpha_2\in\{0,\pi\}$), the modulus of one of the magnetizations is equal to 2, i.e. maximal, and that for the other magnetization vanishes, in agreement with our above observation that one cannot obtain a full spin and full pseudo-spin magnetization at the same time. When $\alpha_1,\alpha_2\notin\{0,\pi\}$, the modulus of the magnetization is between 0 and 2 and this description is not valid anymore.

At filling $\nu=1$, there is only one sub-LL that is filled and thus only one spinor $|f\rangle$ and one entanglement parameter $\alpha$. In that case, the spin and pseudo-spin magnetization magnitudes are both proportional to $\cos\alpha$\cite{Lian2017} : $|\mathbf{M_S}|=|\mathbf{M_P}|=\cos\alpha$ such that there is no entanglement for $\alpha=0,\pi$ where spin and pseudo-spin magnetizations are maximal ($|\mathbf{M_S}|=|\mathbf{M_P}|=1$) and maximal entanglement for $\alpha=\pi/2$ where both spin and pseudo-spin magnetization vanish identically. At $\nu=0$, however, in the case of no entanglement ($\alpha_1,\alpha_2\in\{0,\pi\}$), because the two spinors must be orthogonal, both spin and pseudo-spin cannot be maximal at the same time leading to maximal pseudo-spin magnetization and vanishing spin polarization, or vice versa. 

\begin{figure}[t]
    (a)\includegraphics[width=7cm]{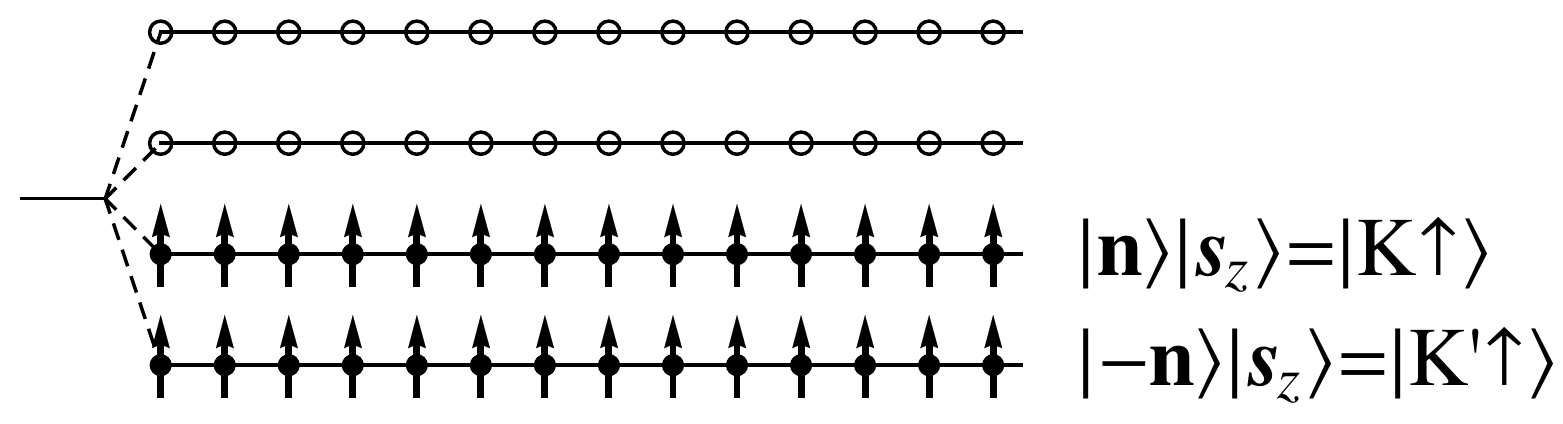}
	(b)\includegraphics[height=3cm]{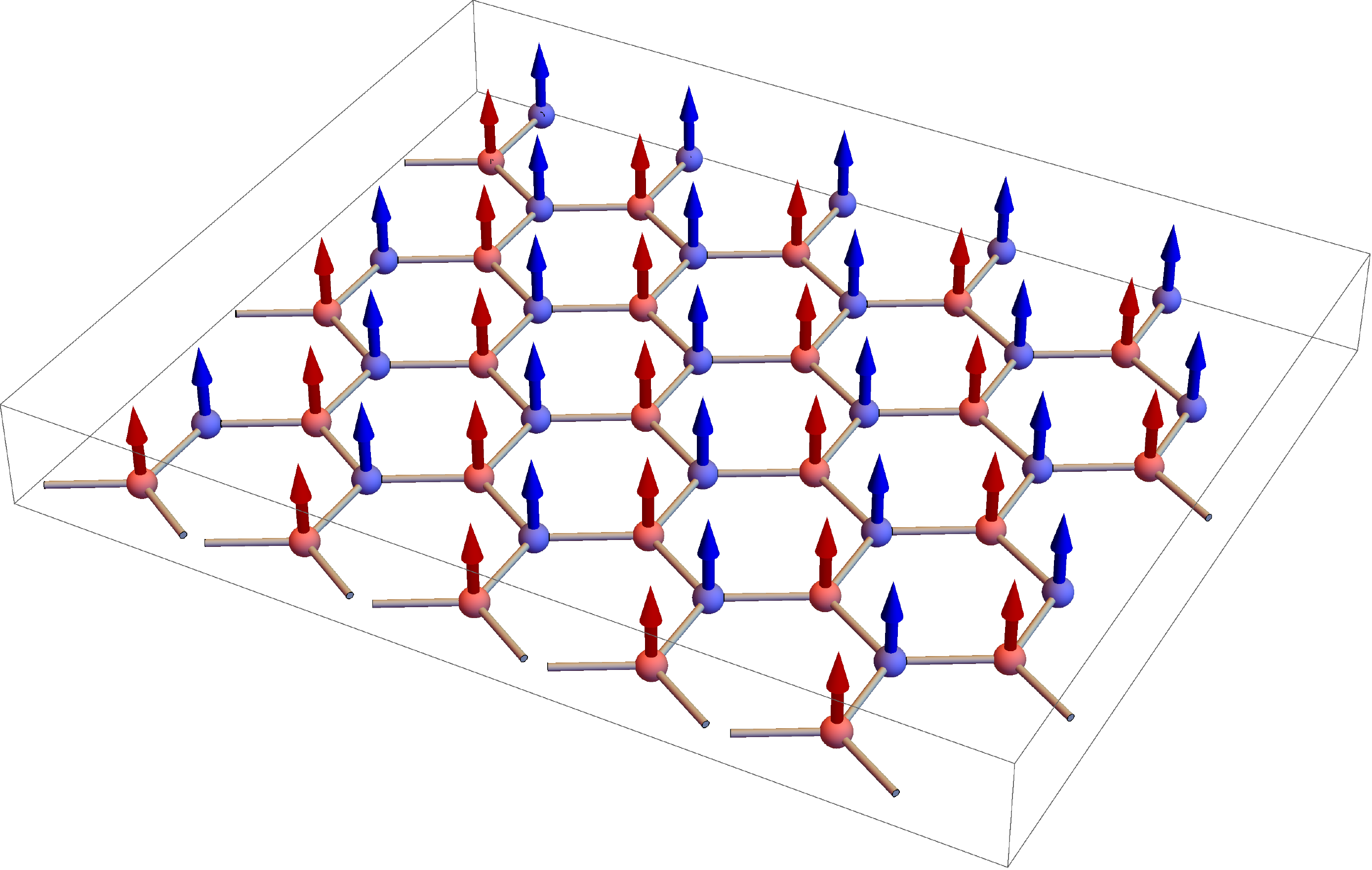}
	(c)\includegraphics[height=2.5cm]{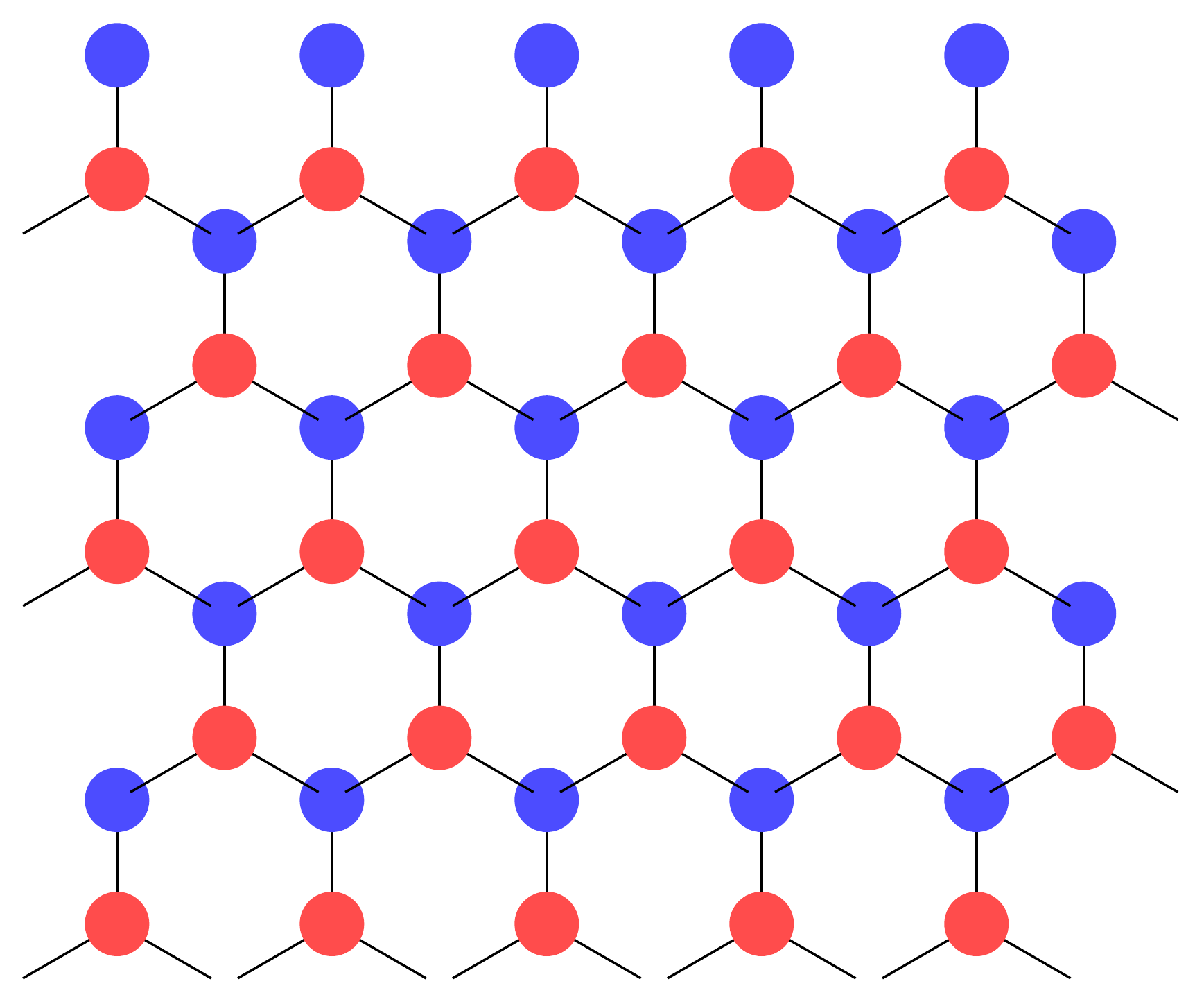}
	\caption{\textbf{Ferromagnetic phase}. (a) Filled sub-LLs with the corresponding spinors. The arrows correspond to the real spin polarization. (b) Spin magnetization and (c) electronic density on the A and B sublattices. All the spins point along the positive $z$ direction, while both sublattices are occupied equally.}
	\label{fig:FerroBG}
\end{figure}

In order to illustrate these facts and to understand in more detail this type of entanglement, let us consider two types of QHFM ground states that can be realized in general formalism of SU(4) ferromagnetism using this parametrization for the spinors : the (spin) ferromagnetic phase which is disentangled and the \textit{anti-ferromagnetic} phase which is maximally entangled.
The ferromagnetic phase is disentangled ($\alpha_1=0$ and $\alpha_2=\pi$) and is reached for $\theta_P=0$ and $\theta_S=0$ while the other angles remain free. The spinors have the expression 
\begin{align}
	|f_1\rangle&=|\mathbf{n}_z\rangle|\mathbf{s}_z\rangle=|\text{K}\uparrow\rangle, \\
	|f_2\rangle&=|-\mathbf{n}_z\rangle|\mathbf{s}_z\rangle=|\text{K'}\uparrow\rangle,
\end{align}
where both spins point towards the north pole of the spin Bloch sphere : $\mathbf{s}_z=(0,0,1)$ and the occupation of the spin-valley branches is depicted in Fig. \ref{fig:FerroBG}(a). The pseudo-spin points in opposite directions at the poles of the pseudo-spin Bloch sphere $\pm\mathbf{n}_z=(0,0,\pm1)$, such that each electron lives on one sublattice and has its spin pointing towards the direction of the magnetic field, which leads to a ferromagnetic phase. Figure \ref{fig:FerroBG} shows (b) the spin magnetization on the A and B sublattices and (c) the electronic density which is identical on both sublattices. Notice that this is a schematic illustration for the electrons of the $n=0$ LL, where each electron occupies a surface on the order of $l_B^2$. Each site is therefore only occupied by a small fraction of an electron. 

\begin{figure}[t]
    (a)\includegraphics[width=7cm]{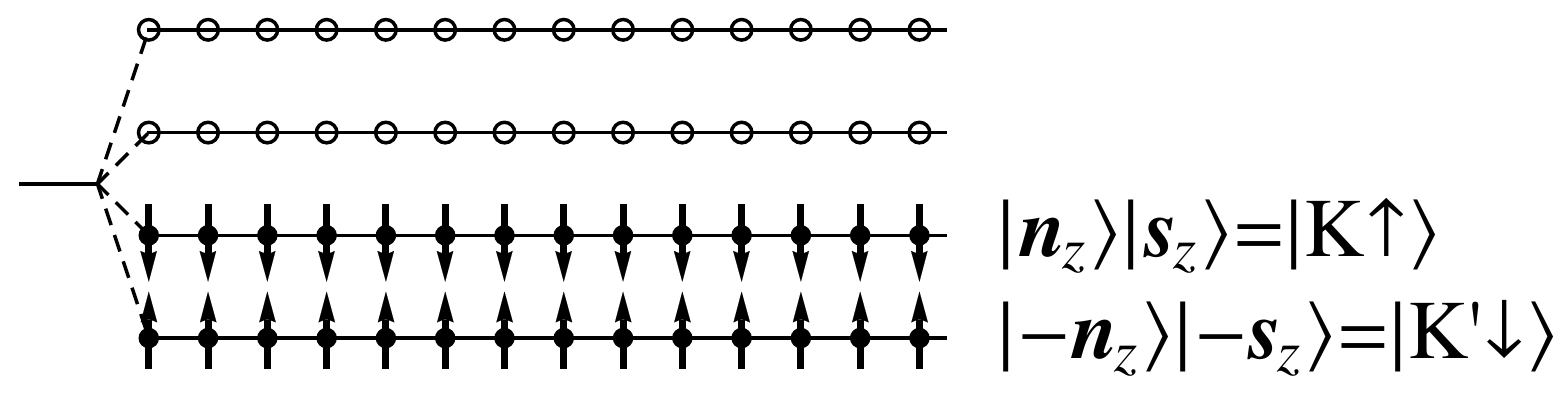}
	(b)\includegraphics[height=3cm]{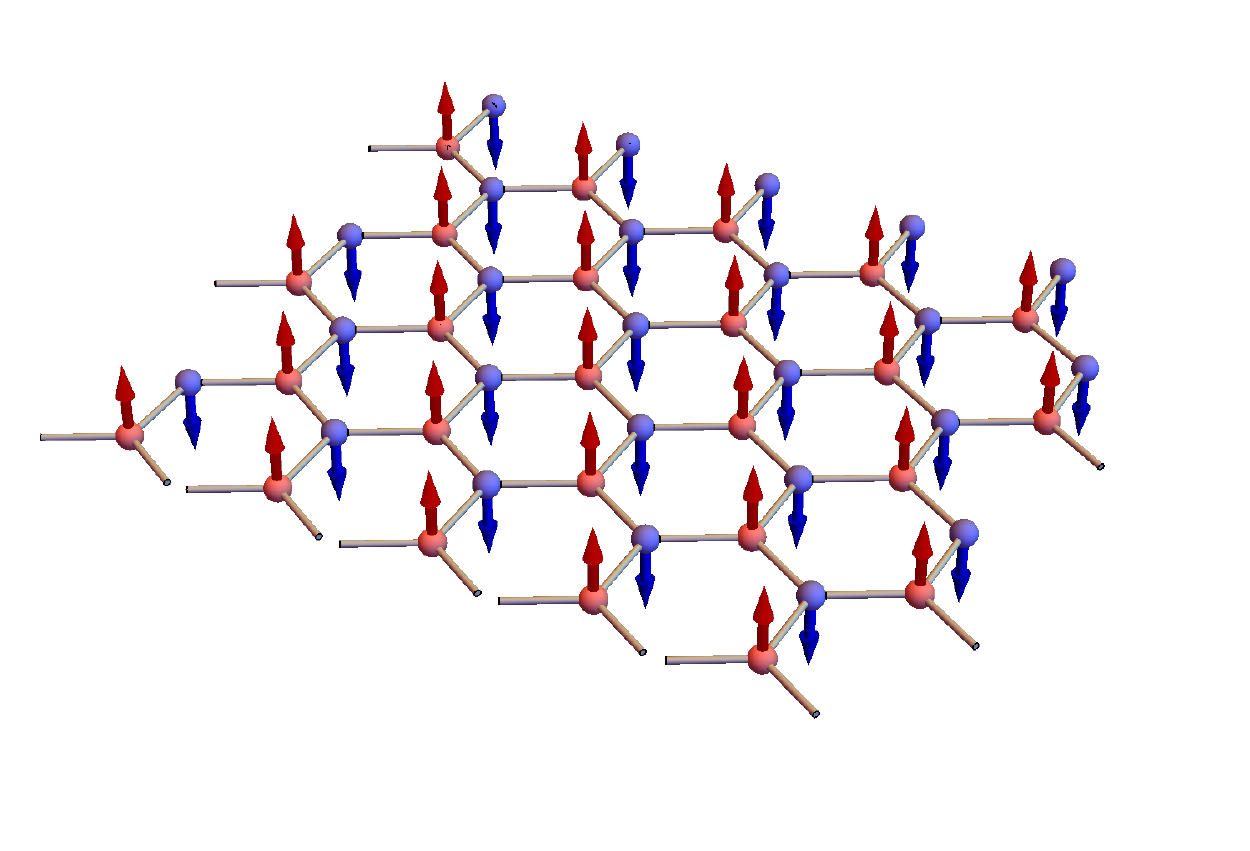}
	(c)\includegraphics[height=2.5cm]{background/DensityFerro.pdf}
	\caption{\textbf{Anti-ferromagnetic phase}. (a) Filled sub-LLs with the corresponding spinors. The arrows correspond to the real spin polarization. (b) Spin magnetization and (c) electronic density on the A and B sublattices. The spins on the sublattice A point along the positive $z$ direction, while the spins on the sublattice B point along the negative $z$ direction forming thus an anti-ferromagnetic pattern.}
	\label{fig:AFBG}
\end{figure}

In contrast to the ferromagnetic phase, the anti-ferromagnetic phase is maximally entangled ($\alpha_1=\alpha_2=\frac{\pi}{2}$) and is reached for $\theta_P=\pi/2$. After using the associated unitary rotation $U=\frac{1}{\sqrt{2}}\begin{pmatrix}1 & -1 \\ 1 & 1\end{pmatrix}$ given by Eq. (\ref{eq:unitary}) between the spinors, we obtain the state 
\begin{align}
	|f_1\rangle&=|\mathbf{n}_z\rangle|\mathbf{s}_z\rangle=|\text{K}\uparrow\rangle, \\
	|f_2\rangle&=|-\mathbf{n}_z\rangle|-\mathbf{s}_z\rangle=|\text{K'}\downarrow\rangle.
\end{align}
Figure \ref{fig:AFBG} shows the spin magnetization on the A and B sublattices. One notices that the pseudo-spin spinors point at opposite poles of the Bloch sphere, $|f_1\rangle$ to the north pole and $|f_2\rangle$ to the south pole, such that each electron is situated in a particular valley. The electrons described by $|f_1\rangle$ therefore occupy a single sublattice whereas those represented by $|f_2\rangle$ occupy the other sublattice. Since both spinors also have an opposite spin, one obtains an anti-ferromagnetic pattern, which is a remarkable consequence of SU(4) ferromagnetism, where \textit{one can turn a ferromagnetic into an anti-ferromagnetic phase simply by a unitary transformation}.

This phase is maximally entangled and has thus a total spin and pseudo-spin magnetization that are equal to zero. We can see that going from the ferromagnetic phase to the anti-ferromagnetic phase simply consists of changing the spin in one valley.

\subsection{Phase diagram of the ground state}

\label{sec:phaseDiagQHFM}

Using the Grassmannian description of the half-filled LL and the more general parametrization of the spinors described in Sec. \ref{sec:parametrization}, we recover Kharitonov's results\cite{Kharitonov2012} for the Quantum Hall ferromagnetic (QHFM) ground state. We review in this section the four phases in the presence of the symmetry breaking terms $u_\perp$ and $u_z$. In order to make connection with experimentally measurable quantities, we focus on the spin magnetization and the density of these states on the A and B sublattices,
\begin{align}
    \rho_{A,B}&=\frac{1}{2}\left(\langle f_1|(\tau_0\pm\tau_z)|f_1\rangle+\langle f_2|(\tau_0\pm\tau_z)|f_2\rangle\right) \\
    \mathbf{M_S}_{A,B}&=\frac{1}{2}\left(\langle f_1|\bm{\sigma}(\tau_0\pm\tau_z)|f_1\rangle+\langle f_2|\bm{\sigma}(\tau_0\pm\tau_z)|f_2\rangle\right)
\end{align}
which stems from the fact that in the $n=0$ LL in graphene we have the property that the sublattice index is locked to the valley index $(A,B)=(K,K')$.

\begin{figure}[h]
	\begin{center}
		\includegraphics[width=8cm]{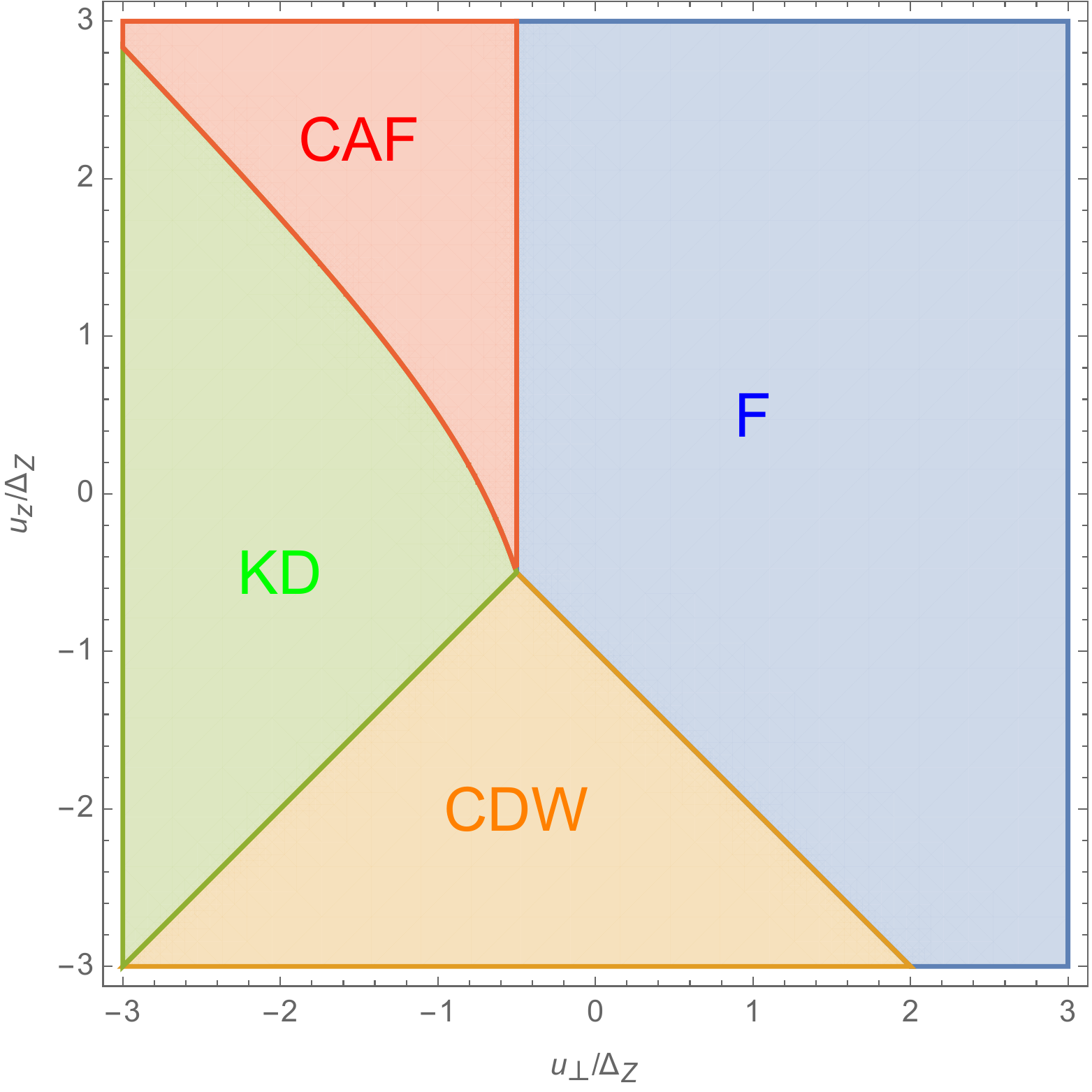}
	\end{center}
	\caption{Quantum Hall Ferromagnetism phase diagram of the half-filled LL in the presence of the symmetry breaking terms $u_\perp$ and $u_z$. The four phases are : ferromagnetic (F), charge density wave (CDW), Kekulé distortion (KD) and canted anti-ferromagnetic (CAF).}
	\label{fig:PhaseDiagQHFM}
\end{figure}

Using the parametrization given by Eqs. (\ref{eq:param1}) and (\ref{eq:param2}), we find that the anisotropy energy of the QHFM ground state given by Eq. (\ref{eq:AnisoEn}) has the expression 
\begin{align}
E_A=&\frac{1}{2}\cos\alpha_1\cos\alpha_2[(u_\perp-u_z)\sin^2\theta_P+2(u_\perp+u_z)] \nonumber \\
&+\frac{1}{2}(u_\perp-u_z)\sin^2\theta_P(\sin\alpha_1\sin\alpha_2\cos\varphi+1) \nonumber \\
&-u_\perp-\Delta_Z(\cos\alpha_1-\cos\alpha_2)\cos\theta_S
\end{align}
with $\varphi=\beta_1+\beta_2-2\varphi_P$. We minimize with respect to the angles to find the valid phases and the positivity of the eigenvalues of the Hessian matrix with respect to the angles give us the domain of validity of the different phases. We obtain the following four phases in agreement with Kharitonov's results : ferromagnetic (F), charge density wave (CDW), Kekul\'e distortion (KD) and canted anti-ferromagnetic (CAF) such that their domains of validity are shown in Fig. \ref{fig:PhaseDiagQHFM}. The F-CDW transition happens at $u_\perp+u_z=-\Delta_Z$, the KD-CDW transition happens at $u_\perp=u_z$, the CAF-F transition is located at $u_\perp=-\Delta_Z/2$ while the CAF-KD transition line is given by $u_\perp+u_z=\Delta_Z^2/2u_\perp$. The four transition lines meet at the point $(u_\perp,u_z)=(-\Delta_Z/2,-\Delta_Z/2)$.

Let us comment on the difference with the phase diagram at $\tilde{\nu}=1$ obtained in Ref. \onlinecite{Lian2017}. At $\tilde{\nu}=1$, there is only one electron per Landau site in the $n=0$ LL such that this phase described only by the spinor $|f_1\rangle$ given by Eq. (\ref{eq:Schmidt1}). In the absence of entanglement which tends to reduce the spin and pseudo-spin magnetization, the ground state is (spin) ferromagnetic. In addition, the pseudo-spin magnetization can lie at the pole or at the equator of the Bloch sphere depending on the relative value of $u_\perp$ and $u_z$. One obtains thus a charge density wave or a Kekul\'e distortion phase that come along with a ferromagnetic ordering. For example, in the CDW phase at $\nu=1$, all electrons are located on the same sublattice with the same spin orientation. In the $\nu=0$ case, this is not possible since the ground state is described by two spinors which are necessarily orthogonal, which means that if they are on the same sublattice, they must have opposite spin. More generally, they must be orthogonal either in spin or pseudo-spin space. Considering for example the ferromagnetic state with both valleys $K$ and $K'$ occupied : this excludes an additional ordering in the pseudo-spin degree of freedom that would require only the $K$ valley (CDW) or a superposition of $K$ and $K'$ (Kekul\'e) to be occupied. The ground state in absence of entanglement can thus be ferromagnetic, Kekul\'e or charge density wave, but not both of them at the same time. We obtain also an entangled phase at $\nu=0$, the canted-antiferromagnetic phase, for which  the spin and pseudo-spin magnetization lies between 0 and 2.

\subsubsection{Ferromagnetic phase}

We have already introduced the ferromagnetic phase in the previous section that is obtained for $\theta_s=\theta_P=0$ and no entanglement. However, we have seen that it is possible to find physically equivalent states by mixing the spinors with a unitary transformation given by Eq. (\ref{eq:unitary}) which amounts to rotating the pseudo-spin from the $z$ axis to an arbitrary direction $\mathbf{n}$ such that we find a more general expression for the states 
\begin{align}
	|f_1'\rangle&=|\mathbf{n}\rangle|\mathbf{s}_z\rangle \label{eq:FM1} \\
	|f_2'\rangle&=|-\mathbf{n}\rangle|\mathbf{s}_z\rangle. \label{eq:FM2}
\end{align}
We can see that the energy is minimized by aligning both spins in the same direction such that the pseudo-spins point in opposite direction. This direction for the pseudo-spins can be arbitrarily chosen -- the ferromagnetic phase remains the same -- such that this state has a remaining SU(2) rotation symmetry in the pseudo-spin space and a U(1) spin symmetry for rotations around $\mathbf{s}_z$. While the total pseudo-spin magnetization vanishes $\mathbf{M}_P=0$, the phase is ferromagnetic in spin space because both spins point in the same direction, and the spin magnetization equals $\mathbf{M_S}=2\mathbf{s}_z$. In the absence of the Zeeman term, both spins point along any direction of the spin Bloch sphere and the magnetization would be $\mathbf{M_S}=\mathbf{s}$. Because the two spinors of this state have opposite magnetization of  $\mathbf{M_P}_{f_1}=-\mathbf{M_P}_{f_2}=\mathbf{n}$, and the term proportional to $|\langle f_1|\bm{\tau}|f_2\rangle|^2$ in the anisotropic energy is always negative, this means that the three components of the vector $\mathbf{t}(P)$ given by Eq. (\ref{eq:tiP}) are negative, $\mathbf{t}(P)=(-1,-1,-1)$. This implies that the energy of this state reads
\begin{align}
    E_A^\text{F}=-N_\phi(u_z+2u_\perp+2\Delta_Z)
\end{align}
The opposite alignment of the pseudo-spin of the two spinors implies thus a negative cost in anisotropic energy and this state is thus generally favored in regions of positive $u_z$ or $u_\perp$. We can see that this phase is realized at the center of the phase diagram, namely when $u_\perp$ and $u_z$ are small compared to the Zeeman term $\Delta_Z$. One way to realize this phase is thus to change the magnitude of $u_\perp$ and $u_z$ relative to $\Delta_Z$ such that any point in the phase diagram gets closer to the center $(u_\perp,u_z)=(0,0)$. This can be achieved by tilting the magnetic field\cite{Kharitonov2012,Young2014}. In fact, orbital interaction effects such as $u_\perp$ and $u_z$ are proportional to the out-of-plane component of the magnetic field as can be seen in Eq. (\ref{eq:aniso_orbital}), while the Zeeman term is proportional to the total magnetic field. It has been shown experimentaly that such a spin polarized phase is not realized in the $n=0$ LL in graphene with a magnetic field perpendicular to the graphene plane\cite{Young2012}, however it has been realized in an experiment with a very high tilted magnetic fields\cite{Young2014}. Such a phase was also realized by screening Coulomb interactions using a high-dielectric substrate, which thus reduces $u_\perp$ and $u_z$ compared to $\Delta_Z$\cite{Veyrat2020}.

A salient feature of the ferromagnetic phase is the presence of two spin-polarized edge states that counter-propagate around the sample\cite{Abanin2006,Abanin2007,Young2014,Veyrat2020} analogously to the quantum spin Hall effect in topological insulators.

The spin magnetization and the electronic density on the A and B sublattices are shown in Fig. \ref{fig:FerroBG}.(b) and (c) respectively and have the expression 
\begin{align}
    \rho_A&=\rho_B=1 \\
    \mathbf{M_S}_A&=\mathbf{M_S}_B=\mathbf{s}_z.
\end{align}
Indeed, each atomic site or alternatively each valley is homogeneously occupied by spin-up particles.

\subsubsection{Charge density wave phase}

\begin{figure}[t]
    (a)\includegraphics[width=8cm]{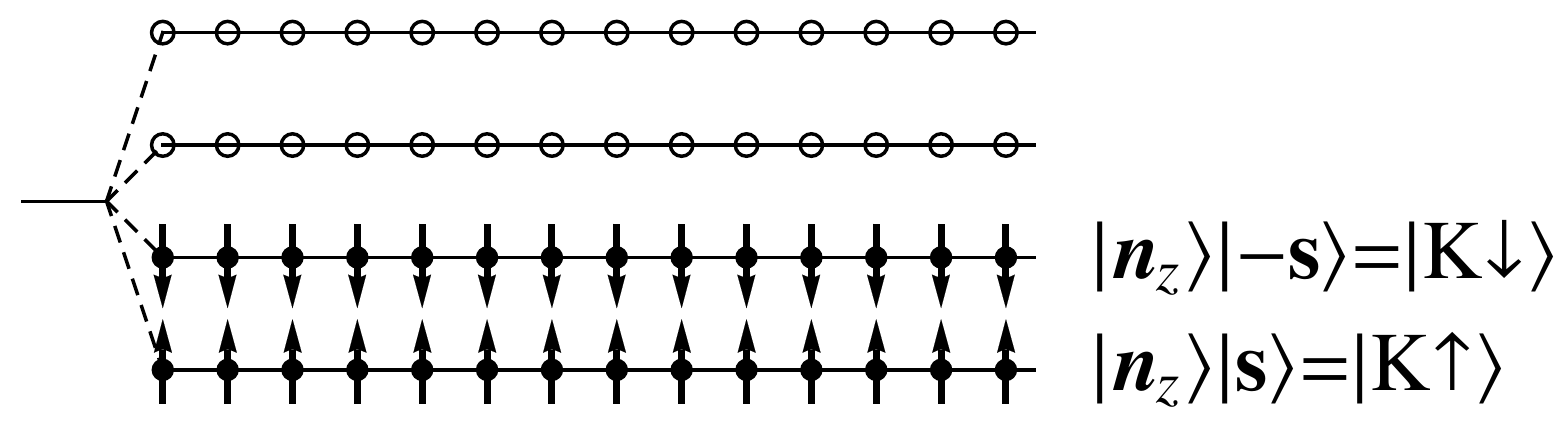} \\
	(b)\includegraphics[height=2.5cm]{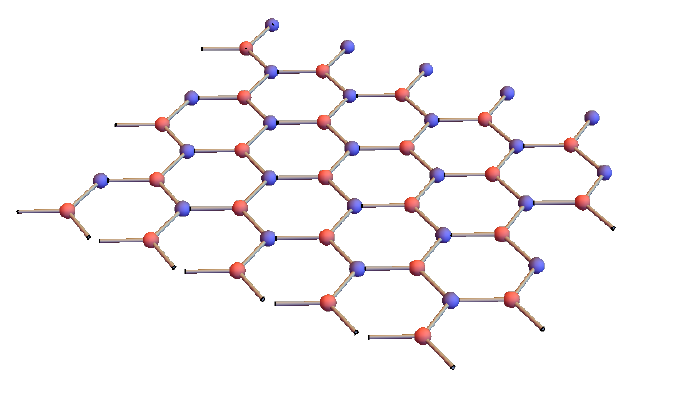}
	(c)\includegraphics[height=2.5cm]{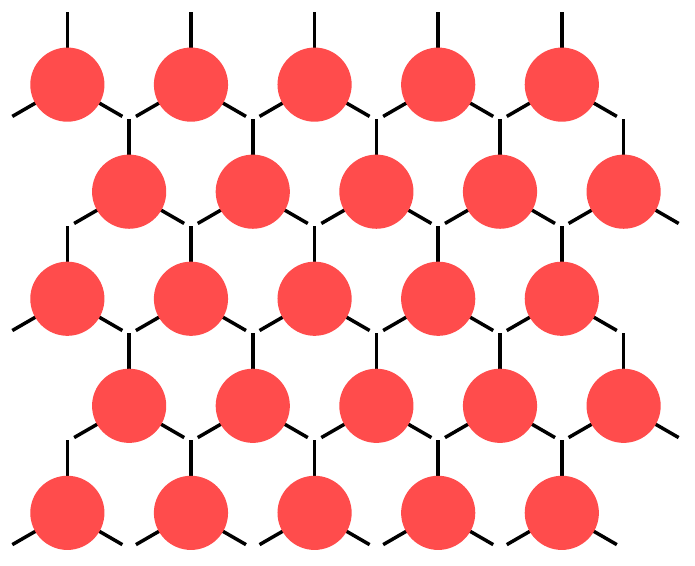}
	\caption{\textbf{Charge density wave phase}. (a) Filled sub-LLs with the corresponding spinors. The arrows correspond to the real spin polarization. (b) Spin magnetization and (c) electronic density on the A and B sublattices. The total spin vanishes equally, while only one sublattice is occupied.}
	\label{fig:CDWBG}
\end{figure}

The charge density wave (CDW) phase has $\alpha_1=\alpha_2=0$ and $\theta_P\in\{0,\pi\}$ and the spinors are 
\begin{align}
	|f_1\rangle&=|\pm\mathbf{n}_z\rangle|\mathbf{s}\rangle \\
	|f_2\rangle&=|\pm\mathbf{n}_z\rangle|-\mathbf{s}\rangle,
\end{align}
which is schematically represented in Fig. \ref{fig:CDWBG}(a). 
As opposed to the case of the ferromagnetic background for which the energy was minimized by aligning both spins, the energy of this state is minimized by aligning both pseudo-spins either along $z$ (for $|+\textbf{n}_z\rangle$) or $-z$ (for $|-\textbf{n}_z\rangle$). Since the Hamiltonian for the anisotropic terms (\ref{eq:SB}) depends only on the square of the $z$ component of the pseudo-spin, we encounter here a residual $Z_2$ symmetry associated with the orientation of the pseudo-spin. As for the physical spin, for both electrons they point in arbitrary but opposite directions. This state 
has zero spin magnetization, $\mathbf{M}_\mathbf{S}=0$ [see Fig. \ref{fig:CDWBG}(b)], and is thus insensitive to the Zeeman interaction.

Because both spinors point at the same pole of the pseudo-spin Bloch sphere, all electrons reside on a single sublattice which gives rise to a charge density wave pattern as shown in Fig. \ref{fig:CDWBG}(c). This above-mentioned $Z_2$ symmetry in pseudo-spin, which corresponds to the occupation of the A or B sublattice, is thus spontaneously broken down by the occupation of a single lattice along with an SU(2) rotation symmetry in spin space. The pseudo-spin magnetization of the two spinors is identical $\mathbf{M_P}_{f_1}=\mathbf{M_P}_{f_2}=\pm\mathbf{n}_z$, and we have $\mathbf{M_P}=\pm2\mathbf{n}_z$ for the total pseudo-spin polarization,
while the cross term $|\langle f_1|\bm{\tau}|f_2\rangle|^2$ vanishes because the spinors are orthogonal in spin space, which means that the vector $\mathbf{t}(P)$ has only a non-zero component along $z$ : $\mathbf{t}(P)=(0,0,1)$. Its energy equals therefore
\begin{align}
    E_A^\text{CDW}=N_\phi u_z.
\end{align}
Thus, aligning the spinors at the poles of the pseudo-spin Bloch sphere costs the energy $u_z$, and this state will thus be favored for large negative values of $u_z$. In fact, for negative $u_z$, the interaction between two neighboring electrons is attractive which favors electron occupation of the same sublattice.

The electronic density and spin magnetization shown in  Figs. \ref{fig:CDWBG}(b) and (c) are given by
\begin{align}
    \rho_A&=2,\quad \rho_B=0, \\
    \mathbf{M_S}_A&=\mathbf{M_S}_B=0.
\end{align}

\subsubsection{Kekul\'e distortion phase}

\begin{figure}[t]
    (a)\includegraphics[width=8cm]{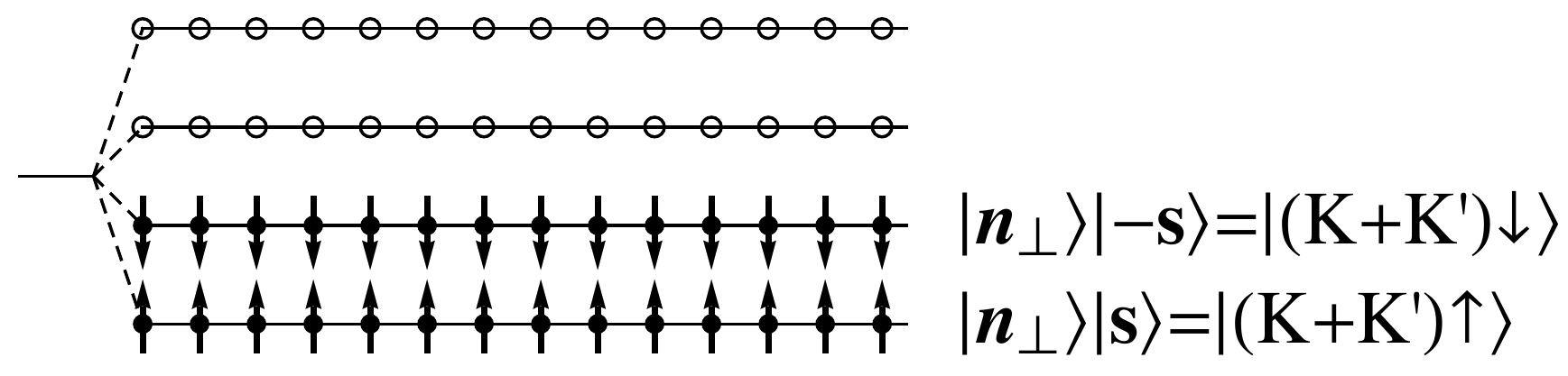} \\
	(b)\includegraphics[height=2.5cm]{background/SpinCDW.pdf}
	(c)\includegraphics[height=2.5cm]{background/DensityFerro.pdf}
	\caption{\textbf{Kekul\'e distortion phase}. (a) Filled sub-LLs with the corresponding spinors. The arrows correspond to the real spin polarization which have different orientation. The notation $(K+K')$ indicates that the state is a superposition of the K and K' valleys with relative phase $e^{i\varphi_P}$. (b) Spin magnetization and (c) electronic density on the A and B sublattices. The total spin vanishes, while both sublattices are occupied equally.}
	\label{fig:KDBG}
\end{figure}

The Kekulé distortion (KD) phase has $\alpha_1=\alpha_2=0$ and $\theta_P=\pi/2$ and the spinors are 
\begin{align}
	|f_1\rangle&=|\mathbf{n}_\perp\rangle|\mathbf{s}\rangle \\
	|f_2\rangle&=|\mathbf{n}_\perp\rangle|-\mathbf{s}\rangle
\end{align}
and depicted in Fig. \ref{fig:KDBG}(a). 
This state is similar to the CDW state except for the fact that the pseudo-spin points in the $xy$ plane of the Bloch sphere $\mathbf{n}_\perp=(\cos\varphi_P,\sin\varphi_P,0)$ and thus the occupation of the A and B sublattices is equal as can be seen in Fig. \ref{fig:KDBG}(c). The superposition of the electrons spinors over the two valleys creates a KD pattern which enlarges the elementary unit cell by a factor of 3\cite{Hou2007}. Such an enlargement of the unit cell was indeed observed experimentally in the $n=0$ LL at $\nu=0$ by Li \textit{et al.}\cite{Li2019} in a STM measurement.

This state has a SU(2) spin symmetry and a U(1) symmetry for the pseudo-spin orientation in the plane. The pseudo-spin magnetization equals $\mathbf{M_P}=2\mathbf{n}_\perp$,
while the total spin also vanishes $\mathbf{M_S}=0$ as in the CDW case [see Fig. \ref{fig:KDBG}(b)]. The pseudo-spin magnetization is identical for both spinors $\mathbf{M_P}_{f_1}=\mathbf{M_P}_{f_2}=\pm\mathbf{n}_\perp$, while the cross term $|\langle f_1|\bm{\tau}|f_2\rangle|^2$ is also zero. The vector $\mathbf{t}(P)$ has thus the expression : $\mathbf{t}(P)=(\cos^2\varphi_P,\sin^2\varphi_P,0)$  The energy of this state is given by
\begin{align}
    E_A^\text{KD}=N_\phi u_\perp,
\end{align}
and it is thus realized for negative values of $u_\perp$. One thus easily understands the transition line between the CDW and the KD phase which is at $u_\perp=u_z<-|\Delta_Z|/2$, where the pseudo-spin SU(2) symmetry is restored, as one can also immediately see from Eq. (\ref{eq:SB}). The spin magnetization and electronic density are shown in Fig. \ref{fig:KDBG} and are equal to 
\begin{align}
    \rho_A&=\rho_B=1, \\
    \mathbf{M_S}_A&=\mathbf{M_S}_B=0
\end{align}

\subsubsection{Canted anti-ferromagnetic phase}

\begin{figure}[t]
    (a)\includegraphics[width=7cm]{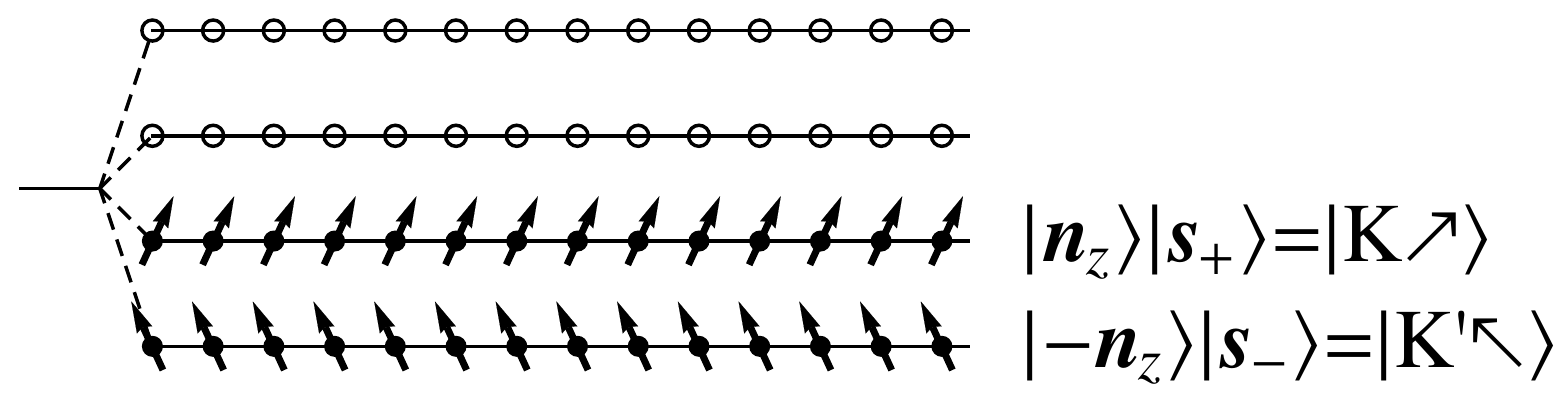}
	(b)\includegraphics[height=2.5cm]{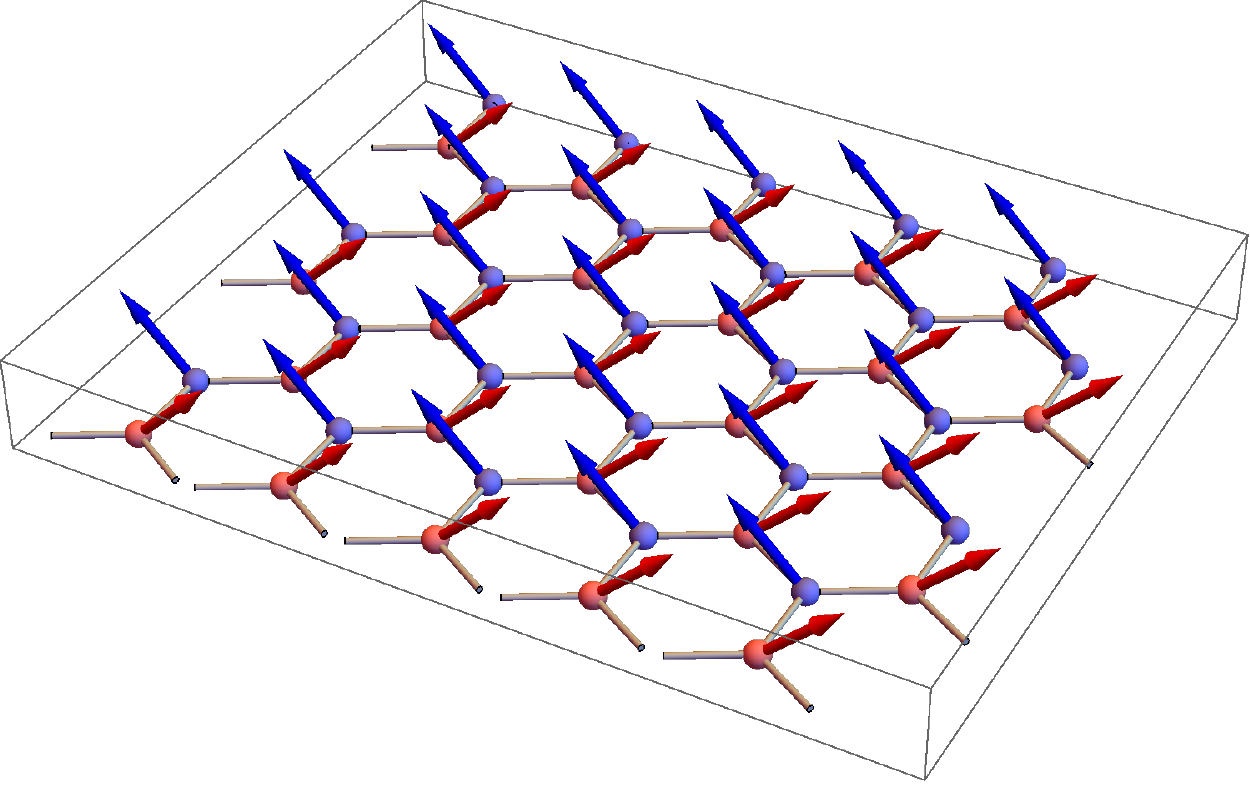}
	(c)\includegraphics[height=2.5cm]{background/DensityFerro.pdf}
	\caption{\textbf{Canted anti-ferromagnetic phase}. (a) Filled sub-LLs with the corresponding spinors. The arrows correspond to the real spin polarization. (b) Spin magnetization and (c) electronic density on the A and B sublattices. The total spin vanishes has opposite canting angle on the A and B sublattices.}
	\label{fig:CAFBG}
\end{figure}

The canted-antiferromagnetic (CAF) phase is reached for $\alpha_2=\pi-\alpha_1$, $\theta_P=\pi/2$, $\cos\varphi=1$ and $\theta_S=0$. The spinors are 
\begin{align}
	|f_1\rangle&=\cos\left(\frac{\alpha}{2}\right)|\mathbf{n}_\perp\rangle|\mathbf{s}_z\rangle+e^{i\beta_1}\sin\left(\frac{\alpha}{2}\right)|-\mathbf{n}_\perp\rangle|-\mathbf{s}_z\rangle \label{eq:CAF1} \\
	|f_2\rangle&=\sin\left(\frac{\alpha}{2}\right)|\mathbf{n}_\perp\rangle|-\mathbf{s}_z\rangle+e^{i(2\phi_P-\beta_1)}\cos\left(\frac{\alpha}{2}\right)|-\mathbf{n}_\perp\rangle|\mathbf{s}_z\rangle \label{eq:CAF2}
\end{align}
with $\alpha\equiv\alpha_1$ and $\beta\equiv\beta_1$ and 
\begin{equation}
    \cos\alpha=-\frac{\Delta_Z}{2u_\perp}.
    \label{eq:alphaCAF}
\end{equation}
We can see that this phase presents a non-zero entanglement as discussed in Sec. \ref{sec:parametrization}. After rearranging, and operating an SU(2) transformation among the spinors, we obtain Kharitonov's expression for the canted-antiferromagnetic spinors,
\begin{align}
	|f_1'\rangle&=|\mathbf{n}_z\rangle|\mathbf{s}_-\rangle \label{eq:spCAF1}\\
	|f_2'\rangle&=|-\mathbf{n}_z\rangle|\mathbf{s}_+\rangle ,\label{eq:spCAF2}
\end{align}
where 
\begin{align}
    |\mathbf{s}_+\rangle&=\begin{pmatrix}\cos\frac{\alpha}{2} \\\sin\frac{\alpha}{2}e^{i(\beta-\varphi_P)}
    \end{pmatrix}, \\
    |\mathbf{s}_-\rangle&=\begin{pmatrix}\cos\frac{\alpha}{2} \\ -\sin\frac{\alpha}{2}e^{i(\beta-\varphi_P)}\end{pmatrix}.
\end{align}
Each spinor, represented schematically in Fig. \ref{fig:CAFBG}(a),
corresponds to a sublattice and the electrons on each sublattice point in different directions forming a canted anti-ferromagnetic pattern [see Fig. \ref{fig:CAFBG}(b)]. This phase has a vanishing pseudo-spin magnetization $\mathbf{M_P}=0$ and a reduced spin magnetization $\mathbf{M_S}=2\cos\alpha\mathbf{s}_z$. At the border with the ferromagnetic phase, for $u_\perp=-\Delta_Z/2$, the canting angle is equal to $\alpha=0$ which corresponds to the ferromagnetic phase, and entanglement ($\alpha\neq 0$) builds up continuously when entering the CAF phase. 
The transition between the CAF and the F states is therefore a second-order transition. As $u_\perp/\Delta_Z$ increases, the canting angle increases and reaches $\pi/2$ at infinity, forming thus an anti-ferromagnetic phase which is maximally entangled with $\mathbf{M_P}=\mathbf{M_S}=0$.  

The vector $\mathbf{t}(P)$ has thus the expression $\mathbf{t}(P)=-(\cos^2\alpha,\cos^2\alpha,1)$ whereas for the ferromagnetic case we had $\mathbf{t}(P)=-(1,1,1)$. We can thus see that "entanglement" reduces the anisotropic energy. The energy of the canted-anti ferromagnetic phase is thereby 
\begin{align}
    E^\text{CAF}_A&=-N_\phi(\cos^2\alpha u_\perp+u_z+2\Delta_Z\cos\alpha) \\
    &=N_\phi\left(-u_z+\frac{\Delta_Z^2}{2u_\perp}\right).
\end{align}
This phase is thus favored for negative values of $u_\perp$ compared to the ferromagnetic phase. 

The spin magnetization and the electronic density on the A and B sublattices are shown on Fig. \ref{fig:CAFBG}. In Kharitonov's expression for the spinors, because the spinors point at the poles of the pseudo-spin Bloch sphere, each spinor corresponds to a distinct sublattice and the spin magnetization equals :
\begin{align}
    \mathbf{M_S}_A=\begin{pmatrix}
    -\sin\alpha\cos\beta \\
    -\sin\alpha\sin\beta\\
    \cos\alpha
    \end{pmatrix}&, \quad
    \mathbf{M_S}_B=\begin{pmatrix}
    \sin\alpha\cos\beta \\
    \sin\alpha\sin\beta \\
    \cos\alpha
    \end{pmatrix} \\
    \rho_A&=\rho_B=1
\end{align}
[see Fig. \ref{fig:CAFBG}(c)],
where we have set $\varphi_P=0$ for simplicity. The spins on the A and B sublattices have opposite orientations in the $xy$ plane while they have the same magnitude along the $z$ direction. Experimental evidence have shown that this state may be realized in graphene at $\nu=0$\cite{Young2014}. 

When the Zeeman coupling is negligible compared to the valley anisotropic energies, we have $\cos\alpha\sim0$ and this state is nearly anti-ferromagnetic with the spin oriented in the plane. In the limit $\Delta_Z=0$, the CAF state becomes completely anti-ferromagnetic and the SU(2) spin symmetry is restored. In that case, the spinors are :
\begin{align}
	|f_1\rangle&=|\mathbf{n}_z\rangle|\mathbf{s}\rangle \\
	|f_2\rangle&=|-\mathbf{n}_z\rangle|-\mathbf{s}\rangle,
\end{align}
and the transition between the F state and AF occurs at $u_\perp=0$.
 
\subsection{Phase transitions}

At the transitions between the different phases, we can observe some symmetry restoration or a continuous phase transition. The simplest phase transition is the transition between the F and CAF phases which is of second order, as already mentioned above, because the spin magnetization continuously interpolates from full polarization along the $z$ direction towards a progressive canting of the spins. At the KD-CDW and F-CDW transitions, we find two different SU(2) symmetry restorations, whereas at the CAF-KD transition, there is no symmetry restoration nor continuous parameter. In this section, we describe the symmetries at the KD-CDW and F-CDW transitions.

\subsubsection{KD-CDW transition}

\label{sec:symmetry_uz=up}

At the transition line between the CDW and KD phase located at $u_z=u_\perp$, the anisotropic Hamiltonian reads
\begin{align}
    H_A=\int d^2r\left[U_\perp \mathbf{P}^2(\mathbf{r})-\Delta_ZS_z(\mathbf{r})\right]
\end{align}
with $\mathbf{P}^2=P^2_x+P^2_y+P^2_z$. The Hamiltonian commutes thus with $P_x$, $P_y$ and $P_z$ such that the full SU(2) pseudo-spin symmetry is restored. This is reminiscent of the case at $\nu=\pm 1$ \cite{Lian2016,Lian2017}, with the difference that there the system is ferromagnetic also in the spin
sector. Here, however, the spin magnetization vanishes in both the CDW and the KD phases such that,
at this line. the spinors are invariant under SU(2)$\times$SU(2) spin and valley rotations and read 
\begin{align}
	|f_1\rangle&=|\mathbf{n}\rangle|\mathbf{s}\rangle \\
	|f_2\rangle&=|\mathbf{n}\rangle|-\mathbf{s}\rangle.
\end{align}

\subsubsection{F-CDW transition}

\label{sec:F-CDW}

At the transition line between the F and CDW phases located at $u_\perp+u_z=-\Delta_Z$, the energy of the ground state is continuous and equals $E_{\text{F-CDW}}=N_\phi u_z$. However, in this situation, it is slightly more complicated to unveil the symmetry restoration at the transition, which is given by a rotation that involves both the spin and the pseudo-spin degrees of freedom and thus entanglement. In order to appreciate this symmetry restoration, consider the spinors $|f_1\rangle$ at each side of the transition 
\begin{align}
    &|f_1\rangle_\text{F}=|\mathbf{n}\rangle|\mathbf{s}_z\rangle=(\cos\frac{\theta}{2}, 0,\sin\frac{\theta}{2}e^{i\varphi},0), \\
    &|f_1\rangle_\text{CDW}=|\mathbf{n}_z\rangle|\mathbf{s}\rangle=(\cos\frac{\theta}{2}, \sin\frac{\theta}{2}e^{i\varphi},0,0),
\end{align}
in the basis $(K\uparrow,K\downarrow,K'\uparrow,K'\downarrow)$, where we observe a duality transformation that exchanges the spin and pseudo-spin at the transition. A similar observation can be made for the spinors
\begin{align}
    &|f_2\rangle_\text{F}=|-\mathbf{n}\rangle|\mathbf{s}_z\rangle, \\
    &|f_2\rangle_\text{CDW}=|\mathbf{n}_z\rangle|-\mathbf{s}\rangle
\end{align}
on each side of the transition. As a consequence, the matrices 
\begin{align}
    \gamma_\mu=\begin{pmatrix}1&& \\ &s_\mu & \\ &&1\end{pmatrix},
    \label{eq:generatorsF-CDW}
\end{align}
where $s_\mu$ are the three Pauli matrices, form an su(2) subalgebra of su(4) and generate the rotations that transform precisely the F spinors to the CDW spinors. Consider the associated rotation matrix 
\begin{align}
    R=\begin{pmatrix}1&0&0&0 \\ 0&\cos\alpha&\sin\alpha e^{-i\beta}&0 \\ 0&\sin\alpha e^{i\beta}&-\cos\alpha &0 \\ 0&0&0&1\end{pmatrix},
    \label{eq:R}
\end{align}
which we apply 
to the projector, and then compute the energy of this state at the transition. The projector is transformed as $P'=RPR^\dagger$, and we find that the vector $\mathbf{t}(P')$ and the Zeeman term have the expression 
\begin{align}
    \mathbf{t}(P')&=(-\cos^2\alpha,-\cos^2\alpha,-\cos2\alpha) \\
    \text{Tr}[\sigma_zP']&=2\cos^2\alpha.
\end{align}
At the transition, the energy of the state 
\begin{align}
   E/N_\phi&=u_\perp(t_x(P')+t_y(P')+\text{Tr}[\sigma_z P']) \nonumber \\
   &+u_z(t_y(P')+\text{Tr}[\sigma_zP']) \\
   &= u_z
\end{align}
is thus independent of the angles $\alpha$ and $\beta$, which determine the transformation.
The energy of the state therefore remains unchanged at the transition upon mixing of the levels $|K\downarrow\rangle$ and $|K'\uparrow\rangle$ and is thus invariant under the  SU(2) subgroup generated by the matrices (\ref{eq:generatorsF-CDW}).

\section{Skyrmions}

\label{sec:skyrmions}

Now that we have identified the different QHFM phases as a function of the parameters $\Delta_Z$, $u_\perp$ and $u_z$, let us discuss the possible skyrmions which are hosted by these types of QHFMs. Quite generally, in QHFM systems, an additional charge -- be it an electron or a hole -- can be dressed by a spin, or in our case by a spin-valley texture to minimize the exchange energy. This texture is precisely the skyrmion. It is localized at a certain position and retrieves, far away from its center, the lowest-energy ferromagnetic background, which we have identified in the previous section. Notice that these QHFM backgrounds do not fully constrain the type of skyrmion that we may encounter, namely the spin-valley polarization at the skyrmion center, which, as we discuss below, is described by spinors that are orthogonal to those representing the background. The aim of the present section is therefore to characterize the different skyrmions, compatible with the backgrounds, as a function of the same parameters $\Delta_Z$, $u_\perp$ and $u_z$ and to obtain the relevant phase diagram. Notice furthermore that these symmetry-breaking terms also determine the skyrmion size -- while the skyrmions are scale-invariant in the SU(4) limit described by the leading non-linear sigma term in the Hamiltonian, the symmetry-breaking terms in Eq. (\ref{eq:SB}) have a tendency to form skyrmions of small size. This tendency is however balanced by a Coulomb interaction that arises from higher gradient terms in the non-linear sigma model, as we discuss below.

\subsection{Non-linear sigma model}

The electrons in the half-filled LL of graphene are described by a Grassmannian field $Z(\mathbf{r})$, the generalization to a position-dependent field of the Grassmannian describing QHFM introduced in Sec. \ref{sec:Grassmannian},
\begin{align}
Z(\mathbf{r})=(Z_1,Z_2)=\begin{pmatrix}
Z_{11}&Z_{12}\\
Z_{21}&Z_{22}\\
Z_{31}&Z_{32} \\
Z_{41}&Z_{42}
\end{pmatrix}=(Z_{\alpha n})
\end{align}
where the spinors $Z_1$ and $Z_2$ are the spinors describing the two electrons with components $Z_k=(Z_{\alpha k})$. Because the two electrons are indistinguishable, the Grassmannian field is invariant under local SU(2) transformations 
\begin{align}
    Z'(\mathbf{r})=Z(\mathbf{r})U(\mathbf{r}).
    \label{eq:SU2}
\end{align}
This field is subject to the normalisation condition 
\begin{align}
    Z^\dagger(\mathbf{r}) Z(\mathbf{r})=\mathbb{1}
    \label{eq:norm}
\end{align}
at every point such that $Z_1$ and $Z_2$ are two normalized and orthogonal fields at any position $\mathbf{r}$. The order parameter has the expression 
\begin{align}
    P(\mathbf{r})=Z(\mathbf{r})Z^\dagger(\mathbf{r})=Z_1(\mathbf{r})Z_1^\dagger(\mathbf{r})+Z_2(\mathbf{r})Z^\dagger_2(\mathbf{r}),
\end{align}
which remains invariant under transformations given by Eq. (\ref{eq:SU2}). The non-linear sigma model (NLSM) energy is given by\cite{Yang2006,Arovas1999,Kharitonov2012} 
\begin{align}
    E_\text{NLSM}=\rho_s\int d^2r\text{Tr}\left[\bm{\nabla}P\bm{\nabla}P\right],
\end{align}
in terms of the spin stiffness 
\begin{equation}
\rho_s=\frac{1}{16\sqrt{2\pi}}\frac{e^2}{\varepsilon l_B} .
\end{equation}
When expressed as a function of the Grassmannian field, the energy of the skyrmion is given by 
\begin{align}
    E_\text{NLSM}&=2\rho_s\int d^2r\text{Tr}[(\mathbf{D}Z)^\dagger\mathbf{D}Z] \\
    &=2\rho_s\int d^2r\left[(\mathbf{D}Z_1)^\dagger\mathbf{D}Z_1+(\mathbf{D}Z_2)^\dagger\mathbf{D}Z_2
    \right],
    \label{eq:NLSM}
\end{align}
where we have introduced the covariant derivative 
\begin{equation}
    D_\mu Z=\partial_\mu Z-A_\mu Z=(1-ZZ^\dagger)\partial_\mu Z, 
\end{equation}
with the connection defined as 
\begin{align}
    A_\mu&=Z^\dagger\partial_\mu Z.
\end{align}
The topological charge density is equal to\cite{Yang2006,Arovas1999} 
\begin{align}
    \rho_\text{topo}(\mathbf{r})&=\frac{1}{2\pi i}\varepsilon_{\mu\nu}\text{Tr}[P\partial_\mu P\partial_\nu P], \\
    &=\frac{1}{2\pi i}\varepsilon_{\mu\nu}\text{Tr}[(\mathbf{D}_\mu Z)^\dagger (\mathbf{D}_\nu Z)],
\end{align}
while the total topological charge is 
\begin{align}
    Q=\int d^2r \rho_\text{topo}(\mathbf{r}).
\end{align}

In the next order in the gradient expansion of the NLSM\cite{Moon1995}, on obtains the expression for the Coulomb interaction of the skyrmion 
\begin{align}
    E_C=\frac{1}{2}\int d^2r\rho_\text{topo}(\mathbf{r})V(\mathbf{r}-\mathbf{r}')\rho_\text{topo}(\mathbf{r}'),
    \label{eq:Coulomb_en}
\end{align}
which depends only on the size of the skyrmion as we will see later.

Up to now, in the absence of the anisotropy energy, the model is approximately SU(4) symmetric. The quantum Hall ferromagnetic background breaks this symmetry down to U(4)/U(2)$\times$U(2) and the presence of a skyrmion breaks this symmetry even further with no preferred orientation of the spinors. In order to find which skyrmion is realized, we introduce the same anisotropic energies as for the ferromagnetic background.
The energy of the skyrmion originating from the anisotropic energy is given by 
\begin{align}
	E_A[Z]=\int \frac{d^2r}{2\pi l_B^2}&\left\{u_zt_z(P)+u_\perp\left[t_x(P)+t_y(P)\right]\right. \nonumber \\
	&\left.-\Delta_ZM_{S_z}\right\}-E_A[f],
	\label{eq:aniso}
\end{align}
with $t_i(P)$ given by 
\begin{align}
    t_i(P)=\langle Z_1|\tau_i|Z_1\rangle\langle Z_2|\tau_i|Z_2\rangle-|\langle Z_1|\tau_i|Z_2\rangle|^2,
\end{align}
in agreement with Eq. (\ref{eq:tiP2} and where we have subtracted the anisotropy energy $E_A[f]$ originating from the background in order to obtain the excess energy of the skyrmion.

The total energy of the skyrmion is thus the sum of the NLSM energy, the anisotropic energy and the Coulomb energy,
\begin{align}
    E_\text{sk}=E_\text{NLSM}+E_A+E_C.
\end{align}

\subsection{Solution for the skyrmion}

\begin{figure}[h]
	\includegraphics[width=8cm]{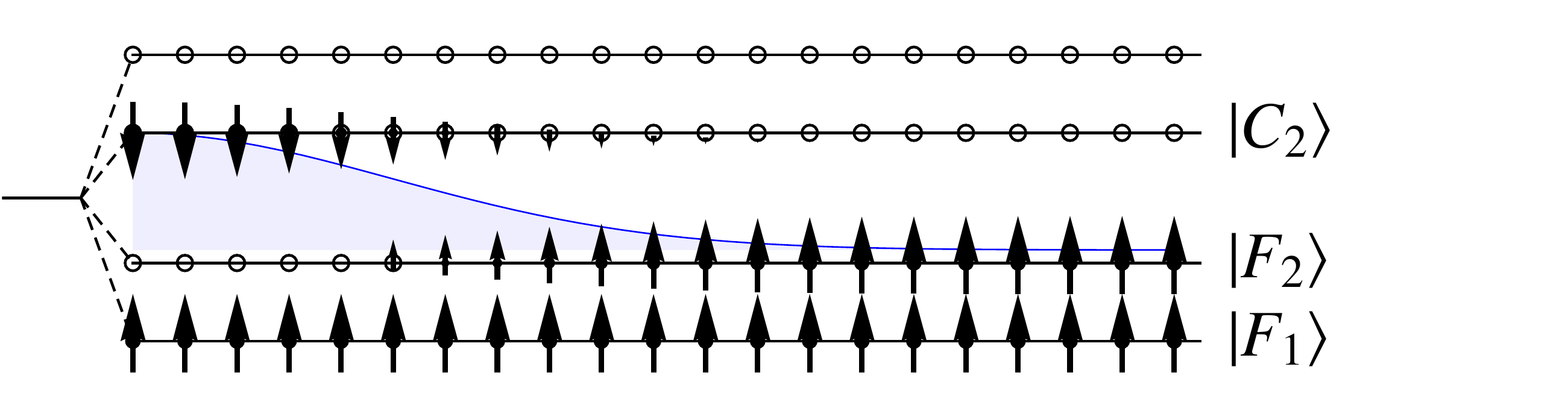}
	\caption{Graphic representation of the skyrmion spinors corresponding to the filling of the Landau levels. At $\mathbf{r}\rightarrow\infty$, the wavefunction has weight on the spinors $|F_1\rangle$ and $|F_2\rangle$, while at $\mathbf{r}=0$, it has weight on $|F_1\rangle$ and $|C_2\rangle$.}
	\label{fig:LLSk}
\end{figure}

In order to find solutions for the skyrmion that minimize the non-linear sigma model energy (\ref{eq:NLSM}), which constitutes the leading energy scale,
we start from the inequality\cite{MacFarlane1979,Lian2017} 
\begin{align}
	\int d^2r(D_\mu Z_k\pm i\varepsilon_{\mu\nu}D_\nu Z_k)^\dagger (D_\mu Z_k\pm i\varepsilon_{\mu\lambda}D_\lambda Z_k)\geq 0.
\end{align}
Upon summation over every spinor labelled by $k$, we obtain that 
\begin{align}
E_{NLSM}\geq 4\pi\rho_S|Q|,
\end{align}
such that the energy is bounded below to $4\pi\rho_S$ times the topological charge. The equality $E_{NLSM}=4\pi\rho_S|Q|$ is reached when 
\begin{align}
\bar{D}Z=0\quad\Rightarrow\quad(1-ZZ^\dagger)\bar{\partial} Z=0\quad\text{for}\quad Q>0, \label{eq:skyrmion1} \\
DZ=0\quad\Rightarrow\quad(1-ZZ^\dagger)\partial Z=0\quad\text{for}\quad Q<0, \label{eq:skyrmion2}
\end{align}
where $\bar{D}=\frac{1}{2}(D_x+iD_y)$, $D=\frac{1}{2}(D_x-iD_y)$ $\bar{\partial}=\partial/\partial z^*=\frac{1}{2}(\partial_x+i\partial_y)$ and $\partial=\partial/\partial z$. The solutions for the skyrmions are thus holomorphic (anti-holomorphic) function for a positive (negative) charge. For simplicity, we focus only on skyrmions of charge $Q=+1$. General solutions of Eq. (\ref{eq:skyrmion1}) are found\cite{Sasaki1983,Din1984} by constructing two linearly independent, orthogonal holomorphic spinors and then normalize them. Moreover, for $r\rightarrow\infty$, where $r=0$ is the center of the skyrmion, the spinors must reach the expression for the spinors $F_1$ and $F_2$ compatible with the ferromagnetic background. As we discuss in more detail below, the spinors $F_1$ and $F_2$ represent two completely filled sub-LLs and are thus related to $f_1$ and $f_2$ by a unitary SU(2) transformation that represents precisely a symmetry of the filled sub-LLs. 
The simplest solution for a skyrmion of charge $Q=+1$ is thus given by\cite{PhdLian} 
\begin{align}
	Z_1&=F_1 \label{eq:spinors_skyrm1}\\
	Z_2&=\frac{\lambda C_2+zF_2}{\sqrt{\lambda^2+r^2}}
	\label{eq:spinors_skyrm2}
\end{align}
where $z=x+iy$, and $C_2$ represent a spinor of one of the empty sub-LLs. This spinor therefore satisfies the conditions $C_2^\dagger C_2=1$ and $F_1^\dagger C_2=0$ in order to obey Eq. (\ref{eq:norm}). We can see that at $\mathbf{r}=0$, the spinors have the expression 
\begin{align}
	Z_1(\mathbf{r}=0)&=F_1 \\
	Z_2(\mathbf{r}=0)&=C_2,
\end{align}
The spinor $C_2$ is thus the \textit{center spinor} of the skyrmion in addition to the ``spectator'' spinor $F_1$. Indeed, one sees from the expression (\ref{eq:spinors_skyrm2}) that the texture only appeals to the spinor $F_2$, which is orthogonal to $F_1$, far away from the skyrmion center. 
The parameter $\lambda$ corresponds to the size of the skyrmion. We can see that $\lambda$ is free so that the skyrmion is scale-invariant in the non-linear sigma model. Figure \ref{fig:LLSk} illustrates the interpolation of the skyrmion spinors as $\mathbf{r}$ moves from the center to infinity. Notice that, as much as $F_1$ for the filled sub-LLs, the spinor $C_1$, with $C_1^\dagger C_2=0$, can be viewed as a ``spectator'' spinor, but now for the empty sub-LLs, since it does not take part in the formation of the spin-valley skyrmion texture.

Notice that we choose $F_2^\dagger C_2=0$, which is not a restrictive choice since 
one can always rearrange, for $F_2^\dagger C_2\neq 0$, the spinor $Z_2$ such that 
\begin{align}
	Z_2=\frac{\lambda' C_2'+(z-z_0)F_2}{\sqrt{\lambda^2+r^2}},
\end{align}
with $z_0=-\lambda F_2^\dagger C_2$, $\lambda'=\lambda^2(1-|C_2^\dagger F_2|^2)$ and 
\begin{align}
	C_2'=\frac{1-F_2F_2^\dagger}{\sqrt{1-|C_2^\dagger F_2|^2}}C_2,
\end{align}
such that $C_2'^\dagger F_2=0$. This change amounts to simply shifting the center of the skyrmion to $\mathbf{r}_0=(x_0,y_0)$. For simplicity, in the following, we will only consider skyrmions centered at $\mathbf{r}=0$ such that the condition $F_2^\dagger C_2=0$ is satisfied. 

The topological charge density of the skyrmion given by Eqs. (\ref{eq:spinors_skyrm1}) and (\ref{eq:spinors_skyrm2}) is 
\begin{align}
    \rho_\text{topo}(\mathbf{r})=\frac{\lambda^2}{\pi(\lambda^2+r^2)^2},
    \label{eq:topo_charge}
\end{align}
which indeed corresponds to a charge $Q=+1$ skyrmion.


Let us now construct an \textit{ansatz} for the skyrmion and understand the number of parameters that describe the skyrmion embedded in a ferromagnetic background. Remember that the latter is described by two spinors $f_1$ and $f_2$ that describe the filled sub-LLs. Similarly, we can describe the two empty sub-LLs by the spinors $c_1$ and $c_2$, which are orthogonal to the ferromagnetic background. As we have already seen above, the skyrmion texture (\ref{eq:spinors_skyrm2}) involves two arbitrary sub-LLs : the spinor $F_2$, which is a superposition of $f_1$ and $f_2$, is retrieved at large distances from the center, where it is given by $C_2$, which is orthogonal to the ferromagnetic background and thus a superposition of $c_1$ and $c_2$. Formally, these choices can be described with the help of SU(2) unitary matrices of the type 
\begin{equation}
U(\theta,\varphi,\alpha)=\begin{pmatrix}
\cos\frac{\theta}{2}e^{i\alpha} & -\sin\frac{\theta}{2}e^{-i(\alpha+\varphi)}\\
\sin\frac{\theta}{2}e^{i(\alpha+\varphi)} & \cos\frac{\theta}{2}e^{-i\alpha}
\label{eq:unit}
\end{pmatrix}.
\end{equation}
Indeed, the application of the unitary transformation (\ref{eq:unit}) on the orthogonal spinors $f_1$ and $f_2$ 
\begin{align}
	\begin{pmatrix} F_1 & F_2\end{pmatrix} =\begin{pmatrix} f_1 & f_2\end{pmatrix}U(\theta_f,\varphi_f,\alpha_f)
	\label{eq:angles_infinity}
\end{align}
where $\begin{pmatrix} F_1 & F_2\end{pmatrix}$ is a $4\times 2$ Grassmanian matrix, yields the two orthogonal spinors $F_1$ (the spectator spinor in the skyrmion texture) and $F_2$ (the active player), in terms of the three angles $\theta_f$, $\varphi_f$ and $\alpha_f$. At infinity, the order parameter (the projector $ZZ^\dagger$) remains identical. We can operate the same procedure for the empty levels and create a superposition 
\begin{align}
\begin{pmatrix} C_1 & C_2\end{pmatrix} =\begin{pmatrix} c_1 & c_2\end{pmatrix}U(\theta_c,\varphi_c,\alpha_c),
\label{eq:angles_center}
\end{align}
where $\begin{pmatrix} C_1 & C_2\end{pmatrix}$ is also a $4\times 2$ matrix, to obtain the spinors $C_1$, i.e. the spectator level that remains empty all the time, and $C_2$, which describes the sub-LL that the electric charge is transferred to at the origin. We therefore see that the Gr(2,4) skyrmion is characterized by six parameters, in agreement with the counting presented in Ref. \onlinecite{Yang2006}.

\subsection{Size of the skyrmion}

We have seen that the energy of the skyrmion is the sum of three terms. The NLSM energy imposes the special skyrmion form given in term of holomorphic functions, as we have seen in the previous subsection. However, the NLSM term is scale-invariant, i.e. its energy does not depend on the size parameter $\lambda$ and thus does not affect the skyrmion size. The anisotropic energy yields a cost in energy that is proportional to the size of the skyrmion ($\sim\lambda^2$) and thus tends to \textit{decrease} the skyrmion size. On the other hand, the Coulomb energy involves higher-order gradient terms and therefore favors \textit{larger} skyrmions in order to spread the charge to a larger region and thus reduce the gradients. Hence, the size of the skyrmion is determined by the competition between the anisotropic energy and the Coulomb energy. Using Eq. (\ref{eq:Coulomb_en}), we find  that the expression of the Coulomb energy of a skyrmion with a topological charge given by Eq. (\ref{eq:topo_charge}) is 
\begin{align}
E_C[Z]=\frac{3\pi^2}{64}\frac{e^2}{\varepsilon \lambda}
\end{align}
and thus proportional to $1/\lambda$. If we introduce the skyrmion spinor $\check{Z}=Z|_{\lambda=l_B}$ and the adimensional size parameter $\tilde{\lambda}=\lambda/l_B$, we obtain that the energy of the skyrmion scales as :
\begin{align}
E_{sk}[Z]=E_\text{NLSM}[\check{Z}]+\frac{1}{\tilde{\lambda}}E_C[\check{Z}]+\tilde{\lambda}^2E_A[\check{Z}].
\end{align}
Minimizing with respect to $\tilde{\lambda}$ gives us the size of the skyrmion in units of the magnetic length :
\begin{equation}
\frac{\lambda}{l_B}=\left(\frac{E_C[\check{Z}]}{2E_A[\check{Z}]}\right)^{1/3}
\label{eq:size}
\end{equation}
Therefore, we can see that the skyrmion size increases with the Coulomb energy and decreases with the anisotropic energy. When the anisotropy vanishes, the skyrmion size goes to infinity.

Notice finally that there is a slight drawback in these arguments: due to the algebraic form of the skyrmion (\ref{eq:spinors_skyrm2}), the anisotropy energies show a logarithmic divergence. However, this divergence, which can be healed by introducing e.g. an exponential cutoff \cite{Lian2017}, does not affect the scaling arguments invoked to determine the skyrmion size. 

\section{Skyrmion phase diagram}

\label{sec:phase_diag_skyrm}

Using the formalism established in the previous section, we are now equipped to compute the phase diagram of the skyrmions in the QHFM backgrounds presented in Sec. \ref{sec:phaseDiagQHFM}. As we have seen, the energy of the skyrmion is composed of three terms, the NLSM energy which is scale invariant, the anisotropic energy which breaks the SU(4) symmetry of the system and the Coulomb energy. The NLSM energy minimization states that the skyrmion must be built from holomorphic function compatible with the QHFM background. However, the skyrmion angles $\theta_f$, $\varphi_f$, $\alpha_f$, $\theta_c$, $\varphi_c$ and  $\alpha_c$ defined in Eqs. (\ref{eq:angles_infinity}) and (\ref{eq:angles_center}) remain free. The minimization of the anisotropy energy allows us to fix these angles depending on the values of $u_\perp$ and $u_z$ and thus to characterize the energetically favored \textit{skyrmion type}. Finally, the competition between the anisotropic and the Coulomb energy fixes the \textit{skyrmion size}.
From a technical point of view, we use the expression for the vectors $f_1$ and $f_2$ given in Sec. \ref{sec:phaseDiagQHFM} and construct two other orthogonal spinors $c_1$ and $c_2$, which form an orthonormal basis with respect to the QHFM background. We then mix them to obtain the skyrmion spinors $F_1$, $F_2$ and $C_2$ with the six skyrmion angles. Next, we compute the anisotropic energy as a function of the angles using Eq. (\ref{eq:aniso}) and then minimize the latter to find the different phases and their region of validity.

\begin{figure}[t]
	\includegraphics[height=8cm]{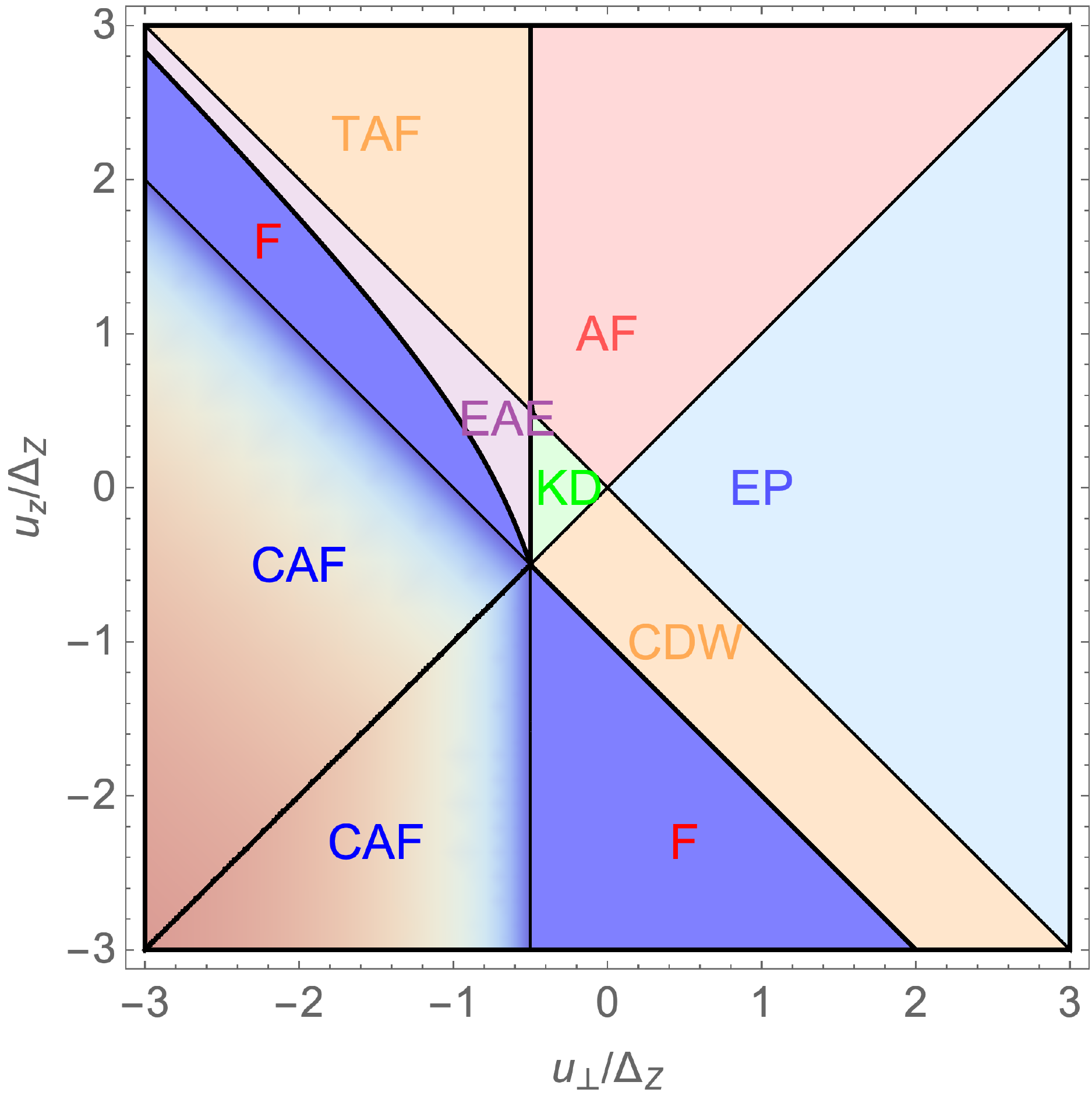}
	\caption{Phase diagram of the skyrmions as a function of $u_\perp$ and $u_z$. AF : anti-ferromagnetic, EP : easy-plane, CDW : charge density wave, KD : Kekul\'e distortion, F : ferromagnetic, CAF : canted anti-ferromagnetic, TAF : tilted anti-ferromagnetic and EAE : Easy-axis entanglement.}
	\label{fig:PhaseDiagSk}
\end{figure}

Figure \ref{fig:PhaseDiagSk} represents our central finding. It
shows the eight different skyrmions phases over the different QHFM backgrounds obtained in the phase diagram in Fig. \ref{fig:PhaseDiagQHFM}. We label the skyrmions according the spin and pseudo-spin magnetization at the center of the skyrmion. For example, the AF skyrmion has an anti-ferromagnetic pattern at its center. In the following, we discuss in detail the different skyrmion types in view of the various QHFM backgrounds. Most saliently, we can distinguish these skyrmion types by spin (and charge) patterns on the two different sublattices that can serve as a fingerprint in the experimental identification of quantum Hall skyrmions in graphene, e.g. by STM techniques. 

In Sec. \ref{sec:skyrmF}, we present the four skyrmions realized in the ferromagnetic background in addition to the symmetry restoration at the transitions. In Sec. \ref{sec:skyrmCDW}, we present the F and the CAF skyrmions in the CDW background. In Sec. \ref{sec:skyrmKD}, we present the F and the CAF skyrmions in the KD background, while in Sec. \ref{sec:skyrmCAF} we discuss the two skyrmions in the CAF background.

\subsection{Ferromagnetic background}

\label{sec:skyrmF}

In the FM background, the spinors have the expression 
\begin{align}
|F_1\rangle&=e^{i\alpha_f}|\mathbf{n}_f\rangle|\mathbf{s}_z\rangle, \label{eq:FerroBG1}\\
|F_2\rangle&=e^{-i\alpha_f}|-\mathbf{n}_f\rangle|\mathbf{s}_z\rangle ,
\end{align}
while the center spinor is 
\begin{align}
|C_2\rangle=e^{i\alpha_c}|\mathbf{n}_c\rangle|-\mathbf{s}_z\rangle,
\end{align}
with 
\begin{align}
	|\mathbf{n}_{f,c}\rangle=\begin{pmatrix}
	\cos\frac{\theta_{f,c}}{2} \\ \sin\frac{\theta_{f,c}}{2}e^{i\varphi_{f,c}}
	\end{pmatrix}.
	\label{eq:FerroBG4}
\end{align}
As we have seen earlier, both spins in $|F_1\rangle$ and $|F_2\rangle$
point in the same direction in the ferromagnetic background, while at the center of the skyrmion, the two spins must point in opposite directions. The spin magnetization at infinity is therefore $|\mathbf{M_S}(\mathbf{r}\rightarrow\infty)|=2$ while at the center we have $|\mathbf{M_S}(\mathbf{r}=0)|=0$. The different skyrmions are thus characterized by the relative pseudo-spin magnetization at the center and at infinity. The anisotropic energy in the ferromagnetic background is equal to 
\begin{align}
    E_A[Z]&=A(u_z\cos\theta_f\cos\theta_c+u_\perp\sin\theta_f\sin\theta_c\cos(\varphi_f-\varphi_c)) \nonumber \\
    &+A(u_z+2u_\perp+2\Delta_Z),
\end{align}
where we have introduced the quantity 
\begin{align}
    A\equiv \int \frac{d^2r}{2\pi l_B^2}\frac{\lambda^2}{\lambda^2+r^2}&=\left(\frac{\lambda}{l_B}\right)^2\ln\left(\frac{\Lambda}{l_B}\right),
\end{align}
which is logarithmically divergent due to the fact that the integrand has an algebraic tail proportional to $1/\lambda$, as mentioned above. We have thus introduced a cut-off $\Lambda$ that will impact the size of the skymion, but its impact is very small. We have also used the expression for the integral 
\begin{align}
    \int \frac{d^2r}{2\pi l_B^2}\frac{r^2}{\lambda^2+r^2}=N_\phi-A,
\end{align}
such that the terms proportional to $N_\phi$ in the skyrmion energy cancel with the background energy and the energy of the skyrmion is only proportional to the quantity $A$.

As shown in Fig. \ref{fig:PhaseDiagSk}, we obtain four skyrmion phases compatible with a FM background: the anti-ferromagnetic skyrmion (AF), the Kekul\'e distortion skyrmion (KD), the charge density wave skyrmion (CDW) and the easy-plane skyrmion (EP). These four phases are characterized by the relative value of $u_\perp$ and $u_z$ and can be separated in two categories: two easy-axis solution for $\theta_f,\theta_c\in\{0,\pi\}$ which is the case of the AF and CDW skyrmions and two easy-plane solutions for $\theta_f=\theta_c=\frac{\pi}{2}$ for the KD and EP skyrmions. In analogy with the QHFM CDW and KD ground states, the easy-axis solution are realized when $|u_z|>|u_\perp|$, such that a pseudo-spin polarization at the poles of the Bloch sphere is favored, whereas the easy-plane solutions are realized for $|u_z|<|u_\perp|$ and the pseudo-spin points at the equator of the Bloch sphere.

\subsubsection{Anti-ferromagnetic skyrmion}

\begin{figure}[t]
    (a)\includegraphics[width=8cm]{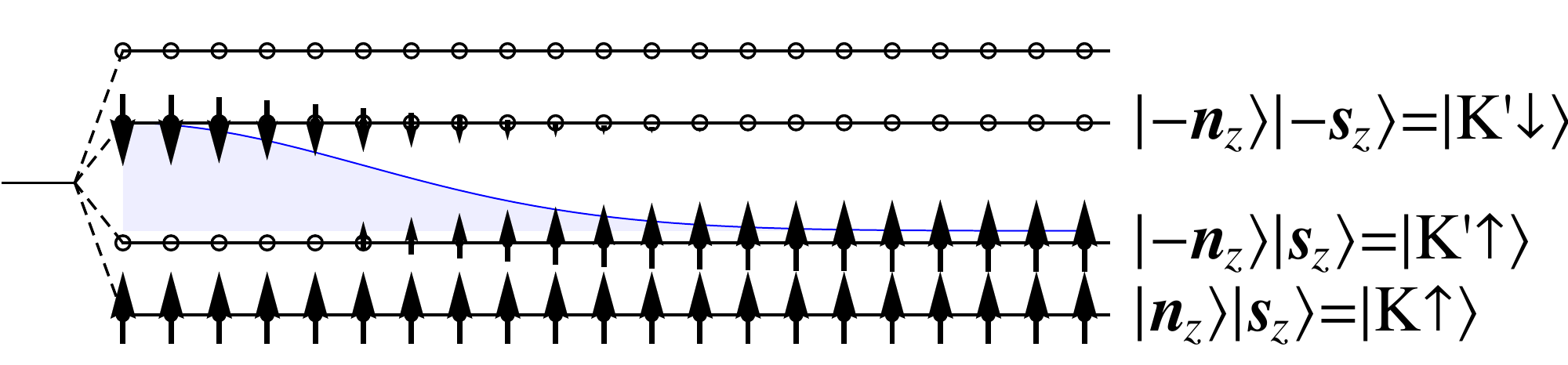} \\
	(b)\includegraphics[width=8cm]{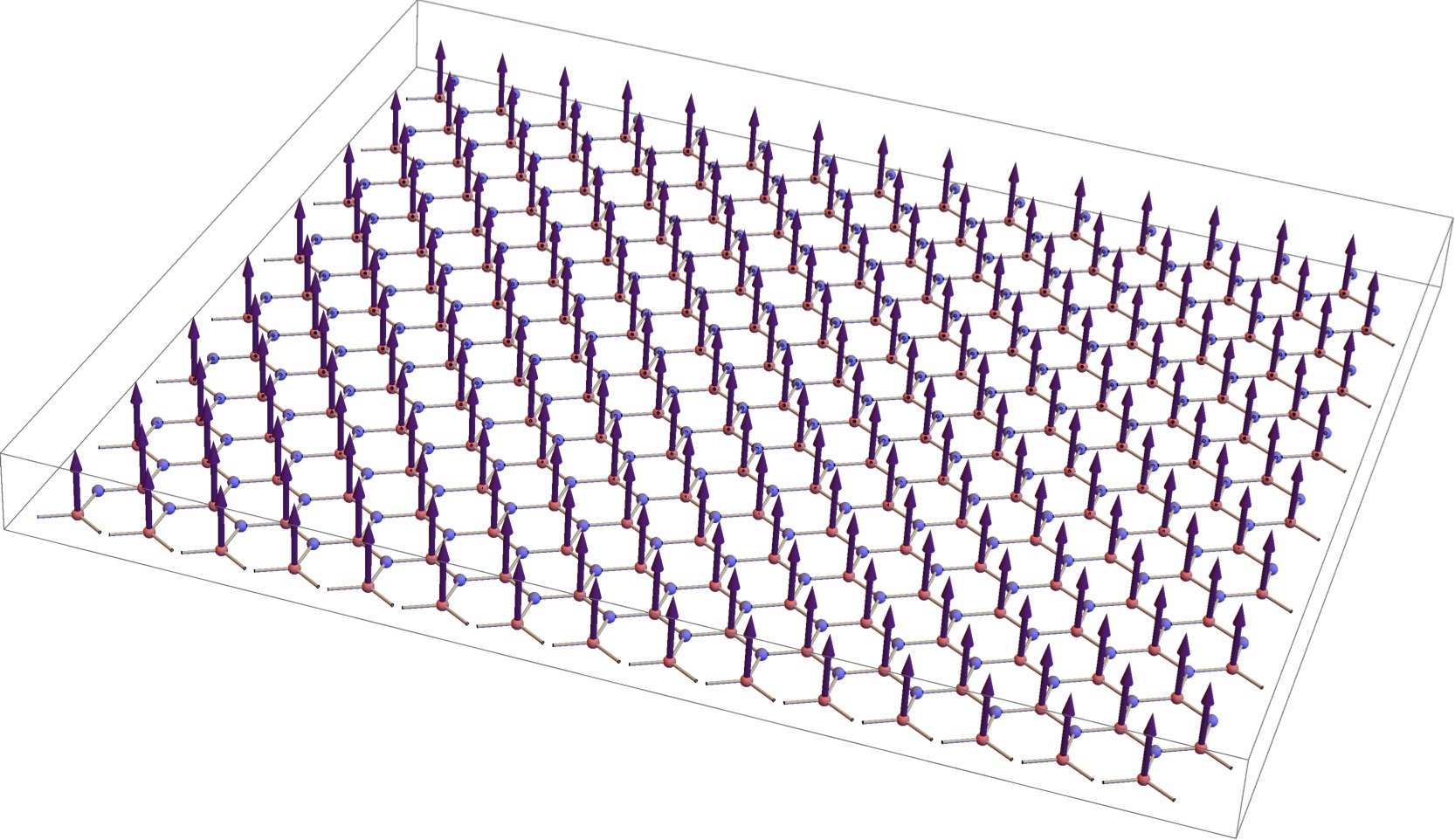} \\
	(c)\includegraphics[width=8cm]{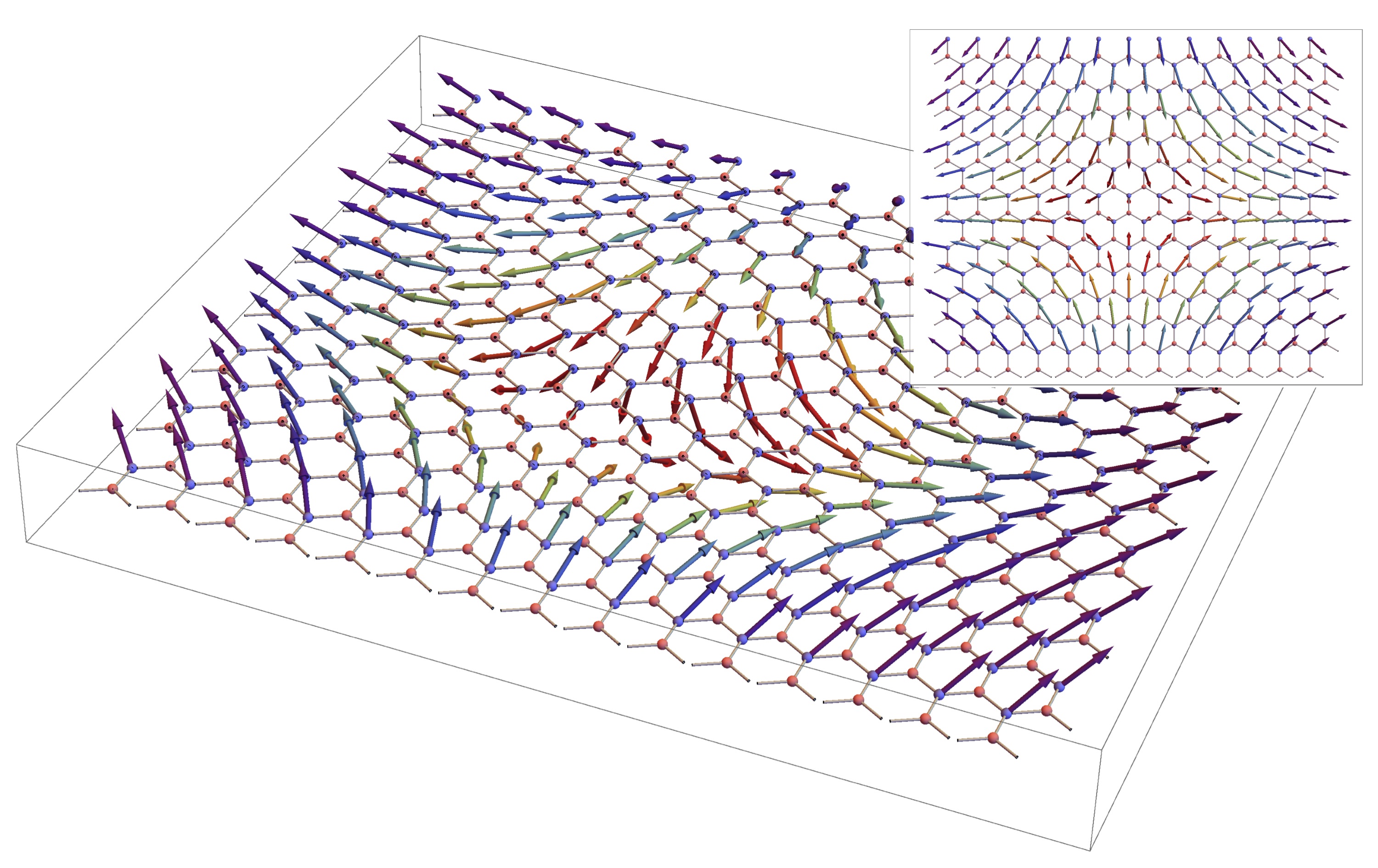}
	
	\caption{\textbf{Antiferromagnetic skyrmion} in the F background. (a) Filled sub-LLs at the center and infinity and the corresponding spinors. (b) Spin magnetization on the A and (c) on the B sublattices. The color coding represents the $z$ component of the magnetization. We can see that the sublattice A remains unaffected by the presence of the skyrmion which lives only on the sublattice B. At infinity, both spins point along the magnetic field, while at the center, the spins point in opposite directions on each sublattice, thus forming an anti-ferromagnetic order. Inset: top view of the spin magnetization on the B sublattice.}
	\label{fig:F-AF}
\end{figure}

The anti-ferromagnetic skyrmion is characterized by the angles $(\theta_f,\theta_c)=(0,\pi)$ or $(\pi,0)$ in Eqs. (\ref{eq:FerroBG1})-(\ref{eq:FerroBG4}), such that $|\mathbf{n}_f\rangle=|-\mathbf{n}_c\rangle=|\mathbf{n}_z\rangle$. The spectator spinor $|Z_1\rangle$ has a pseudo-spin pointing at the north of the Bloch sphere all over the 2D plane, and the spinor $|Z_2\rangle$ has its pseudo-spin pointing at the south. Each spinor is thus identified with one sublattice, and there is a $Z_2$ sublattice symmetry associated with $(\theta_f,\theta_c)=(0,\pi)$ or $(\pi,0)$ that is spontaneously broken. The corresponding sub-LLs are depicted in the diagram \ref{fig:F-AF}(a), and one sees that 
the spinor $|Z_2\rangle$ changes its spin orientation in the texture. Its spin points along the positive $z$ direction at infinity and interpolates to the negative $z$ direction at the center, while remaining in the $K'$ valley all the time. At the center of the skyrmion we have thus 
\begin{align}
	|Z_1(\mathbf{r}=0)\rangle&=|\mathbf{n}_z\rangle|\mathbf{s}_z\rangle \\
	|Z_2(\mathbf{r}=0)\rangle&=|-\mathbf{n}_z\rangle|-\mathbf{s}_z\rangle
\end{align}
which indeed corresponds to an anti-ferromagnetic order. Figures \ref{fig:F-AF}(b) and (c) show the spin magnetization on the A and B sublattices, respectively. These magnetizations bare more information than a plot of the charge density, which remains homogeneous for the AF skyrmion. 
Because each spinor is associated with a sublattice, we can see that the spin magnetization remains unchanged on one sublattice (corresponding to the $K$ valley) and the skyrmion is formed 
only by electrons on the other sublattice (corresponding to the valley $K'$). In contrast, the skyrmion involves a spin rotation from a down-spin at the center to an up-spin at infinity to match the ferromagnetic background. As we have already mentioned in Sec. \ref{sec:parametrization}, the center of the skyrmion can thus show, somewhat unexpectedly an AF pattern at the center since the spin rotation only concerns one of the sublattices. The spin magnetization on each sublattice is given by 
\begin{align}
	\mathbf{M_S}_A&=\mathbf{s}_z, \\
	\mathbf{M_S}_B&=\frac{1}{\lambda^2+r^2}R_z(\alpha)\begin{pmatrix}
	\lambda x \\ -\lambda y \\ r^2-\lambda^2
	\end{pmatrix},
\end{align}
where $R_z(\alpha)$ with $\alpha=\alpha_c+\alpha_f$ is an SO(2) rotation matrix around the axis $z$ such that 
\begin{align}
	R_z(\alpha)=\begin{pmatrix}
	\cos\alpha & -\sin\alpha & 0 \\ \sin\alpha & \cos\alpha & 0 \\ 0 &0&1
	\end{pmatrix}.
\end{align}
The U(1) symmetry associated with the phase factors $\alpha_c$ and $\alpha_f$ implies thus an SO(2) rotation invariance of the skyrmion which is coherent with the isotropy of the $x$ and $y$ directions.

\subsubsection{Charge density wave skyrmion}

\begin{figure}[t]
    (a)\includegraphics[width=7cm]{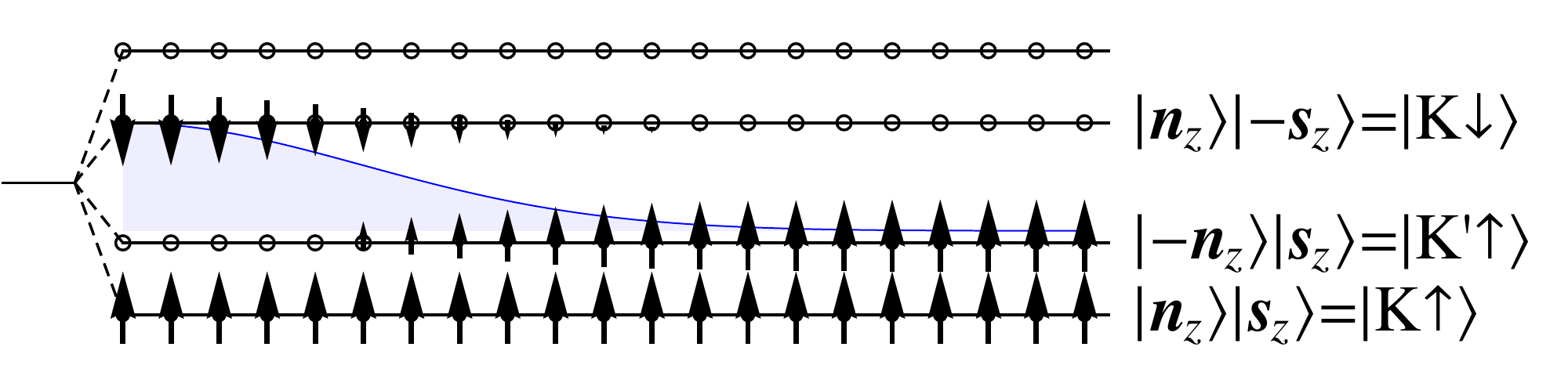} 
	(b)\includegraphics[width=7cm]{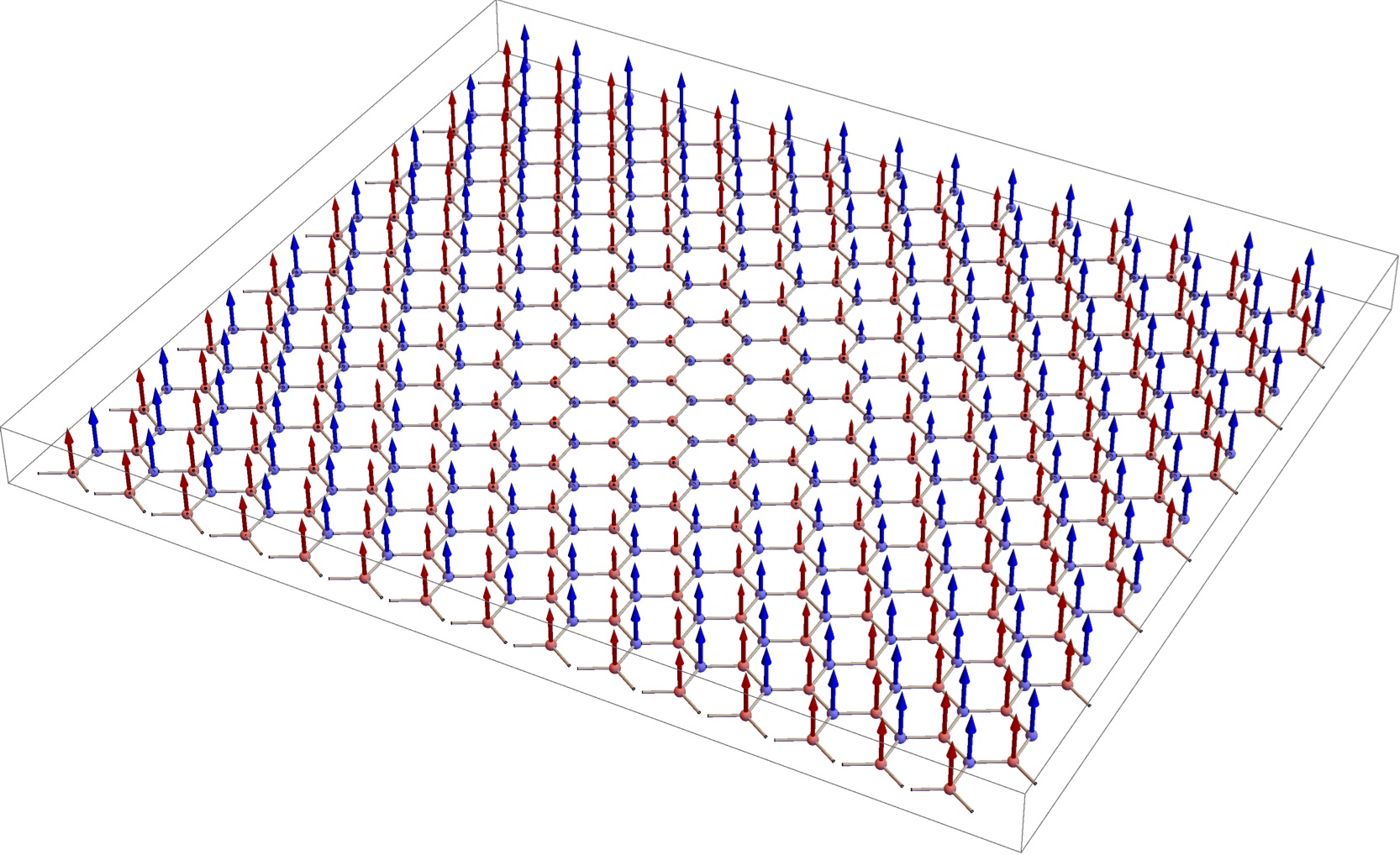}
	(c)\includegraphics[width=4cm]{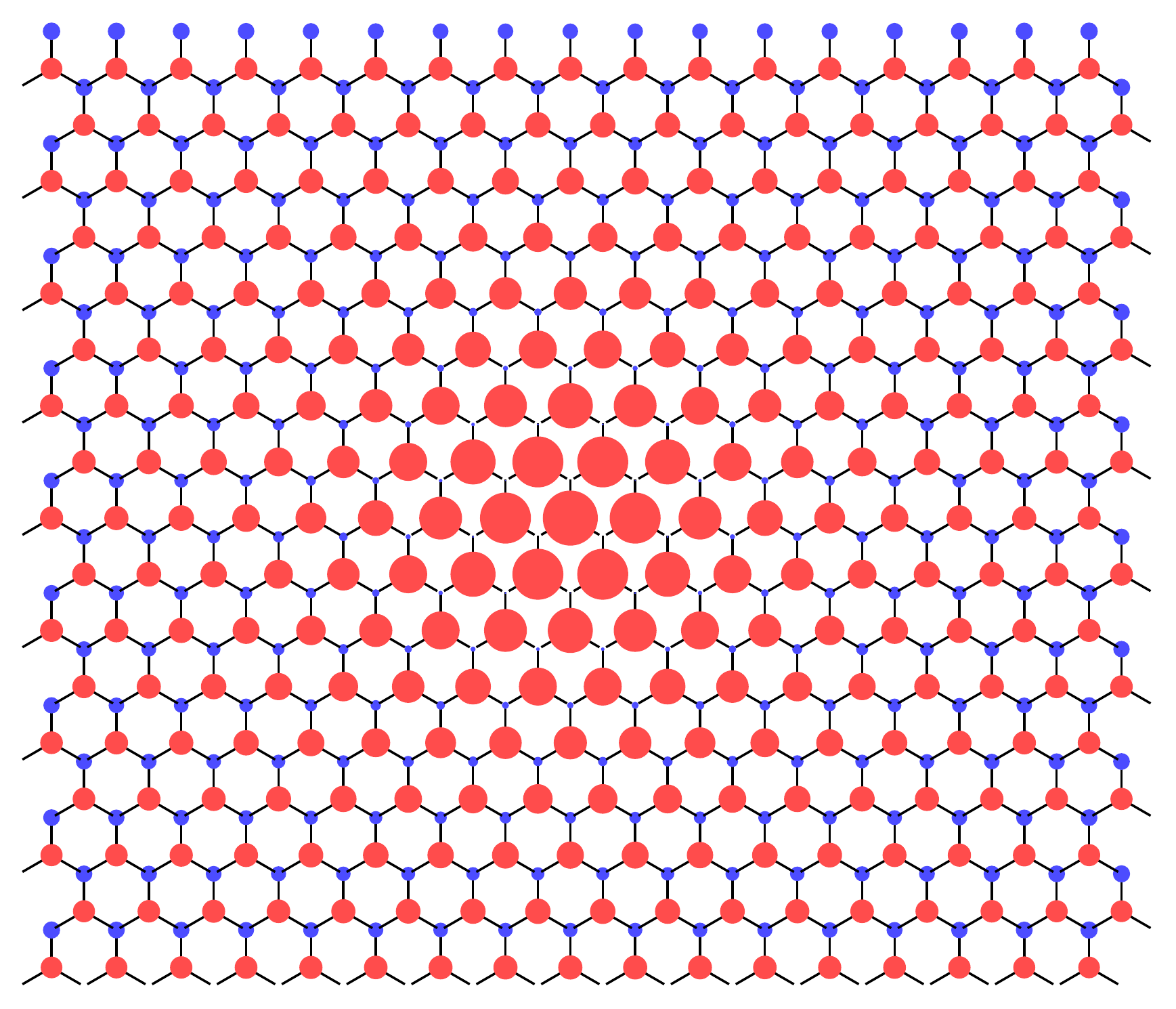}
	\caption{\textbf{Charge density wave skyrmion} in the F background. (a) Filled sub-LLs at the center and infinity and the corresponding spinors. (b) Spin magnetization on the A and B sublattices and (c) electronic density. The spin magnetization is identical in both sublattices and the total spin vanishes at the center. We can see a CDW pattern on panel (b) at the center of the skyrmion where only one sublattice is occupied.}
	\label{fig:F-CDW}
\end{figure}

The CDW skyrmion is characterized by the angles $(\theta_f,\theta_c)=(0,0)$ or $(\pi,\pi)$ in Eqs. (\ref{eq:FerroBG1})-(\ref{eq:FerroBG4}). The difference with the AF skyrmion is that both pseudo-spin spins point towards the same pole at the center of the skyrmion,
\begin{align}
	|Z_1(\mathbf{r}=0)\rangle&=|\mathbf{n}_z\rangle|\mathbf{s}_z\rangle, \\
	|Z_2(\mathbf{r}=0)\rangle&=|\mathbf{n}_z\rangle|-\mathbf{s}_z\rangle,
\end{align}
which are sketched out in Fig. \ref{fig:F-CDW}(a). At the center both electrons are then situated in a single valley and therefore reside on the same sublattice. This yields a CDW pattern as can be seen in Fig. \ref{fig:F-CDW}(c), along with a vanishing spin magnetization since the associated spins point towards different poles on the spin Bloch sphere. The spin magnetization shown in Fig. \ref{fig:F-CDW}(b) is identical in each sublattice,
\begin{align}
	\mathbf{M_S}_A=\mathbf{M_S}_B=\frac{r^2}{\lambda^2+r^2}\mathbf{s}_z.
\end{align}
We can see that the spin magnetization points towards the positive $z$ direction and decreases as we get closer to the center of the skyrmion, where only one sublattice is occupied.

\subsubsection{Kekul\'e distortion skyrmion}

\begin{figure}[t]
    (a)\includegraphics[width=8cm]{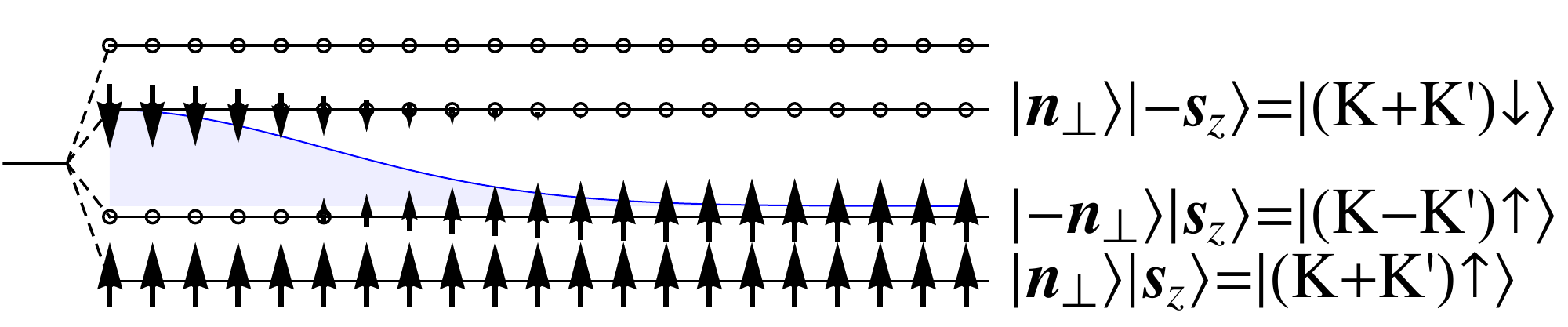} 
	(b)\includegraphics[width=8cm]{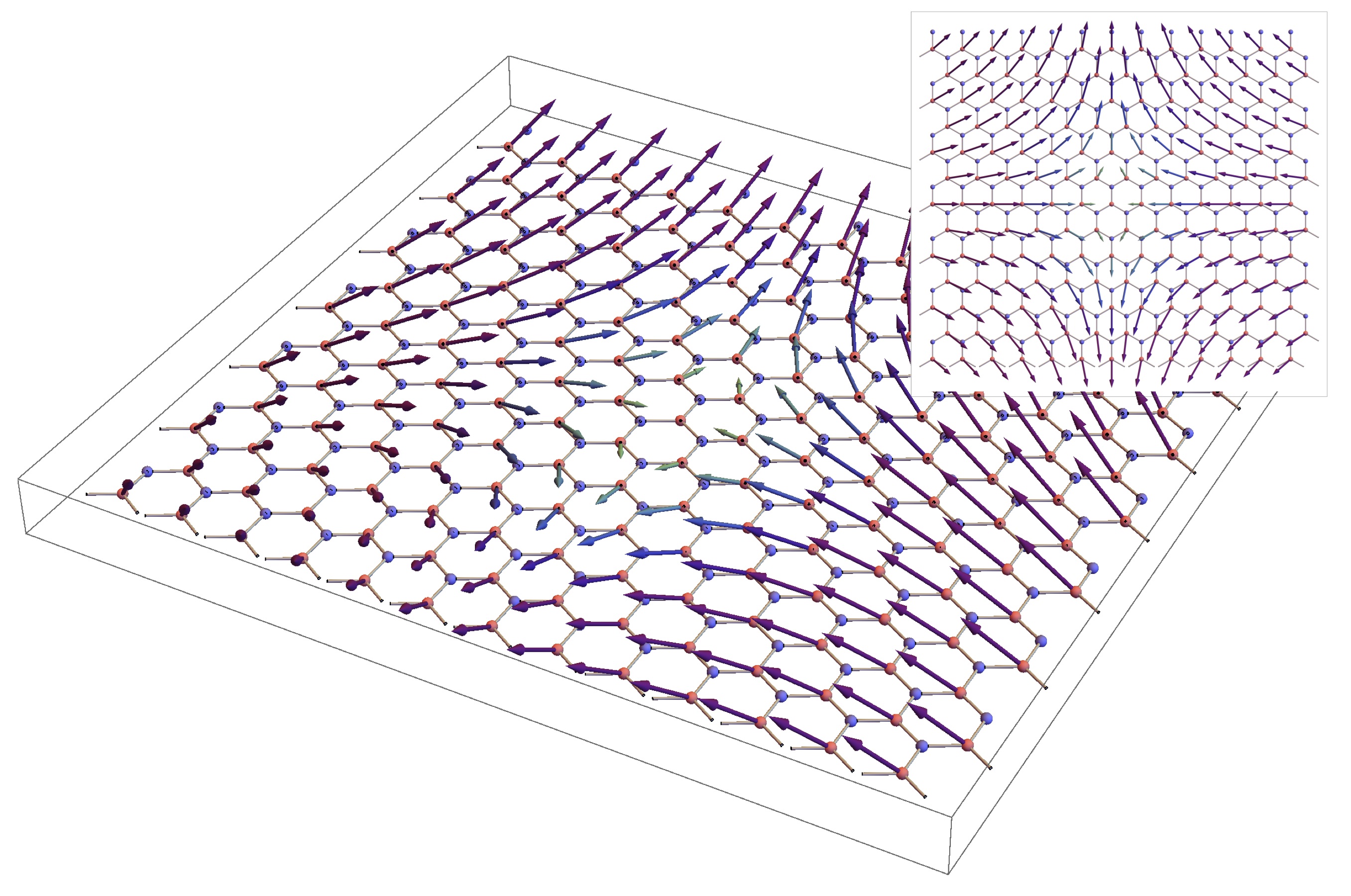}
	(c)\includegraphics[width=8cm]{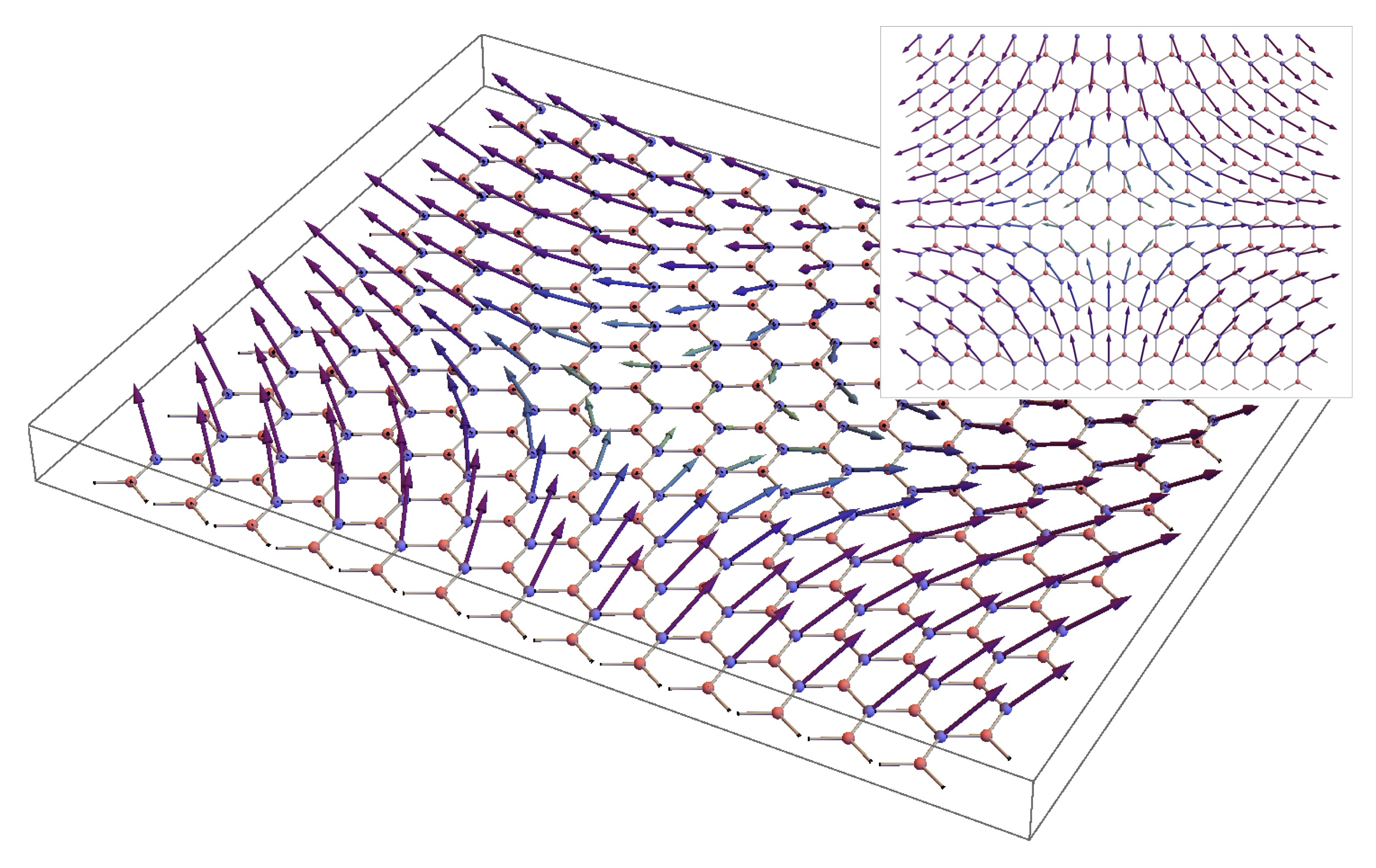}
	\caption{\textbf{Kekul\'e distortion skyrmion} in the F background. (a) Filled sub-LLs at the center and infinity and the corresponding spinors. (b) and (c) Spin magnetization on the A and B sublattices respectively which has opposite direction in the $xy$ plane. Because the pseudo-spin of all spinors lie at the equator of the Bloch sphere, both sublattices are occupied equally and this creates a $1/r$ tail in the spin magnetization in the $xy$ plane.}
	\label{fig:F-KD}
\end{figure}

The KD skyrmion is characterized by $\theta_f=\theta_c=\frac{\pi}{2}$ and $\varphi_f=\varphi_c$ in Eqs. (\ref{eq:FerroBG1})-(\ref{eq:FerroBG4}),. The pseudo-spin is thus oriented along the equator of the pseudo-spin Bloch sphere and we have $|\mathbf{n}_f\rangle=|\mathbf{n}_c\rangle\equiv|\mathbf{n}_\perp\rangle$. At the center, they have the expression [see Fig. \ref{fig:F-KD}(a)]
\begin{align}
	|Z_1(\mathbf{r}=0)\rangle&=e^{i\alpha_f}|\mathbf{n}_\perp\rangle|\mathbf{s}_z\rangle \\
	|Z_2(\mathbf{r}=0)\rangle&=e^{i\alpha_c}|\mathbf{n}_\perp\rangle|-\mathbf{s}_z\rangle,
\end{align}
which is similar to the KD QHFM background. The spin magnetization vanishes on both sublattices at the center while both sublattices are equally populated. Because the charge density remains homogeneous for a KD skyrmion, we plot in Figs. \ref{fig:F-KD}(b) and (c) the magnetizations on each of the two sublattices, respectively, rather than the charge density. 
The spin magnetizations on the A and B sublattices are given by 
\begin{align}
	\mathbf{M_S}_A&=\frac{1}{\lambda^2+r^2}R_z(\alpha+\varphi_f)\begin{pmatrix}
	-\lambda x \\ \lambda y \\ r^2\end{pmatrix} \\
	\mathbf{M_S}_B&=\frac{1}{\lambda^2+r^2}R_z(\alpha+\varphi_f)\begin{pmatrix}
	\lambda x \\ -\lambda y \\ r^2
	\end{pmatrix},
\end{align}
respectively. As opposed to the CDW skyrmion, because the electrons are in a superposition of the valleys, this skyrmion has also a magnetization in the $xy$ plane which has opposite direction on the A and B sublattices such that the total spin in the plane vanishes $M_{S_x}(\mathbf{r})=M_{S_y}(\mathbf{r})=0$. The $r^2/(\lambda^2+r^2)$ dependence of the $z$ component of the magnetization implies that the spin magnetization reaches the ferromagnetic value $|\mathbf{M_S}|=|\mathbf{M_S}_A+\mathbf{M_S}_B|=2$ over a few $\lambda$, while the $r/(\lambda^2+r^2)$ dependence of the $x$ and $y$ component implies that they have a long $1/r$ tail before they reach 0 at infinity. This explains the large pattern of the skyrmion in the $xy$ plane in the insets of Figs. \ref{fig:F-KD} (a) and (b). 

\subsubsection{Easy-plane skyrmion}

\begin{figure}[t]
    (a)\includegraphics[width=8cm]{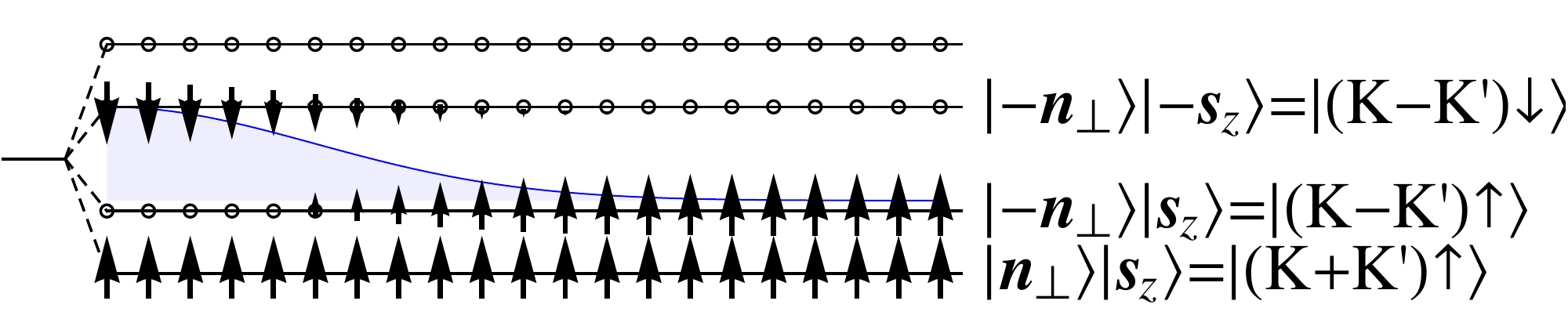} 
	(b)\includegraphics[width=8cm]{Skyrmion/Fbackground/KDBinset.jpg}
	\caption{\textbf{Easy-plane skyrmion} in the F background. (a) Filled sub-LLs at the center and infinity and the corresponding spinors. (b) spin magnetization on the A and B sublattices that is identical such that we only show one sublattice. Because the pseudo-spin of all spinors lie at the equator of the Bloch sphere, both sublattices are occupied equally and this creates a $1/r$ tail in the spin magnetization in the $xy$ plane.}
	\label{fig:F-EP}
\end{figure}

The EP skyrmion is realized for  $\theta_f=\theta_c=\frac{\pi}{2}$ and $\varphi_f=\varphi_c+\pi$ in Eqs. (\ref{eq:FerroBG1})-(\ref{eq:FerroBG4}). At the center the spinors have opposite spin and pseudo-spin 
\begin{align}
	|Z_1(\mathbf{r}=0)\rangle&=e^{i\alpha_f}|\mathbf{n}_\perp\rangle|\mathbf{s}_z\rangle \\
	|Z_2(\mathbf{r}=0)\rangle&=e^{i\alpha_c}|-\mathbf{n}_\perp\rangle|-\mathbf{s}_z\rangle,
\end{align}
as shown in Fig. \ref{fig:F-EP}(a).
This skyrmion is vey similar to the Kekul\'e distortion skyrmion, it also possesses a magnetization in the $xy$ plane. The main difference is that the spin magnetization is identical on both sublattices,
\begin{align}
	\mathbf{M_S}_A=\mathbf{M_S}_B=\frac{1}{\lambda^2+r^2}R_z(\alpha)\begin{pmatrix}
	\lambda x \\ -\lambda y \\ r^2
	\end{pmatrix},
\end{align} 
in contrast to the KD and AF skyrmions, such that this skyrmion has also a total magnetization in the $xy$ plane [see Fig. \ref{fig:F-EP}(b)].

\subsubsection{Symmetry at the transition lines}

\label{sec:symmetryF}

We can see on the phase diagram that the four skyrmion phases in the ferromagnetic background are separated by two transition lines, the line $u_z=u_\perp$ and the line $u_z=-u_\perp$. Notice that, in contrasts to the transitions between the ferromagnetic backgrounds in Fig. \ref{fig:PhaseDiagQHFM}, these transitions are no transitions in the thermodynamic sense since they only delimit the regions in which a certain skyrmion type is energetically favored and because they are obtained by energy calculations of single skyrmions. They may, however, become true thermodynamic transitions if we consider many skyrmions, e.g. when they form skyrmion crystal\cite{Brey1995,Brey1996b,Cote2008}, but this is beyond the scope of the present paper. 

The transition line at $u_z=u_\perp$ has the same properties as studied in Sec. \ref{sec:symmetry_uz=up} where the SU(2) pseudo-spin rotation symmetry is restored. At the transition from the AF to the EP skyrmion, the pseudo-spin in each spinor $|F_1\rangle$, $|F_2\rangle$ and $|C_2\rangle$ rotates from $|\pm\mathbf{n}_z\rangle$ to $|\pm\mathbf{n}_\perp\rangle$ such that at the transition, the spinor $|\mathbf{n}_f\rangle$ can point in any direction, while the center spinor is given by $|\mathbf{n}_c\rangle=|-\mathbf{n}_f\rangle$. At the transition from the KD to the CDW skyrmion, the same scenario happens with the ferromagnetic  background spinor $|\mathbf{n}_f\rangle$ being free while the center spinor is this time given by $|\mathbf{n}_c\rangle=|\mathbf{n}_f\rangle$.

At the transition line $u_z=-u_\perp$, the Hamiltonian is 
\begin{align}
    H_A=\int d^2r\left\{U_\perp \left[P_x^2(\mathbf{r})+P_y^2(\mathbf{r})\right]-P_z^2(\mathbf{r}))-N_\phi\Delta_ZS_z(\mathbf{r})
    \right\}
\end{align} 
and thus commutes with the three operators 
\begin{align}
    \Pi_x(\mathbf{r})&=\Psi^\dagger(\mathbf{r})\tau_x\sigma_z\Psi(\mathbf{r}) \label{eq:transF-CDW1} \\
    \Pi_y(\mathbf{r})&=\Psi^\dagger(\mathbf{r})\tau_y\sigma_z\Psi(\mathbf{r}) \\
    \Pi_z(\mathbf{r})&=\Psi^\dagger(\mathbf{r})\tau_z\Psi(\mathbf{r}),\label{eq:transF-CDW3}
\end{align}
which form an SU(2) subgroup of the SO(5) symmetry group at the transition line $u_z=-u_\perp$ introduced by Wu \textit{et al}\cite{Wu2014a}. The SO(5) symmetry at the transition is realized when $\Delta_Z=0$ and is thus broken down to the SU(2) symmetry generated by the operators (\ref{eq:transF-CDW1})-(\ref{eq:transF-CDW3}) for a finite Zeeman. These operators generate rotations in the pseudo-spin space of opposite sense for the two spin species around axes in the $xy$ plane while it generates rotations of the same sense around the $z$ axis. In the four phases of the ferromagnetic background, the spinors $|F_1\rangle$ and $|F_2\rangle$ are always spin-polarized along $+\mathbf{s}_z$ while the center spinor $|C_2\rangle$ is polarized along $-\mathbf{s}_z$. For example, at the transition between the CDW and EP skyrmion, the spinors $|F_1\rangle$ and $|F_2\rangle$ (with spin up polarizarion) rotate from $|\pm\mathbf{n}_z\rangle$ to $|\pm\mathbf{n}_\perp\rangle$ respectively, while the center spinor (with spin-down polarization) rotates from $|\mathbf{n}_z\rangle$ to $|-\mathbf{n}_\perp\rangle$. Thus, we can see that at the transition, the pseudo-spin of the spin-up electrons is rotated in one direction while the pseudo-spin of the spin-down electrons is rotated in the other direction. The same scenario happens at the AF-KD transition.

At the CDW-EP transition, the ferromagnetic spinor $|\mathbf{n}_f\rangle=|\mathbf{n}_+\rangle$ can point in any direction, while the center spinor is given by $|\mathbf{n}_c\rangle=|\mathbf{n}_-\rangle$ with $\mathbf{n}_\pm=(\pm\sin\theta\cos\varphi,\pm\sin\theta\sin\varphi,\cos\theta)$. On the other hand, at the AF-KD transition, the ferromagnetic spinor is still $|\mathbf{n}_f\rangle=|\mathbf{n}_+\rangle$ while the center spinor is $|\mathbf{n}_c\rangle=|-\mathbf{n}_-\rangle$.

\subsection{Charge density wave background}

\label{sec:skyrmCDW}

\begin{figure}[h]
	\includegraphics[width=4cm]{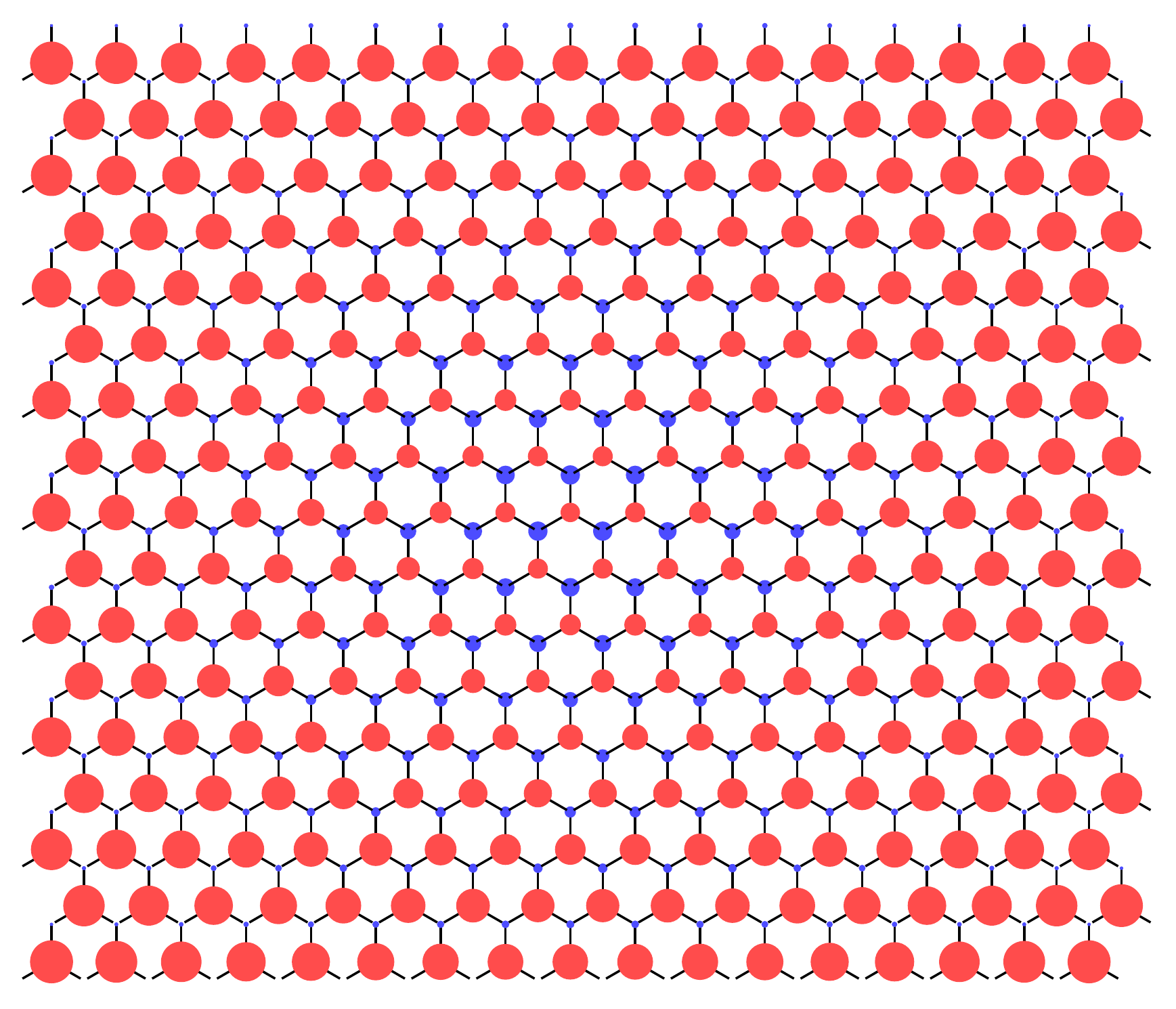}
	\caption{Electronic density on the A and B sublattices of a skyrmion in the charge density wave background.}
	\label{fig:ABdensityCDW}
\end{figure}

In the CDW background both pseudo-spins point in the same direction at the pole of the pseudo-spin Bloch sphere $\mathbf{n}_z$, and both spinors $|f_1\rangle$ and $|f_2\rangle$ have only components on the same sublattice. Therefore, at the center, the pseudo-spin must point in the opposite direction, namely $-\mathbf{n}_z$ such that both sublattices are occupied equally. Moreover, at infinity, because the electrons reside on the same sublattice and the the spins point in opposite directions, the spin magnetization vanishes $\mathbf{M_S}(\mathbf{r}\rightarrow\infty)=0$. Because the pseudo-spin is fixed by the background, the skyrmion angles thus act on the spin,
\begin{align}
	|F_1\rangle&=e^{i\alpha_f}|\mathbf{n}_z\rangle|\mathbf{s}_f\rangle \label{eq:CDWBG1} \\
	|F_2\rangle&=e^{-i\alpha_f}|\mathbf{n}_z\rangle |-\mathbf{s}_f\rangle.
\end{align}
At the center, the pseudo-spin points in the opposite direction such that 
\begin{align}
	|C_2\rangle=e^{i\alpha_c}|-\mathbf{n}_z\rangle |\mathbf{s}_c\rangle,
\end{align}
with 
\begin{align}
	|\mathbf{s}_{f,c}\rangle=\begin{pmatrix}
	\cos\frac{\theta_{f,c}}{2} \\ \sin\frac{\theta_{f,c}}{2}e^{i\varphi_{f,c}}
	\end{pmatrix}.
	\label{eq:sf-sc}
\end{align}
The relative orientation of $|\mathbf{s}_f\rangle$ and $|\mathbf{s}_c\rangle$ thereby determines which skyrmion is realized. We can see that at the center the pseudo-spin components of $|F_1\rangle$ and $|C_2\rangle$ point towards opposite poles so that both sublattices are occupied equally. Figure \ref{fig:ABdensityCDW} shows the electronic density of a skyrmion in the charge density wave background. We can see that at the center, both sublattices are occupied equally while only one sublattice is occupied away from the center.

The anisotropic energy of the skyrmion as a function of the angles is given by
\begin{align}
	E_A[Z]&=-Au_\perp(\cos\theta_f\cos\theta_c+\sin\theta_f\sin\theta_c\cos(\varphi_f-\varphi_c)) \nonumber \\ &-2Au_z-Au_\perp-\Delta_ZA(\cos\theta_f+\cos\theta_c).
\end{align}
We can see in Fig. \ref{fig:PhaseDiagSk} that two types of skyrmions are realized, the ferromagnetic (F) for $u_\perp>-\Delta_Z/2$ and the canted-antiferromagnetic (CAF) skyrmions for $u_\perp<-\Delta_Z/2$. The transition between these two phases is very similar to the transition between the F and CAF backgrounds which happens at the same value of $u_\perp=-\Delta_Z/2$. This transitions is continuous as the canting angle reaches 0 at the transition.

\subsubsection{Ferromagnetic skyrmion}

\begin{figure}[t]
    (a)\includegraphics[width=8cm]{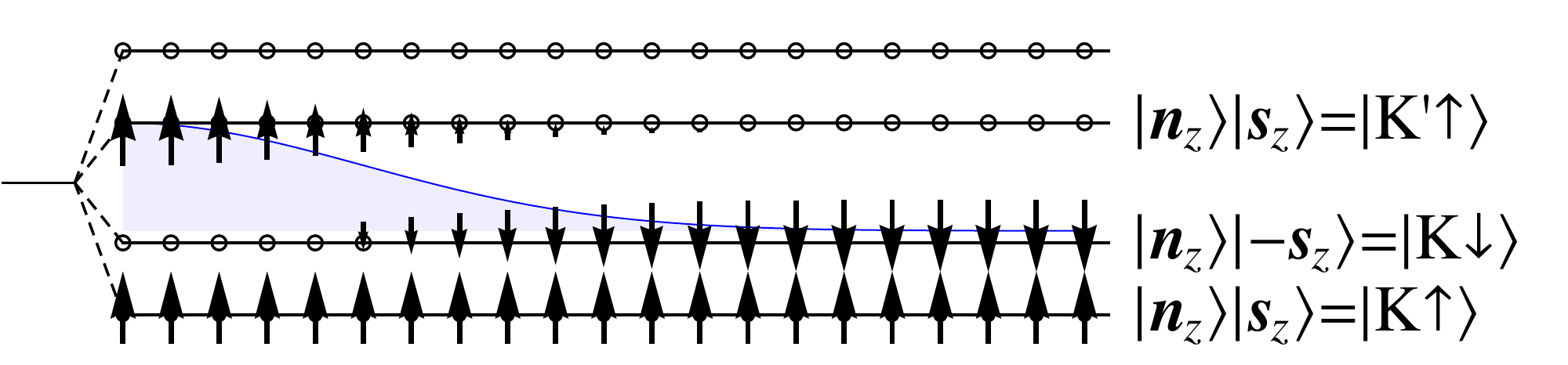}
    (b)\includegraphics[width=8cm]{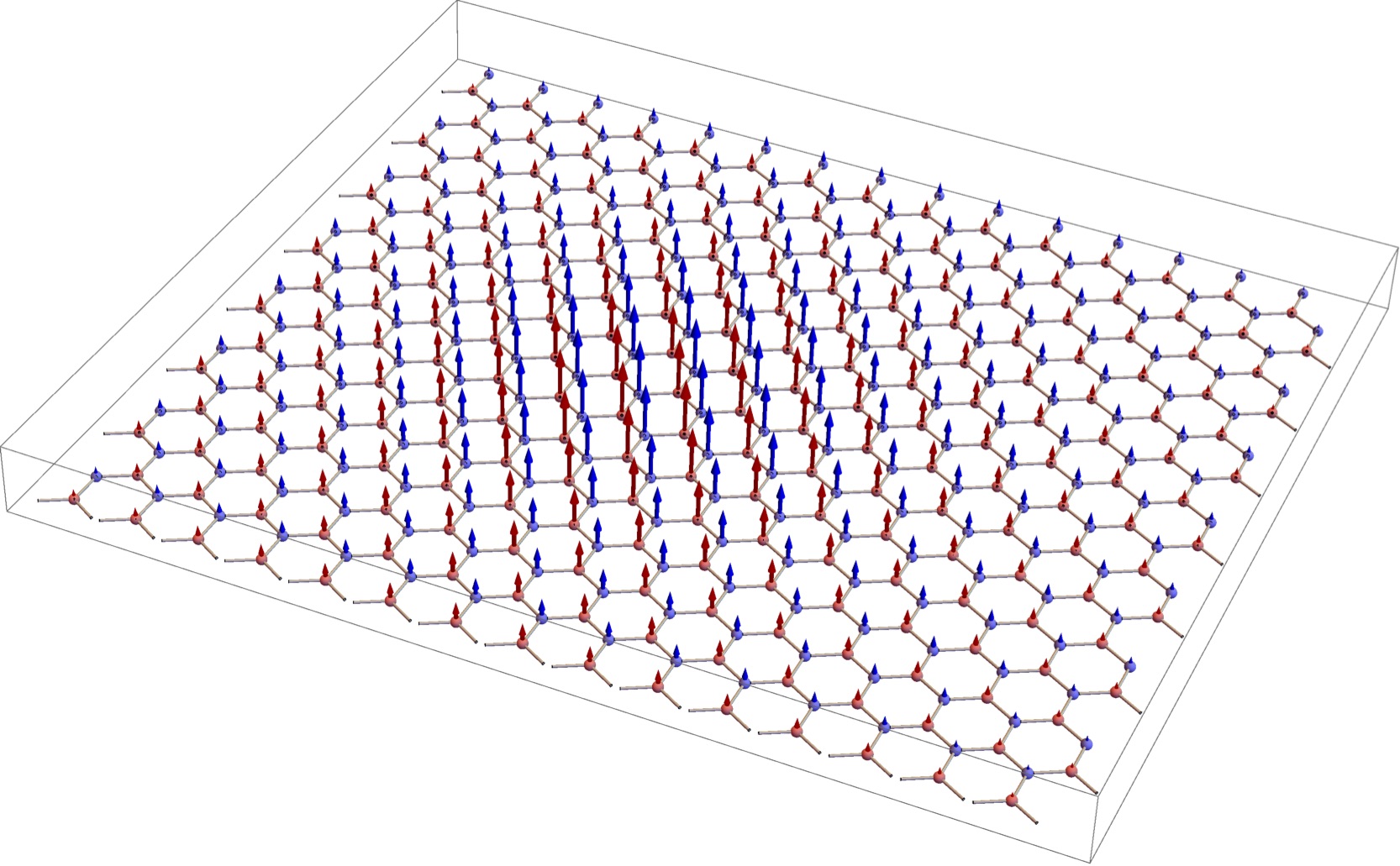}
	\caption{\textbf{Ferromagnetic skyrmion} in the CDW background. (a) Filled sub-LLs at the center and infinity and the corresponding spinors. (b) Spin magnetization on the A (red) and B (blue) sublattices. At infinity, because both spins point in the opposite direction and reside in the same sublattice, the total spin magnetization vanishes, while at the center both spins point in the same direction and reside on different sublattices. }
	\label{fig:CDW-F}
\end{figure} 

The ferromagnetic skyrmion is obtained for $(\theta_f,\theta_c)=(0,0)$ in Eqs. (\ref{eq:CDWBG1})-(\ref{eq:sf-sc}) such that $|\mathbf{s}_f\rangle=|\mathbf{s}_c\rangle=|\mathbf{s}_z\rangle$. At the center of the skyrmion, we have [see Fig. \ref{fig:CDW-F}(a)]
\begin{align}
|Z_1(\mathbf{r}=0)\rangle&=e^{i\alpha_f}|\mathbf{n}_z\rangle |\mathbf{s}_z\rangle \\
|Z_2(\mathbf{r}=0)\rangle&=e^{i\alpha_c}|-\mathbf{n}_z\rangle|\mathbf{s}_z\rangle,
\end{align}
such that both electron point towards the positive $z$ direction realizing thus a local ferromagnetic order. In the absence of Zeeman coupling the ferromagnetic state is realized for any value of $\theta_f=\theta_c$ such that both spins points in the same direction. The spin magnetization shown in Fig. \ref{fig:CDW-F}(b) is therefore identical in both sublattices,
\begin{align}
	\mathbf{M_S}_A=\mathbf{M_S}_B=\frac{\lambda^2}{\lambda^2+r^2}\mathbf{s}_z.
\end{align}

\subsubsection{Canted anti-ferromagnetic skyrmion}

\begin{figure}[t]
    (a)\includegraphics[width=8cm]{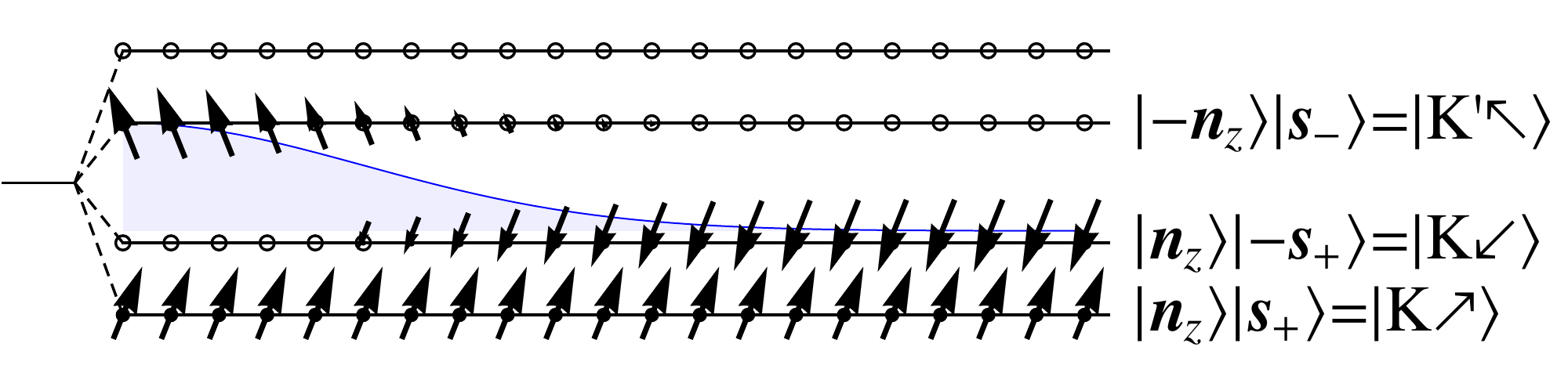}
    (b)\includegraphics[width=8cm]{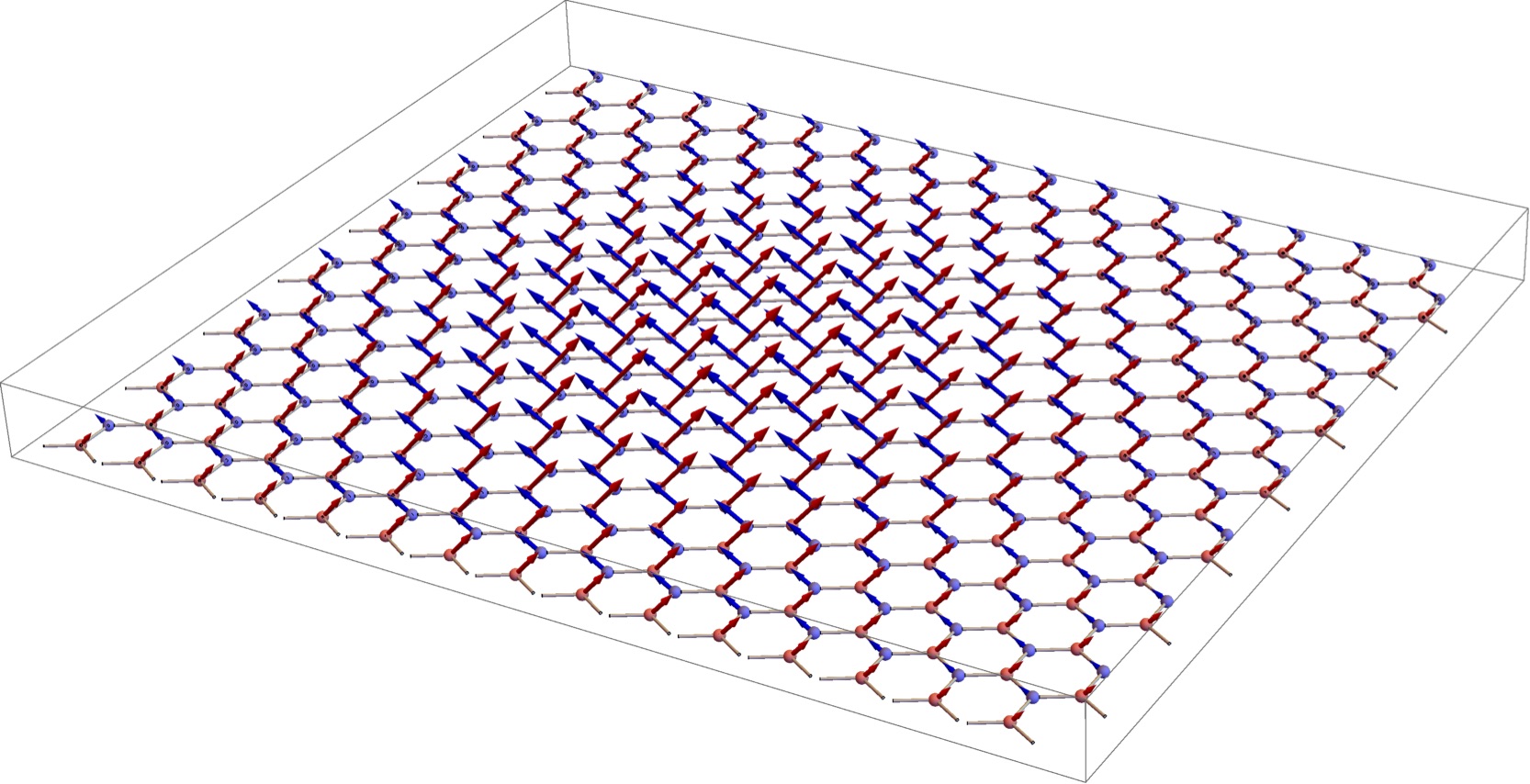}
	\caption{\textbf{Canted anti-ferromagnetic skyrmion} in the CDW background. (a) Filled sub-LLs at the center and infinity and the corresponding spinors. (b) Spin magnetization on the A (red) and B (blue) sublattices. At infinity, because both spins point in the opposite direction and reside in the same sublattice, the total spin magnetization vanishes, while at the center, the spins are canted relative to the direction of the magnetic field and reside on different sublattices.}
	\label{fig:CDW-CAF}
\end{figure} 

Similarly to the transition between the F and the CAF phases of the QHFM background [see phase diagram in Fig. \ref{fig:PhaseDiagQHFM}], the
CAF skyrmions arise for $\theta_f=\theta_c$ and $\phi_c=\phi_f+\pi$ in Eqs. (\ref{eq:CDWBG1})-(\ref{eq:sf-sc}) with 
\begin{align}
    \cos\theta_f=-\frac{\Delta_Z}{2u_\perp}.
\end{align}
The spinors at the centor are given by 
\begin{align}
	|Z_1(\mathbf{r}=0)\rangle&=e^{i\alpha_f}|\mathbf{n}_z\rangle|\mathbf{s}_+\rangle, \\
	|Z_2(\mathbf{r}=0)\rangle&=e^{i\alpha_c}|-\mathbf{n}_z\rangle|\mathbf{s}_-\rangle,
\end{align}
as depicted in Fig. \ref{fig:CDW-CAF}(a), with 
\begin{align}
		|\mathbf{s}_+\rangle=\begin{pmatrix}
		\cos\frac{\theta_f}{2} \\ \sin\frac{\theta_f}{2} e^{i\varphi_f}
		\end{pmatrix},
		\quad \text{and} \quad
		|\mathbf{s}_-\rangle=\begin{pmatrix}
		\cos\frac{\theta_f}{2} \\ -\sin\frac{\theta_f}{2} e^{i\varphi_f}
		\end{pmatrix}.
		\label{eq:s+s-}
\end{align}
At the center, each spinor corresponds to a different sublattice. The spins on the different sublattices have the same orientation along the $z$ axis but opposite orientations in the plane corresponding thus to a canting between the spins in the two sublattices. We can see that at the border with the ferromagnetic skyrmion, $u_\perp=-\Delta_Z/2$, the canting angle $\theta_f$ reaches 0 such that the transition between the ferromagnetic and the canted anti-ferromagnetic skyrmion is continuous. In the absence of the Zeeman term, this skyrmion is anti-ferromagnetic with no preferred spin orientation. In the presence of the Zeeman coupling, it is thus energetically favorable to cant the spin relative to the direction of the magnetic field. The magnetization in the A and B sublattices shown in Fig. \ref{fig:CDW-CAF}(b) equals 
\begin{align}
\mathbf{M_S}_A=\frac{\lambda^2}{\lambda^2+r^2}\mathbf{s}_+, \quad
\mathbf{M_S}_B=\frac{\lambda^2}{\lambda^2+r^2}\mathbf{s}_-,
\end{align}
with $\mathbf{s}_\pm=(\pm\sin\theta_f\cos\varphi_f,\pm\sin\theta_f\sin\varphi_f,\cos\theta_f)$. The total spin magnetization is thus oriented along the $z$ direction equal to
\begin{equation}
\mathbf{M_S}=\mathbf{M_S}_A+\mathbf{M_S}_B=\frac{\lambda^2}{\lambda^2+r^2}2\cos\theta_f\mathbf{s}_z.
\end{equation}

\subsection{Kekul\'e distortion background}

\label{sec:skyrmKD}

Once again, the skyrmions in the KD background are similar to those in the CDW phase. At infinity, the pseudo-spins point in the same direction, but this direction points now at the equator of the Bloch sphere, while the spin magnetization vanishes. In contrast to this the pseudo-spins necessarily point in opposite directions at the skyrmion center. The precise skyrmion type is therefore determined by the orientation of the spins at the center and at infinity. 

The spinors at infinity read
\begin{align}
|F_1\rangle&=e^{i\alpha_f}|\mathbf{n}_\perp\rangle|\mathbf{s}_f\rangle, \label{eq:skyrmKD1} \\
|F_2\rangle&=e^{-i\alpha_f}|\mathbf{n}_\perp\rangle |-\mathbf{s}_f\rangle,
\end{align}
while at the center we have 
\begin{align}
|C_2\rangle=e^{i\alpha_c}|-\mathbf{n}_\perp\rangle|\mathbf{s}_c\rangle, \label{eq:skyrmKD3}
\end{align}
where $|\mathbf{s}_f\rangle$ and $|\mathbf{s}_c\rangle$ are the spin spinors given by Eq. (\ref{eq:sf-sc}). 
The anisotropic energy of the skyrmion is given by
\begin{align}
	E_A[Z]=&-2Au_\perp\nonumber \\
	&-A(1+\cos\theta_f\cos\theta_c+\sin\theta_f\sin\theta_c\cos\varphi)(u_\perp+u_z)\nonumber \\
	&-A\Delta_Z\left(\cos\theta_f+\cos\theta_c\right)
\end{align}

Similarly to the KD background, we obtain a F and a CAF skyrmion because the spin orientations are similar at the center. However, because the pseudo-spin magnetization points at the equator of the Bloch sphere, the spin magnetization on the A and B sublattices possess a component in the $xy$ plane.

\subsubsection{Ferromagnetic skyrmion}

\begin{figure}[t]
    (a)\includegraphics[width=8cm]{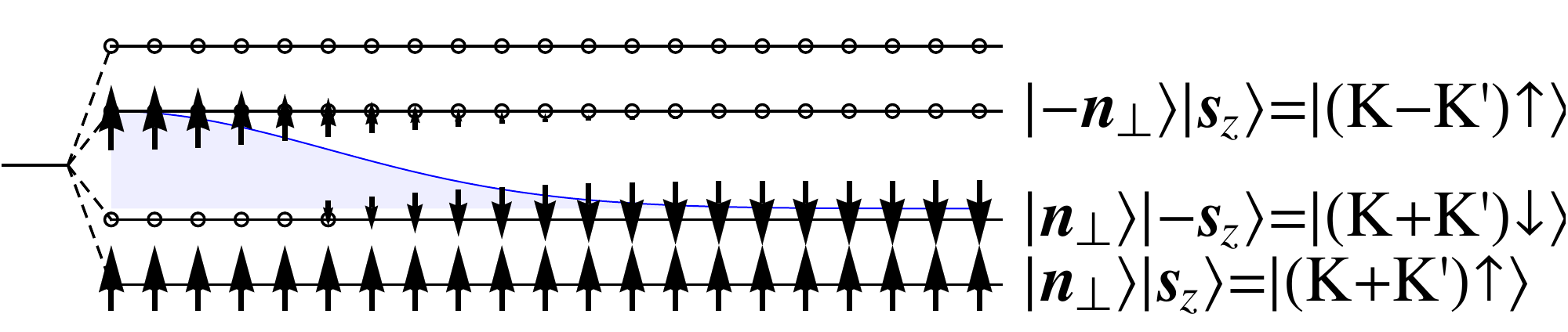}
    (b)\includegraphics[width=7cm]{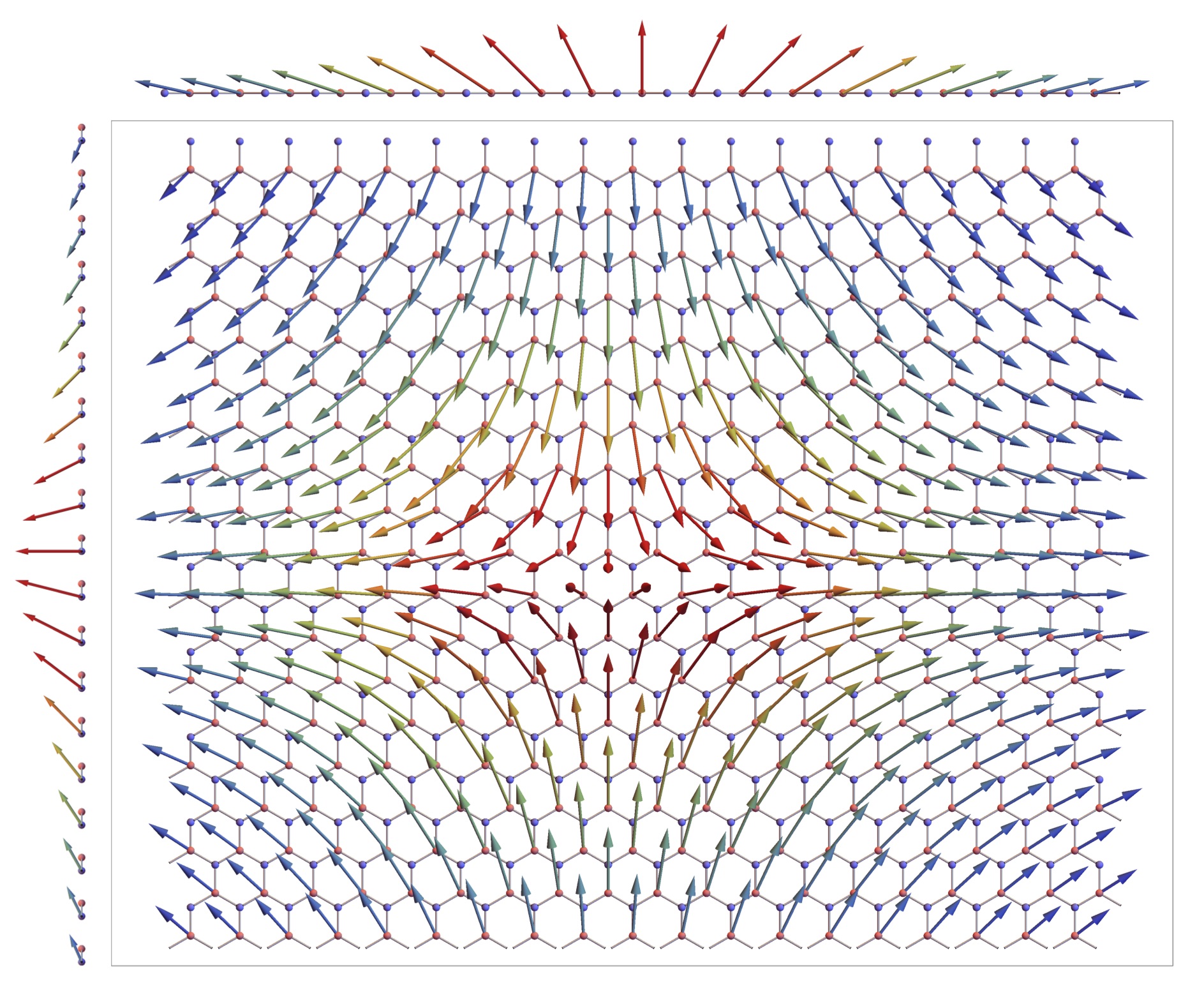}
    (c)\includegraphics[width=7cm]{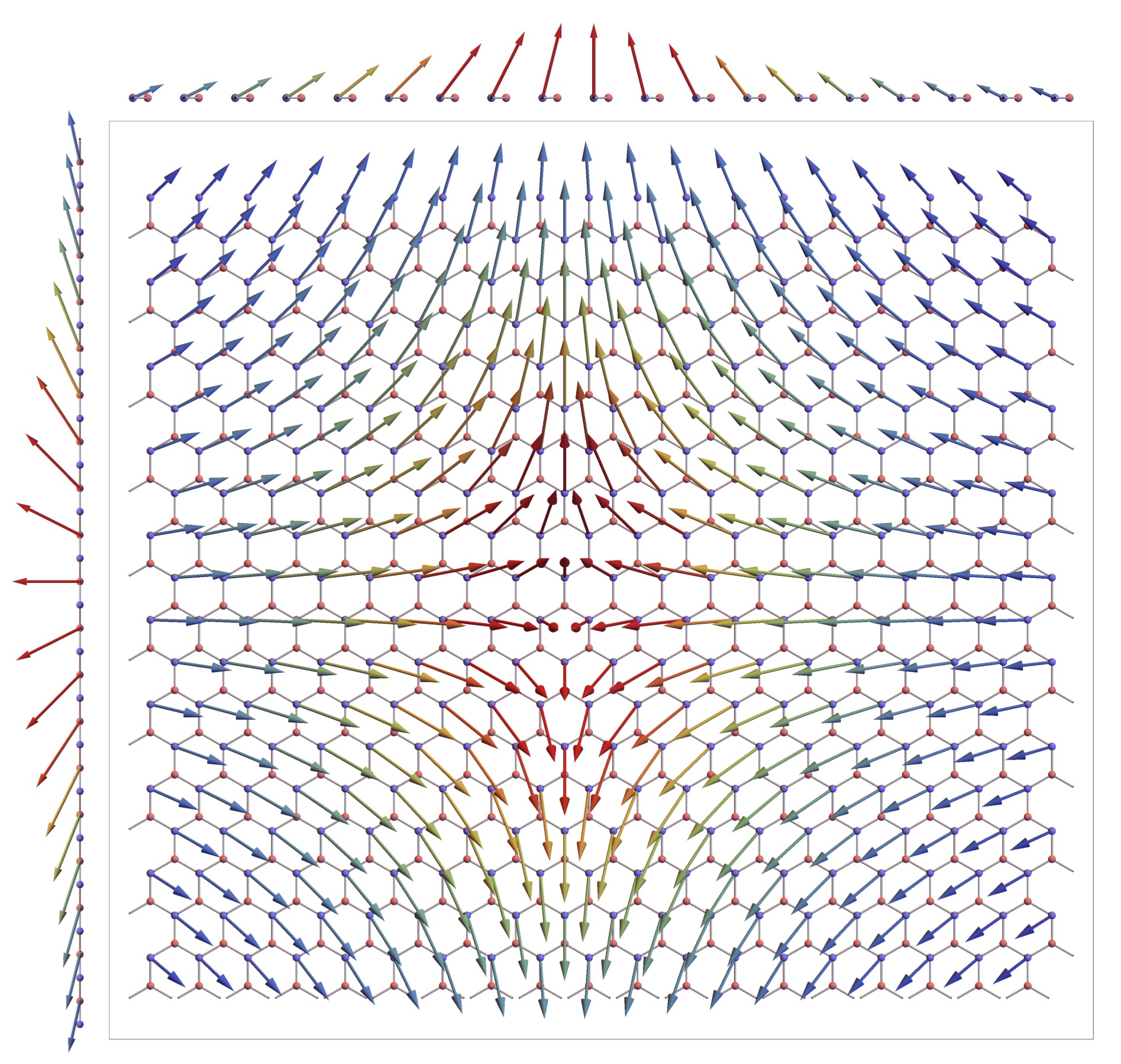}
	\caption{\textbf{Ferromagnetic skyrmion in the KD background} : a) Filled sub-LLs at the center and infinity and the corresponding spinors. b) and c) Top view of the spin magnetization on the A and B sublattices respectively. The side views show the line cuts of the spin magnetization along the axes $x=0$ (left) and $y=0$ (above). The color coding represents the $z$ component of the spin magnetization. We can see that both spin point along the $z$ direction at the center corresponding to a ferromagnetic skyrmion. The spin magnetization possess a $1/r$ tail in the $xy$ plane originating from the fact that the pseudo-spin points at the equator of the Bloch sphere.}
	\label{fig:KD-F}
\end{figure}

The F skyrmion in the KD background is reached for $\theta_f=\theta_c=0$ in Eqs. (\ref{eq:skyrmKD1})-(\ref{eq:skyrmKD3}) and (\ref{eq:sf-sc}). At the center, the spinors have the expression [see Fig. \ref{fig:KD-F}(a)]
\begin{align}
	|Z_1(\mathbf{r}=0)\rangle&=e^{i\alpha_f}|\mathbf{n}_\perp\rangle|\mathbf{s}_z\rangle, \\
	|Z_2(\mathbf{r}=0)\rangle&=e^{i\alpha_c}|-\mathbf{n}_\perp\rangle|\mathbf{s}_z\rangle,
\end{align}
At the center, the spins on each sublattice point towards the $z$ direction forming thus a ferromagnetic pattern. Because the pseudo-spin lie at the equator of the Bloch sphere, both sublattices are occupied equally and we observe a magnetization in the $xy$ plane which has a $1/r$ tail. The magnetization on each sublattice is given by 
\begin{align}
	\mathbf{M_S}_A=\frac{1}{\lambda^2+r^2}R_z(\alpha)\begin{pmatrix}
	-\lambda x \\ \lambda y \\ \lambda^2
	\end{pmatrix}\\
	\mathbf{M_S}_B=\frac{1}{\lambda^2+r^2}R_z(\alpha)\begin{pmatrix}
	\lambda x \\ -\lambda y \\ \lambda^2
	\end{pmatrix}
\end{align} 
and shown in Fig. \ref{fig:KD-F}(b) and (c), respectively. 
One thus notices that the total magnetization in the $xy$ plane vanishes and is oriented along the $z$ axis, 
\begin{equation}
\mathbf{M_S}=\mathbf{M_S}_A + \mathbf{M_S}_B=2\frac{\lambda^2}{\lambda^2+r^2}\mathbf{s}_z.
\end{equation}

\subsubsection{Canted anti-ferromagnetic skyrmion}

\begin{figure}[h!]
    (a)\includegraphics[width=8cm]{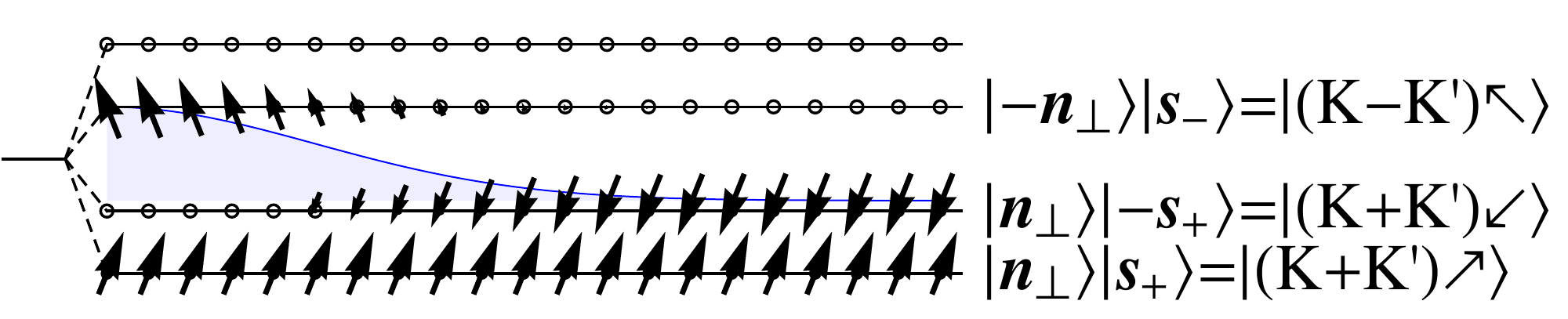}
    (b)\includegraphics[width=6cm]{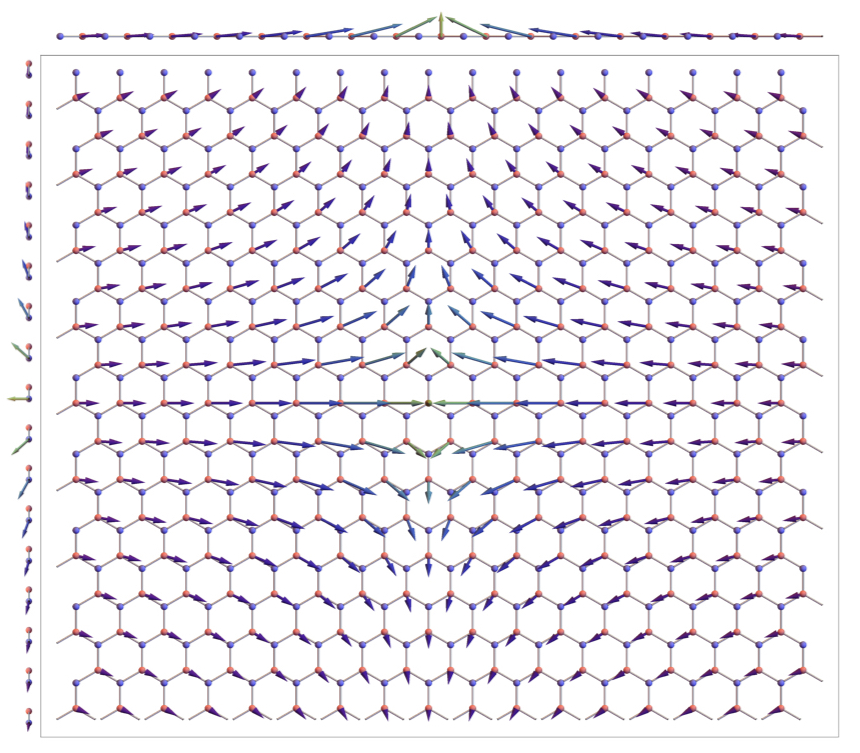}
    (c)\includegraphics[width=6cm]{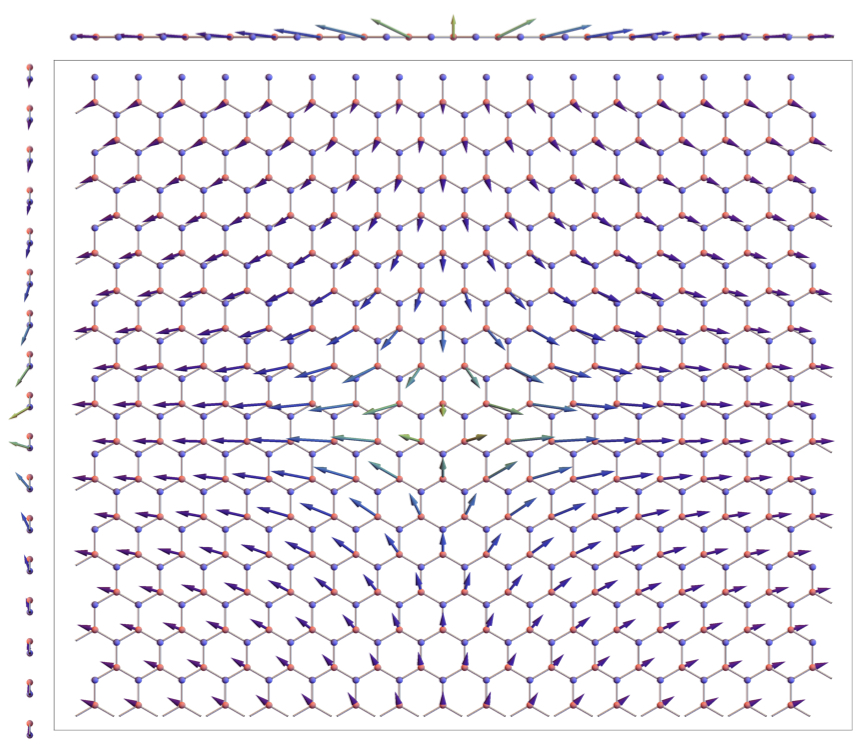}
    (d)\includegraphics[width=6cm]{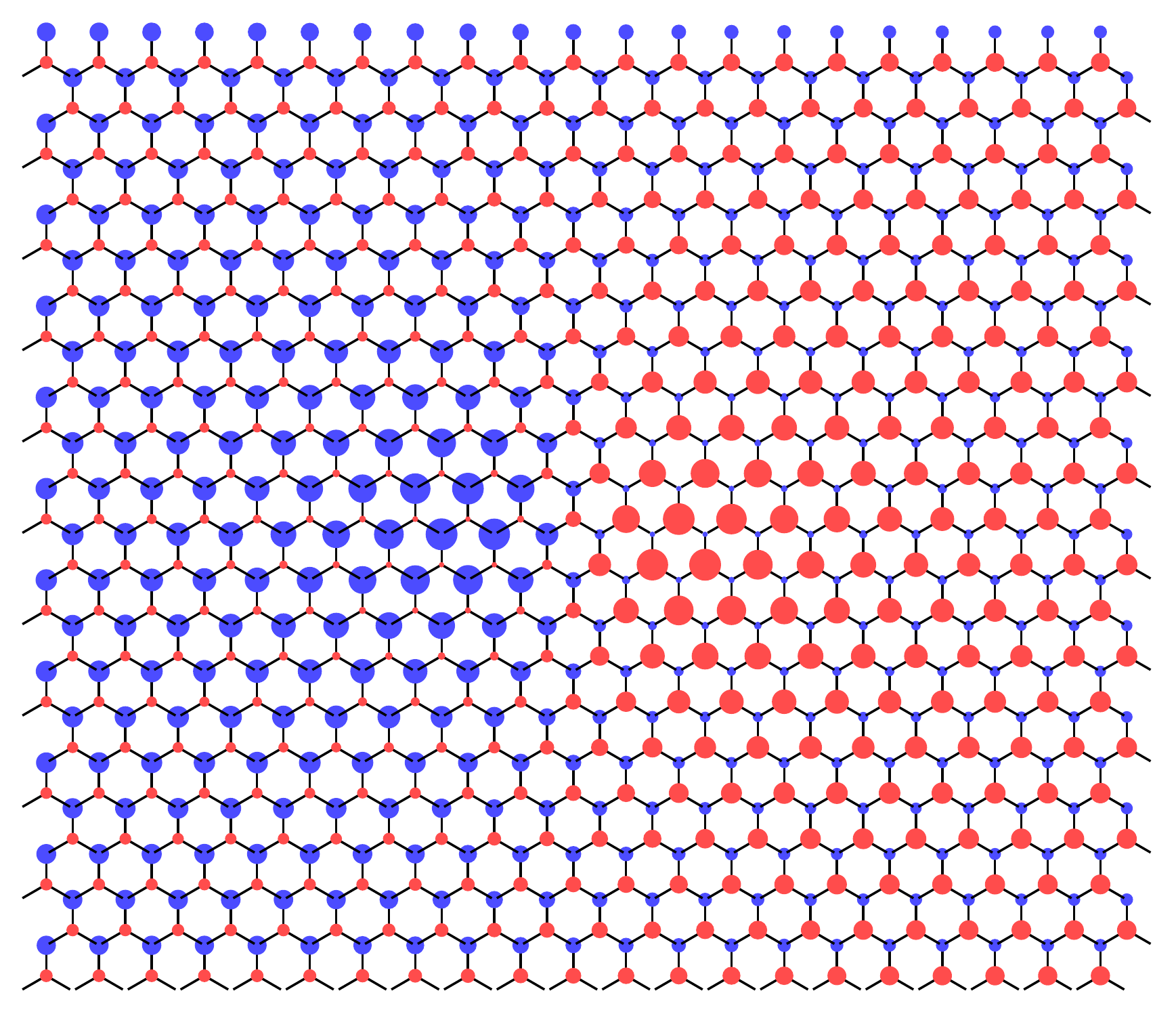}
	\caption{\textbf{Canted anti-ferromagnetic skyrmion in the KD background}. (a) Filled sub-LLs at the center and infinity and the corresponding spinors. (b) and (c) Spin magnetization on the A and B sublattices for $\theta_f=\pi/3$. The side views show the line cuts of the spin magnetization along the axes $x=0$ (left) and $y=0$ (above). Because the pseudo-spin is oriented along the equator of the Bloch sphere, this skyrmion is quite different from the CAF skyrmion in the CDW background. (d) Electronic density on the A and B sublattices.  We observe a double core structure where the pseudo-spin points at the north and south pole of the Bloch sphere corresponding to an imbalance of the A and B sublattice occupation at the cores.}
	\label{fig:KD-CAF}
\end{figure}

The CAF skyrmion is obtained for $\theta_f=\theta_c$ and $\phi_c=\phi_f+\pi$  in Eqs. (\ref{eq:skyrmKD1})-(\ref{eq:skyrmKD3}) and (\ref{eq:sf-sc}) with the canting angle set by 
\begin{align}
    \cos\theta_f=-\frac{\Delta_Z}{(u_\perp+u_z)}
\end{align}
We can see that at the transition with the ferromagnetic skyrmion located at $u_\perp+u_z=0$, we have $\theta_f=\theta_c=0$ which corresponds to the ferromagnetic skyrmion, such that the transition between these two phases is continuous again. We label this skyrmion canted anti-ferromagnetic because the spinors at the center have the expression [see Fig. \ref{fig:KD-CAF}(a)]
	\begin{align}
	|Z_1(\mathbf{r}=0)\rangle&=e^{i\alpha_f}|\mathbf{n}_\perp\rangle|\mathbf{s}_+\rangle, \\
	|Z_2(\mathbf{r}=0)\rangle&=e^{i\alpha_c}|-\mathbf{n}_\perp\rangle|\mathbf{s}_-\rangle, 
	\end{align}
with $\mathbf{s}_+$ and $\mathbf{s}_-$ given by Eq. (\ref{eq:s+s-}). However the spin magnetization of this state is quite different from that in the CAF skyrmion in the CDW background because the pseudo-spin spinors point towards the equator of the Bloch sphere, and the spin magnetizations on the A and B sublattice are
\begin{align}
	\mathbf{M_S}_A=\frac{1}{\lambda^2+r^2}R_z(\varphi_f)
	\begin{pmatrix}
	\lambda(x\cos\alpha-y\sin\alpha) \\ -\lambda(x\sin\alpha+y\cos\alpha)\cos\theta_f \\ \lambda^2\cos\theta_f
	\end{pmatrix} \\
	\mathbf{M_S}_B=\frac{1}{\lambda^2+r^2}R_z(\varphi_f)
	\begin{pmatrix}
	-\lambda(x\cos\alpha-y\sin\alpha) \\ \lambda(x\sin\alpha+y\cos\alpha)\cos\theta_f \\ \lambda^2\cos\theta_f
	\end{pmatrix},
\end{align}
respectively,
where $\alpha=\alpha_f+\alpha_c+\varphi_f-\varphi_P$. Figures \ref{fig:KD-CAF}(b) and (c) show the spin magnetization of the CAF skyrmion on the two different sublattices. Once again, because the pseudo-spin points at a point on the equator of the Bloch sphere, we observe a spin magnetization in the $xy$ plane. The main difference with the F skyrmion is that the total magnetization is oriented along $z$ and is reduced by a factor $\cos\theta_f$ compared to the F skyrmion. Notice that the magnetization along the axis $y'=x\sin\alpha+y\cos\alpha$ is also reduced by a factor $\cos\theta_f$, as compared to the magnetization along the other axis. Due to the spin-valley entanglement, this skyrmion presents also a non-uniform electronic density profile given by
\begin{align}
    \rho_A=1+\frac{\lambda\sin\theta_f}{\lambda^2+r^2}(x\cos\varphi-y\sin\varphi), \\
    \rho_B=1-\frac{\lambda\sin\theta_f}{\lambda^2+r^2}(x\cos\varphi-y\sin\varphi),
\end{align}
which gives rise to a double core structure, as it is shown in Fig. \ref{fig:KD-CAF}(d). This pattern is due to interferences between $C_2$ and $F_2$ as one moves away from the skyrmion center such that the pseudo-spin points in directions close to the south and north pole at the core centers. We thus observe an imbalance in the electronic density at the core centers analogously to a CDW pattern. Such skyrmions are reminiscent of bimerons\cite{Brey1996} in double-layer 2DEGs where the pseudo-spin refers to the layer index instead of the sublattice index. Notice that several skyrmions at $\nu=\pm1$ also present such patterns\cite{Lian2017}.

\subsection{Canted anti-ferromagnetic background}

\label{sec:skyrmCAF}

In the CAF background, the spinors $|f_1\rangle$ and $|f_2\rangle$ given by Eq. (\ref{eq:spCAF1}) and (\ref{eq:spCAF2}) do not have their spin or pseudo-spin in common. Therefore, when mixing them, it is not possible to factor them such as it was done for the other skyrmions. The spinors at infinity are thus a superposition of $|f_1\rangle$ and $|f_2\rangle$,
\begin{align}
|F_1\rangle =& e^{i\alpha_f}\left(\cos\frac{\theta_f}{2}|\mathbf{n}_z\rangle|\mathbf{s}_+\rangle +e^{i\varphi_f}\sin\frac{\theta_f}{2}|-\mathbf{n}_z\rangle|\mathbf{s}_-\rangle\right), \label{eq:skyrmCAF1} \\
\nonumber
|F_2\rangle =& e^{-i\alpha_f}\left(-\sin\frac{\theta_f}{2}e^{-i\varphi_f}|\mathbf{n}_z\rangle|\mathbf{s}_+\rangle\right.\\
&\left.+\cos\frac{\theta_f}{2}|-\mathbf{n}_z\rangle|\mathbf{s}_-\rangle\right),
\end{align}
with $\mathbf{s}_\pm=(\pm\sin\alpha\cos\beta,\pm\sin\alpha\sin\beta,\cos\alpha)$ where $\alpha$ is the canting angle of the QHFM background set by Eq. (\ref{eq:alphaCAF}), while the center spinor is 
\begin{align}\nonumber
|C_2\rangle =& e^{i\alpha_c}\left(-\sin\frac{\theta_c}{2}e^{-i\varphi_c}|\mathbf{n}_z\rangle|-\mathbf{s}_+\rangle \right.\\
&\left.+\cos\frac{\theta_c}{2}|-\mathbf{n}_z\rangle|-\mathbf{s}_-\rangle\right). \label{eq:skyrmCAF3}
\end{align}
The spinors at the center are thus also entangled similarly to the QHFM background.
The anisotropic energy of the skyrmions in the CAF background is 
\begin{align}
E_A[Z]= &-A\frac{\Delta_Z^2}{4u_\perp}\sin\theta_f\sin\theta_c\cos(\varphi) \nonumber \\
\nonumber
&-A\left(u_\perp+u_z-\frac{\Delta_Z^2}{4u_\perp}\right)\cos\theta_f\cos\theta_c\\
&-A\left(u_\perp-u_z+\frac{\Delta_Z^2}{4u_\perp}\right).
\label{eq:anisoCAF2}
\end{align}

\subsubsection{Tilted anti-ferromagnetic skyrmion}

\begin{figure}[t]
    (a)\includegraphics[width=8cm]{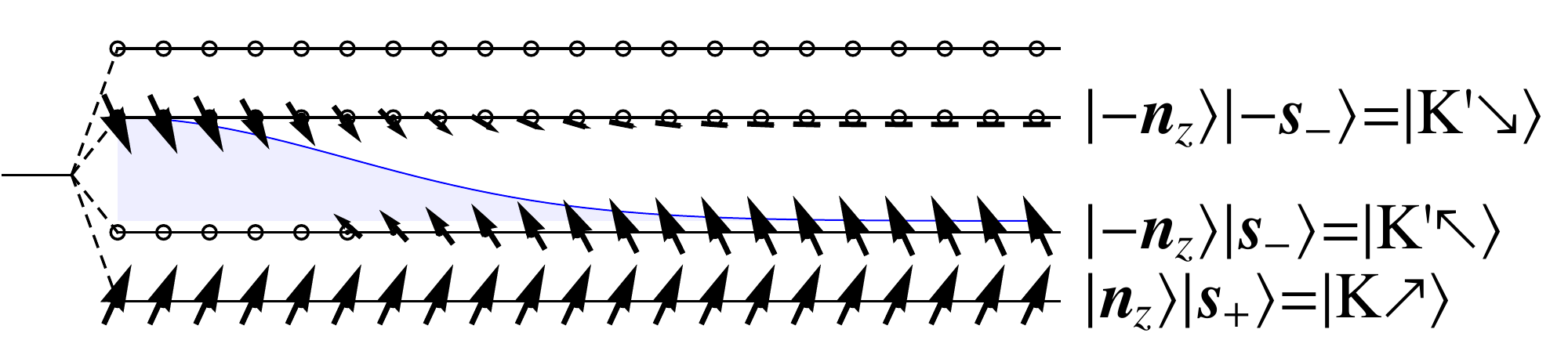}
    (b)\includegraphics[width=7cm]{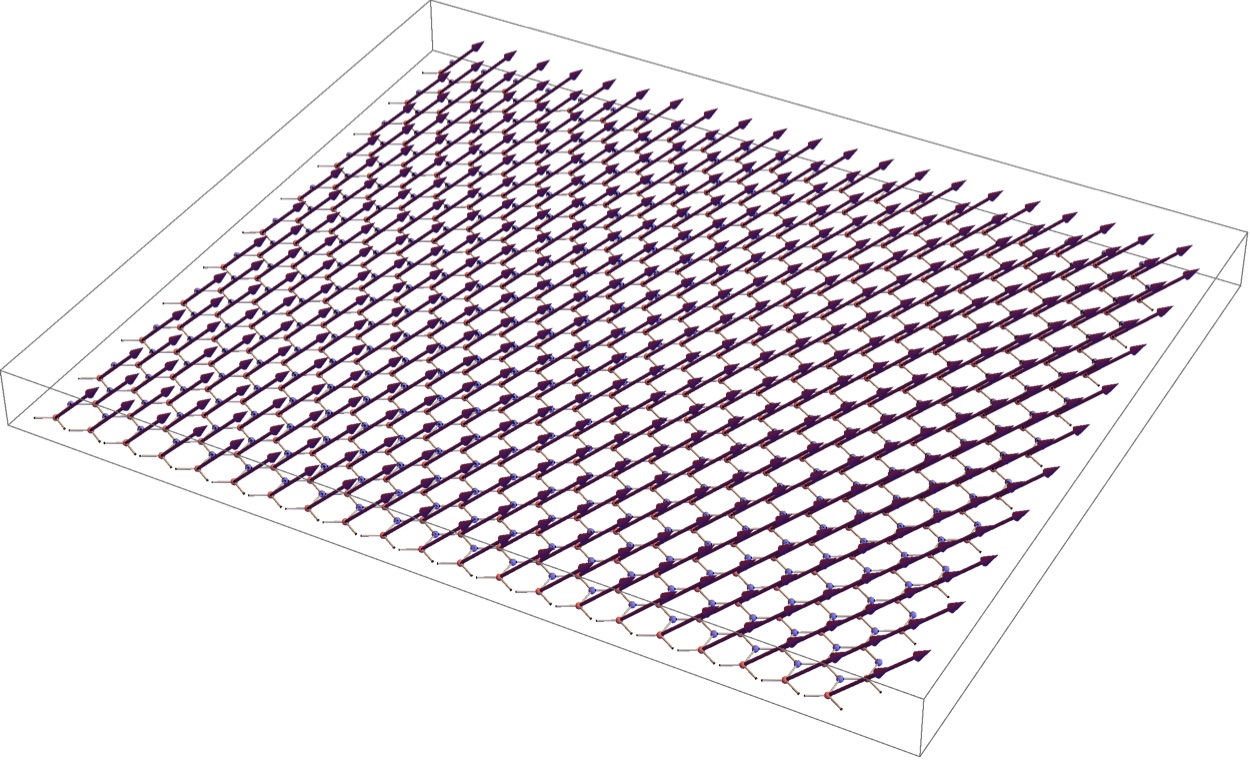}
    (c)\includegraphics[width=7cm]{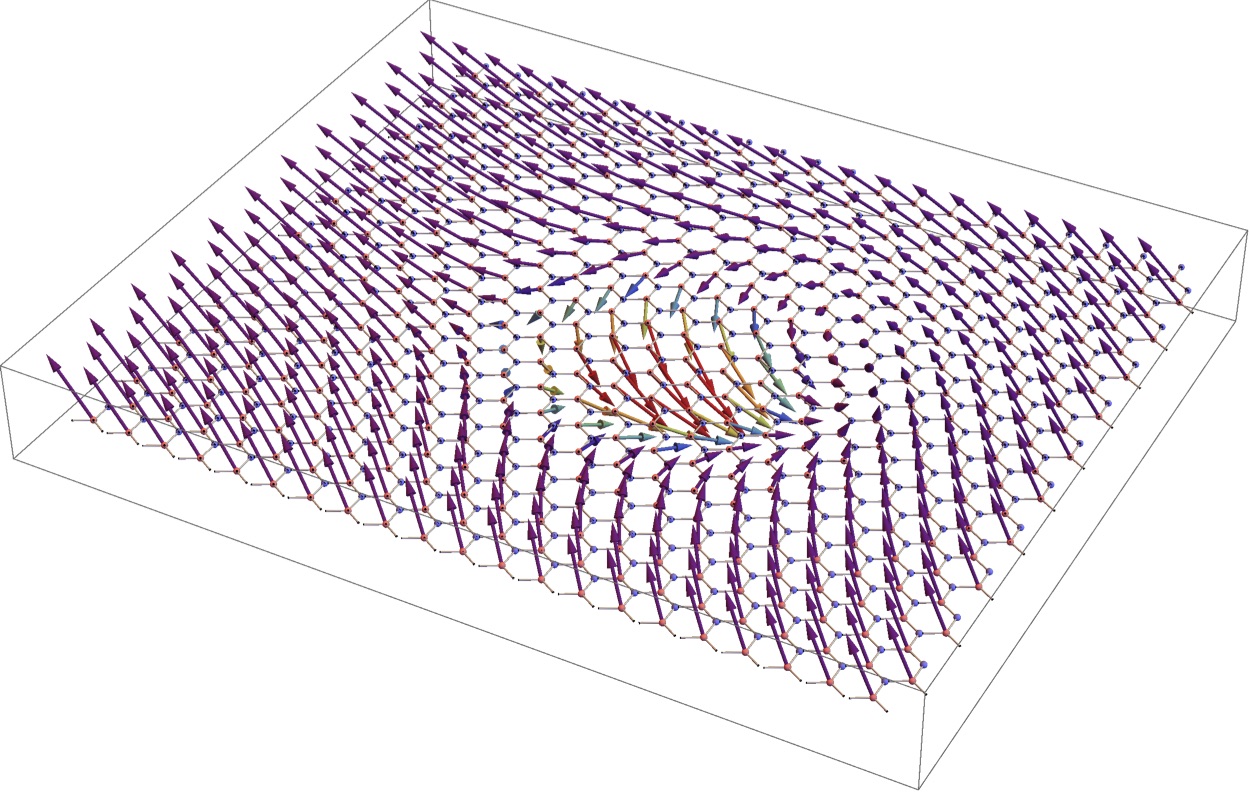}
	\caption{\textbf{Tilted anti-ferromagnetic skyrmion} in the CAF background for $\alpha=\pi/4$, $\gamma=\beta=0$. (a) Filled sub-LLs at the center and infinity and the corresponding spinors. (b) and (c) Spin magnetization on the A and B sublattices for $\alpha=\pi/4$. }
	\label{fig:CAF-TAF}
\end{figure}

The tilted anti-ferromagnetic (TAF) skyrmion is a rather special skyrmion with spin-valley entanglement because it does not have a counterpart in the QHFM background patterns that we encounter here. It
is realized for $\theta_f=\theta_c=0$ in Eqs. (\ref{eq:skyrmCAF1})-(\ref{eq:skyrmCAF3}) such that we have $|F_1\rangle=|f_1\rangle$ and $|F_2\rangle=|f_2\rangle$, while at the center the spinors are equal to 
\begin{align}
	|Z_1(\mathbf{r}=0)\rangle&=e^{i\alpha_f}|\mathbf{n}_z\rangle|\mathbf{s}_+\rangle 
	\label{eq:Z1TAF}\\
	|Z_2(\mathbf{r}=0)\rangle&=e^{i\alpha_c}|-\mathbf{n}_z\rangle|-\mathbf{s}_-\rangle ,
	\label{eq:Z2TAF}
\end{align}
as shown in Fig. \ref{fig:CAF-TAF}(a). 
The spinor $|Z_1\rangle$ has its pseudo-spin pointing towards $+\mathbf{n}_z$ while the spinor $|Z_2\rangle$ has its pseudo-spin pointing towards $-\mathbf{n}_z$ such that each spinor corresponds to a different sublattice. Hence, analogously to the AF skyrmion in the F background, this skyrmion texture involves states that are located only on one sublattice, while the other sublattice remains unaffected. As for the naming, we coin this skyrmion \textit{tilted anti-ferromagnetic} because, at the center, the spin on the sublattice A points towards $\mathbf{s}_+$, while the spin on the sublattice B points towards $-\mathbf{s}_-$. This skyrmion is thus ``tilted'' as opposed to the ``canted'' anti-ferromagnetic skyrmion. At infinity, the spinor $|Z_1\rangle$ still points towards $+\mathbf{s}_+$, while $|Z_2\rangle$ points towards $\mathbf{s}_-$ forming thus a canted ordering. For $\alpha=0$, we recover the expression for the AF skyrmion in the F background such that the skyrmion transition between the F background and the CAF background is continuous as well. The magnetization on the sublattices is given by :
\begin{align}
	\mathbf{M_S}_A=&\mathbf{s}_+ \\
	\nonumber
	\mathbf{M_S}_B=&\frac{r^2-\lambda^2}{\lambda^2+r^2}\mathbf{s}_-\\
	&+2\frac{\lambda}{\lambda^2+r^2}R_z(\beta)
	\begin{pmatrix}
		(x\cos\gamma-y\sin\gamma)\cos\alpha \\
		-(x\sin\gamma+y\cos\gamma)\\ 0
	\end{pmatrix},
\end{align}
with $\gamma=\alpha_c+\alpha_f+\beta$
[see Figs. \ref{fig:CAF-TAF}(b) and (c)]. The energy of the skyrmion is invariant under SO(2) rotations in the plane, however, the presence of a skyrmion spontaneously breaks this invariance with a preferred orientation.

\subsubsection{Easy-axis entanglement skyrmion}

\begin{figure}[t]
    (a)\includegraphics[width=8cm]{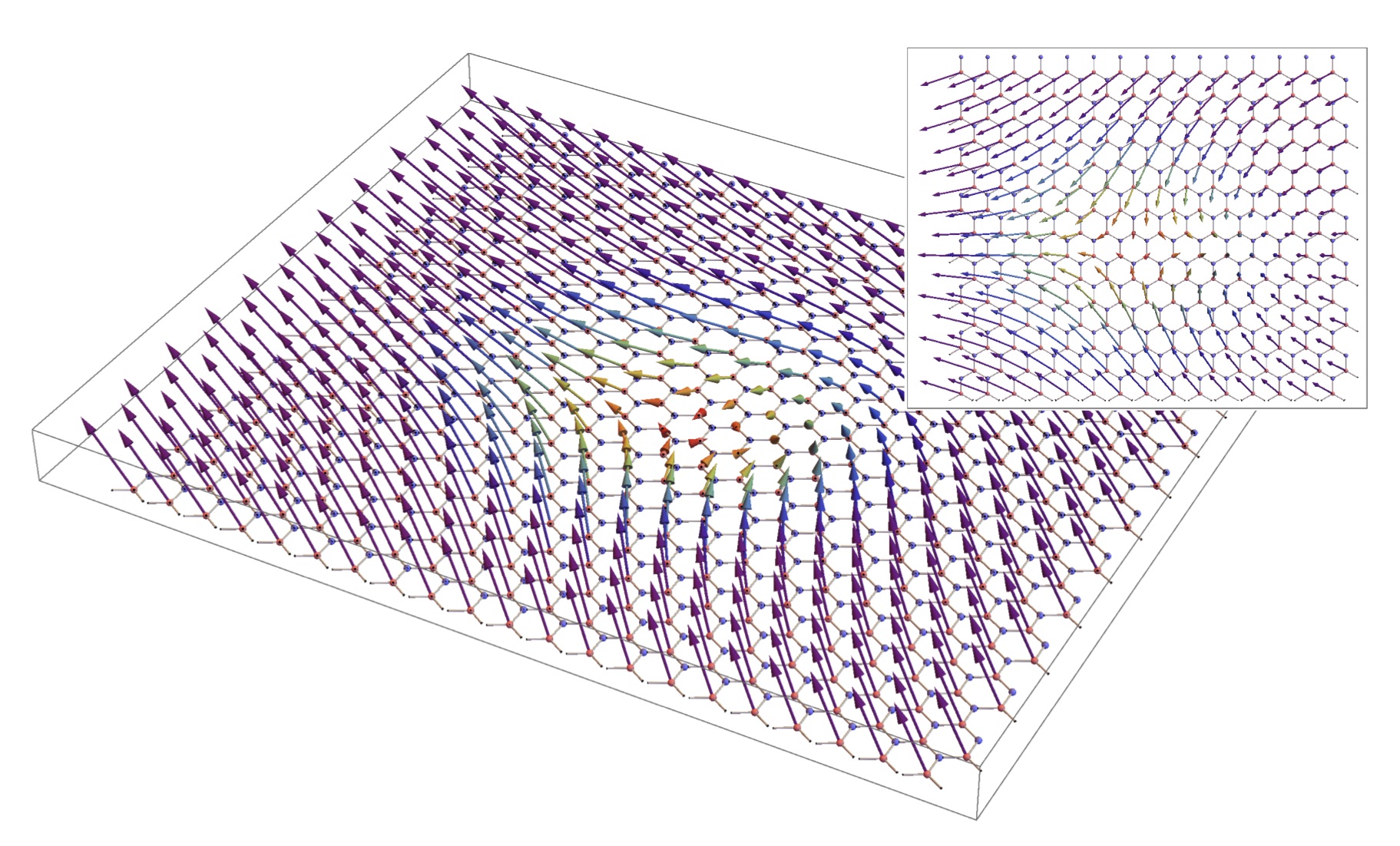}
    (b)\includegraphics[width=8cm]{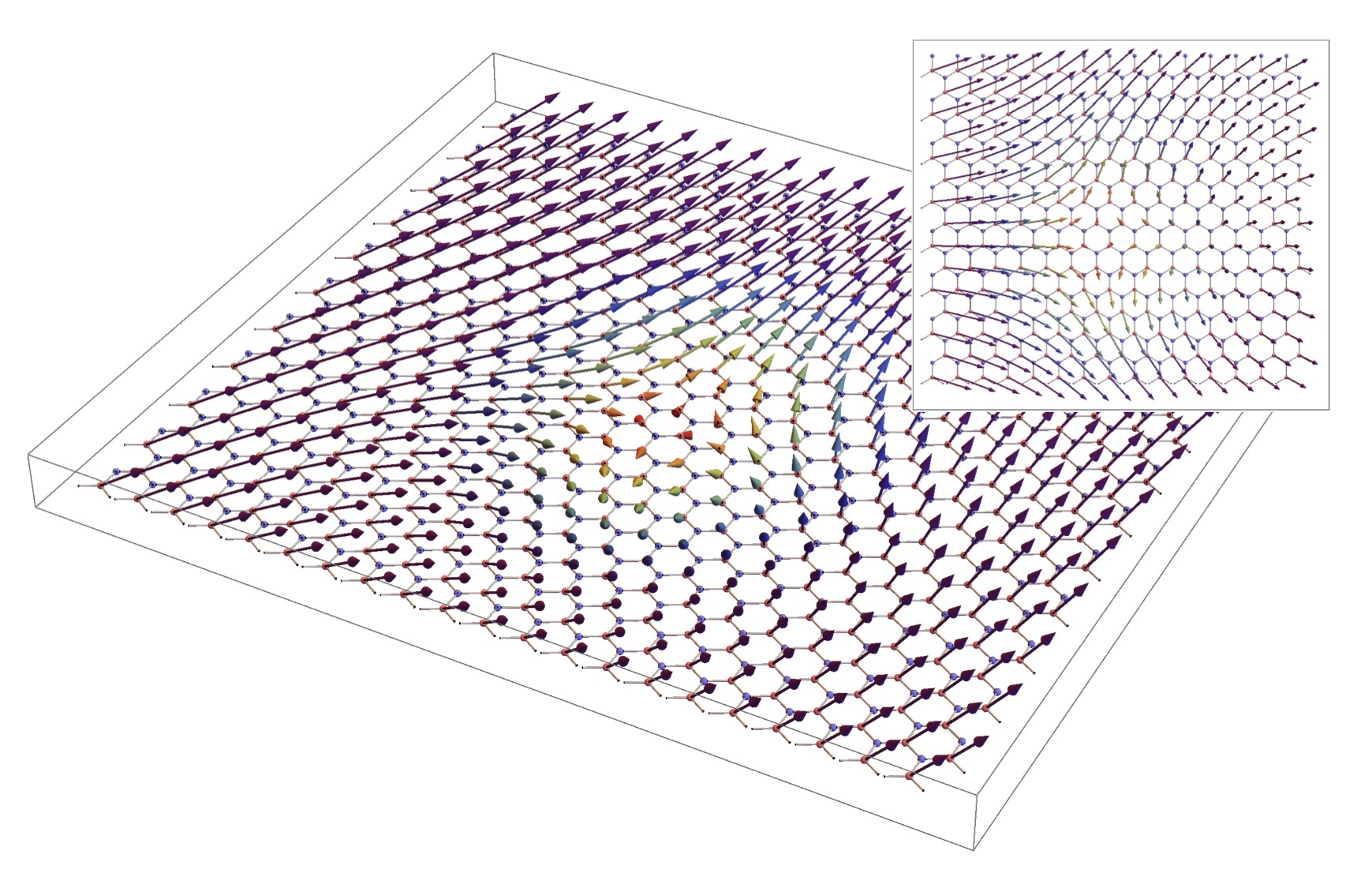}
	\caption{\textbf{Easy-axis entanglement skyrmion} in the CAF background for $\alpha=\pi/4$, $\gamma=\beta=0$. (a) and (b) Spin magnetization on the A and B sublattices. }
	\label{fig:CAF-EAE}
\end{figure}

The easy-axis entanglement (EAE) skyrmion is reached for $\theta_f=\theta_c=\pi/2$ and $\varphi_f=\varphi_c+\pi$ in Eqs. (\ref{eq:skyrmCAF1})-(\ref{eq:skyrmCAF3}). At the center, the spinors are thus superpositions of the two spinors located on the different sublattices with different spin orientation,
\begin{align}
	|Z_1(\mathbf{r}=0)\rangle&=\frac{e^{i\alpha_f}}{\sqrt{2}}(|\mathbf{n}_z\rangle|\mathbf{s}_+\rangle+e^{i\varphi_f}|-\mathbf{n}_z\rangle|\mathbf{s}_-\rangle) \\
	|Z_2(\mathbf{r}=0)\rangle&=\frac{e^{i\alpha_c}}{\sqrt{2}}(e^{-i\varphi_f}|\mathbf{n}_z\rangle|-\mathbf{s}_+\rangle+|-\mathbf{n}_z\rangle|-\mathbf{s}_-\rangle).
\end{align}
The spin magnetizations on the A and B sublattices are given by
\begin{align}\nonumber
	\mathbf{M_S}_A=&\frac{r^2}{\lambda^2+r^2}\mathbf{s}_+\\
	&-2\frac{\lambda}{\lambda^2+r^2}R_z(\beta)
	\begin{pmatrix}
		(x\cos\gamma-y\sin\gamma)\cos\alpha \\
		-(x\sin\gamma+y\cos\gamma)\\ 0
	\end{pmatrix}, \\
	\nonumber
	\mathbf{M_S}_B=&\frac{r^2}{\lambda^2+r^2}\mathbf{s}_-\\
	&+2\frac{\lambda}{\lambda^2+r^2}R_z(\beta)
	\begin{pmatrix}
		(x\cos\gamma-y\sin\gamma)\cos\alpha \\
		-(x\sin\gamma+y\cos\gamma)\\ 0
	\end{pmatrix},
\end{align}
with $\gamma=\alpha_c+\alpha_f+\varphi_f+\beta$, and plotted in Figs. \ref{fig:CAF-EAE}(a) and (b), respectively. We can see that, at the center of the skyrmion ($\mathbf{r}=0$), the spin magnetization vanishes on both sublattices. We do not present a diagram for the sub-LL because, due to the entanglement, there is no clear interpretation of the involved sub-LLs in terms of spin and pseudo-spin indices, and the diagram is therefore given by the generic one in Fig. \ref{fig:LLSk}.

\section{Energy, size and transition lines}

\label{sec:energy_size}

\subsection{Energy}
\label{sec:energy}

\begin{figure}[t]
    (a)\includegraphics[width=8cm]{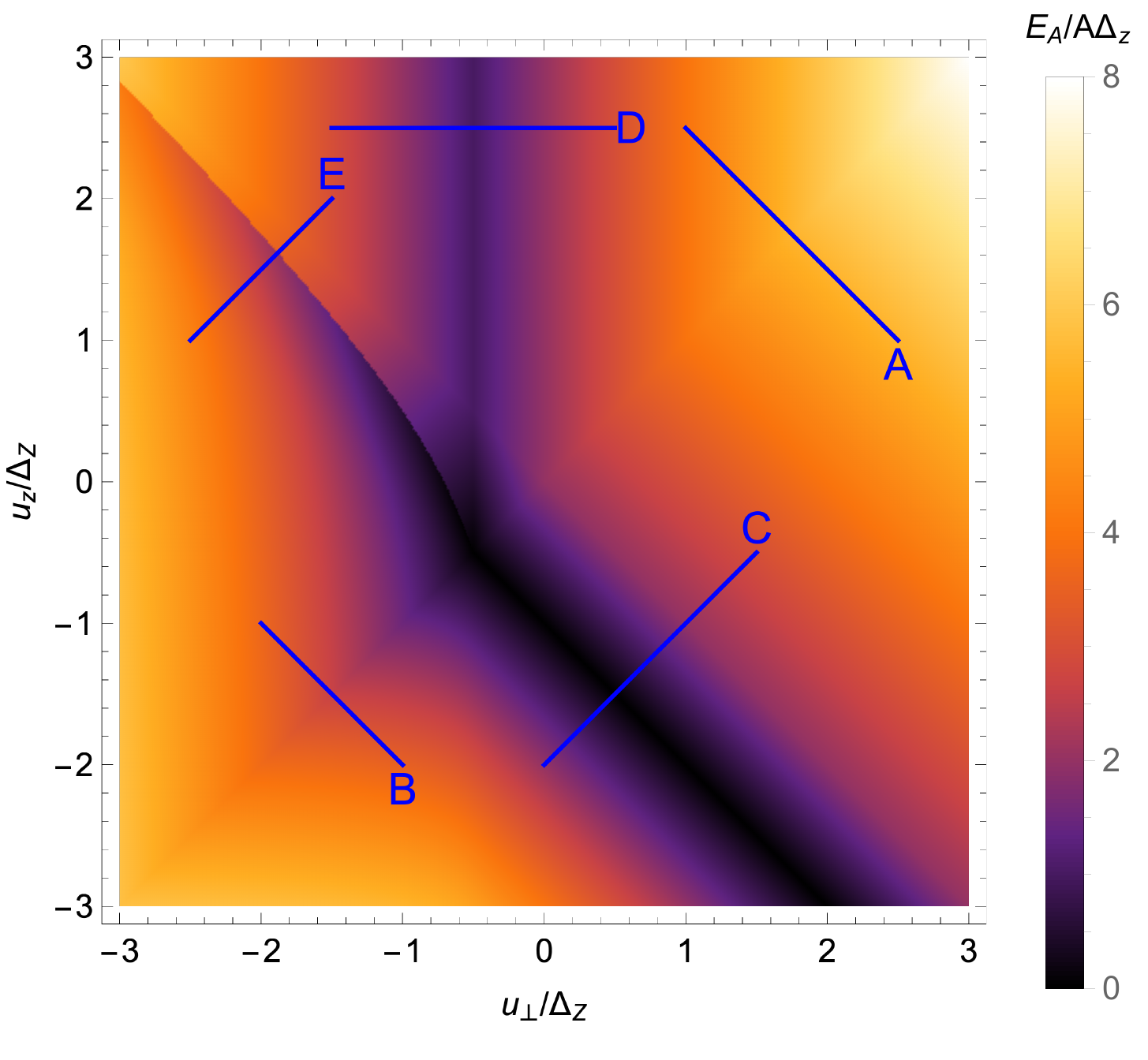}\\
    (b)\includegraphics[width=3.5cm]{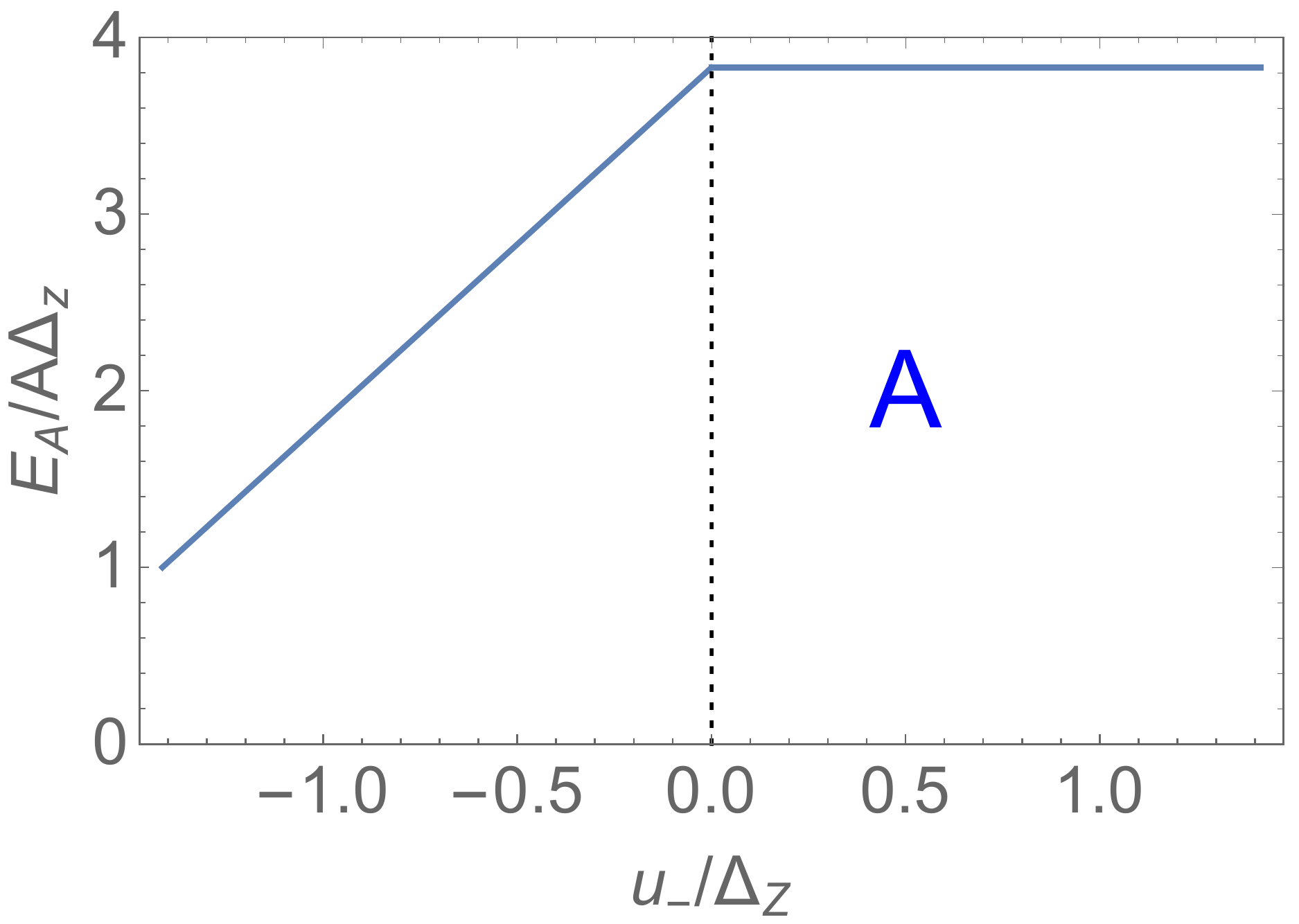}
    (c)\includegraphics[width=3.5cm]{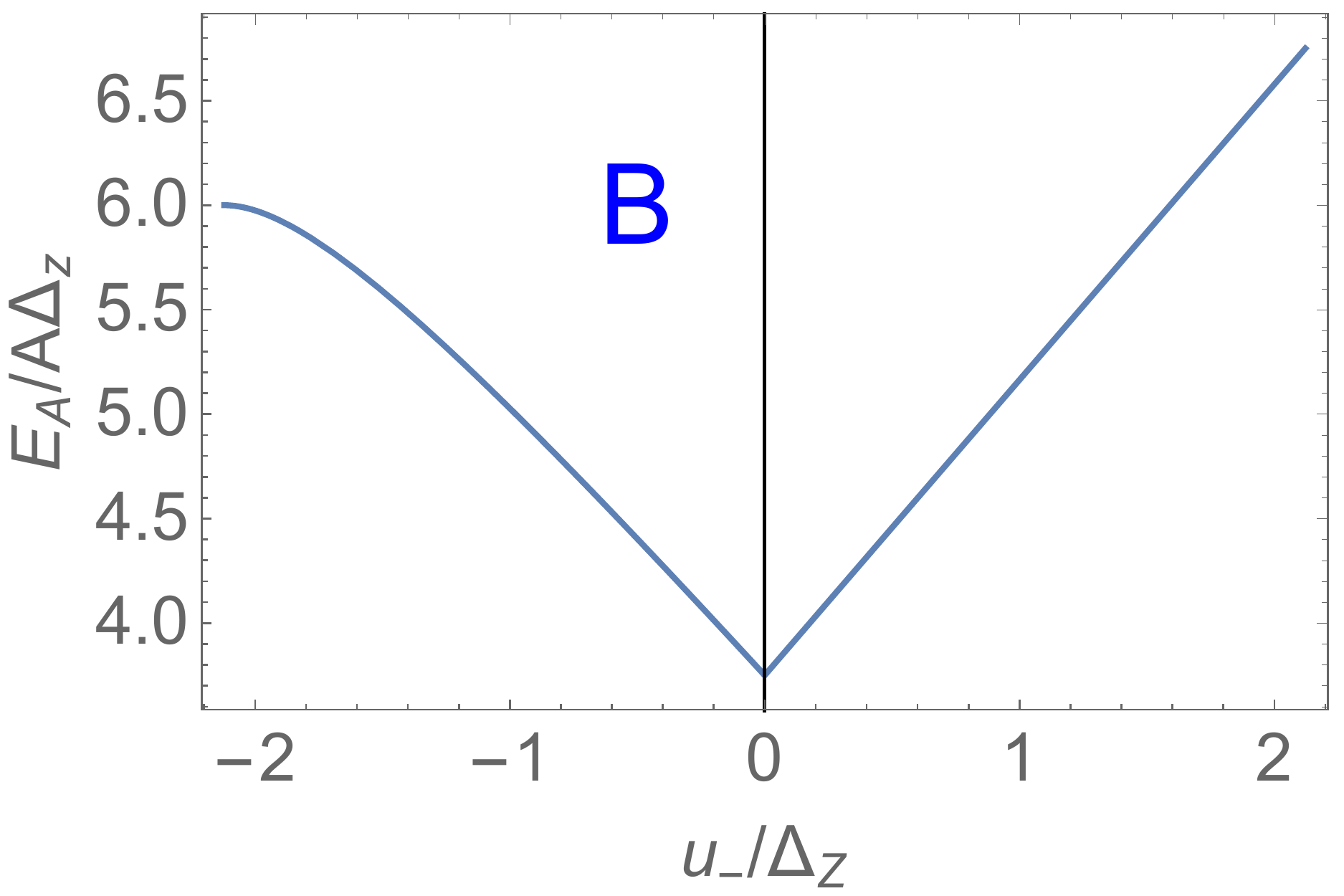}\\
    (d)\includegraphics[width=3.5cm]{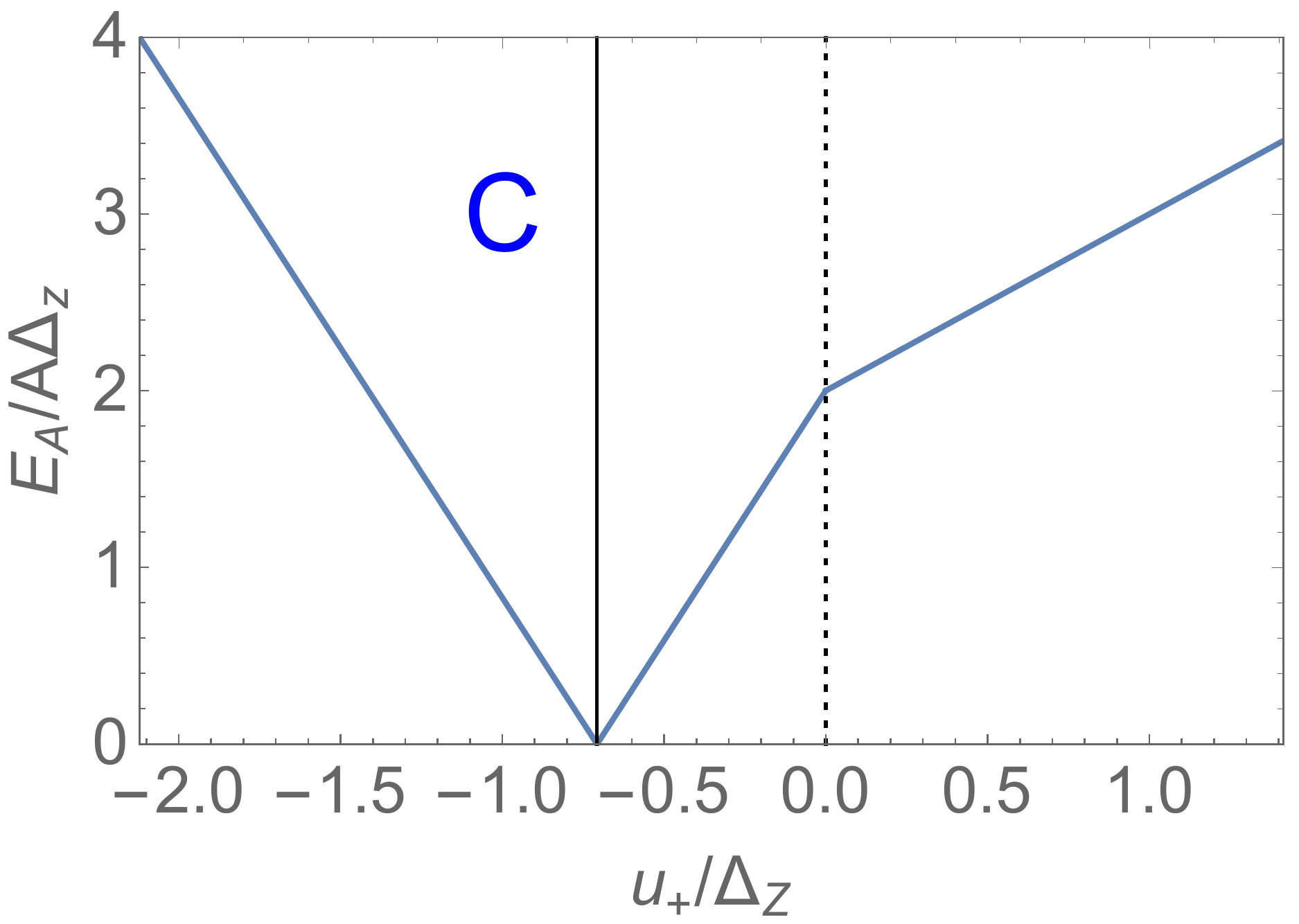}
    (e)\includegraphics[width=3.5cm]{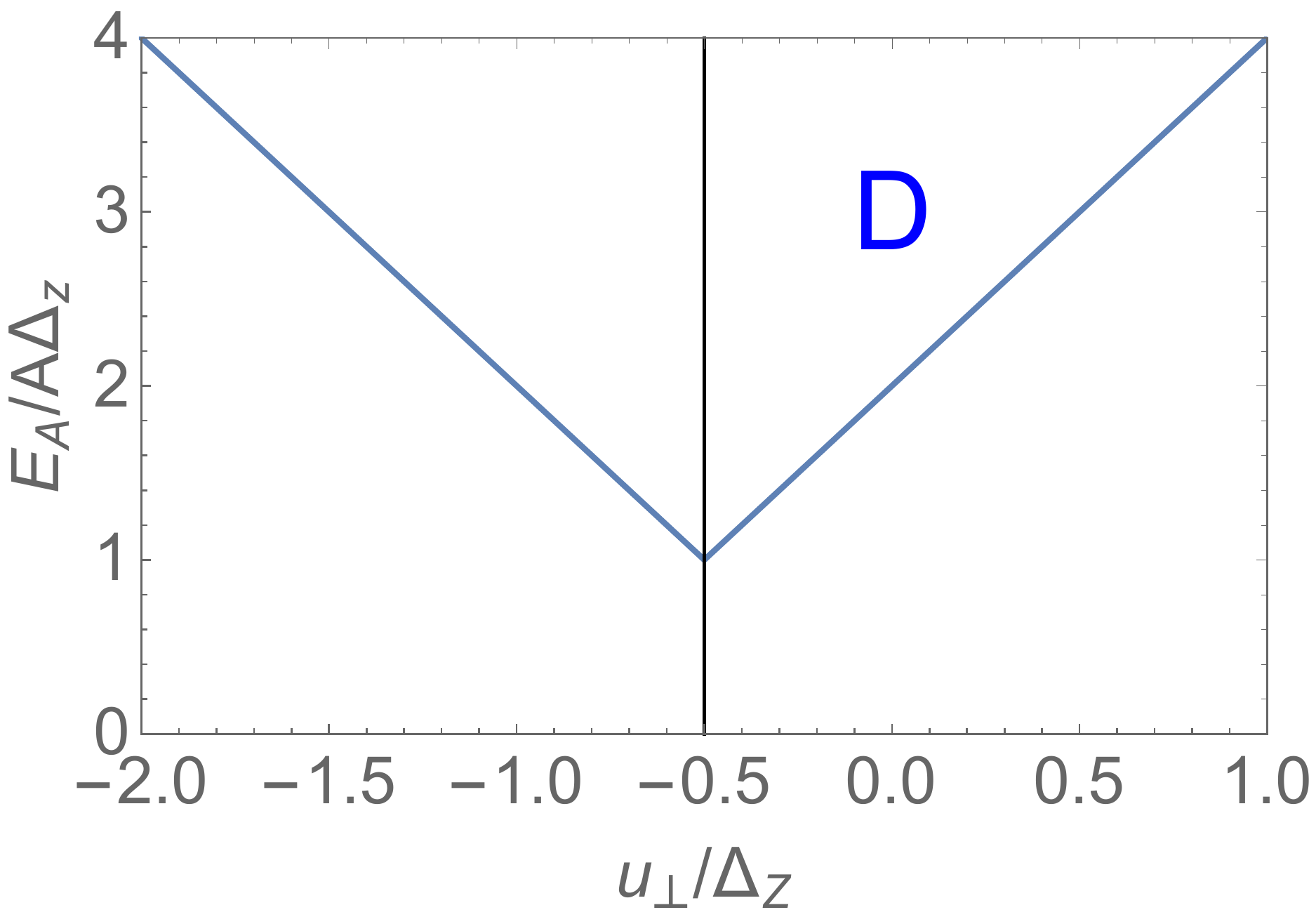}
    (f)\includegraphics[width=3.5cm]{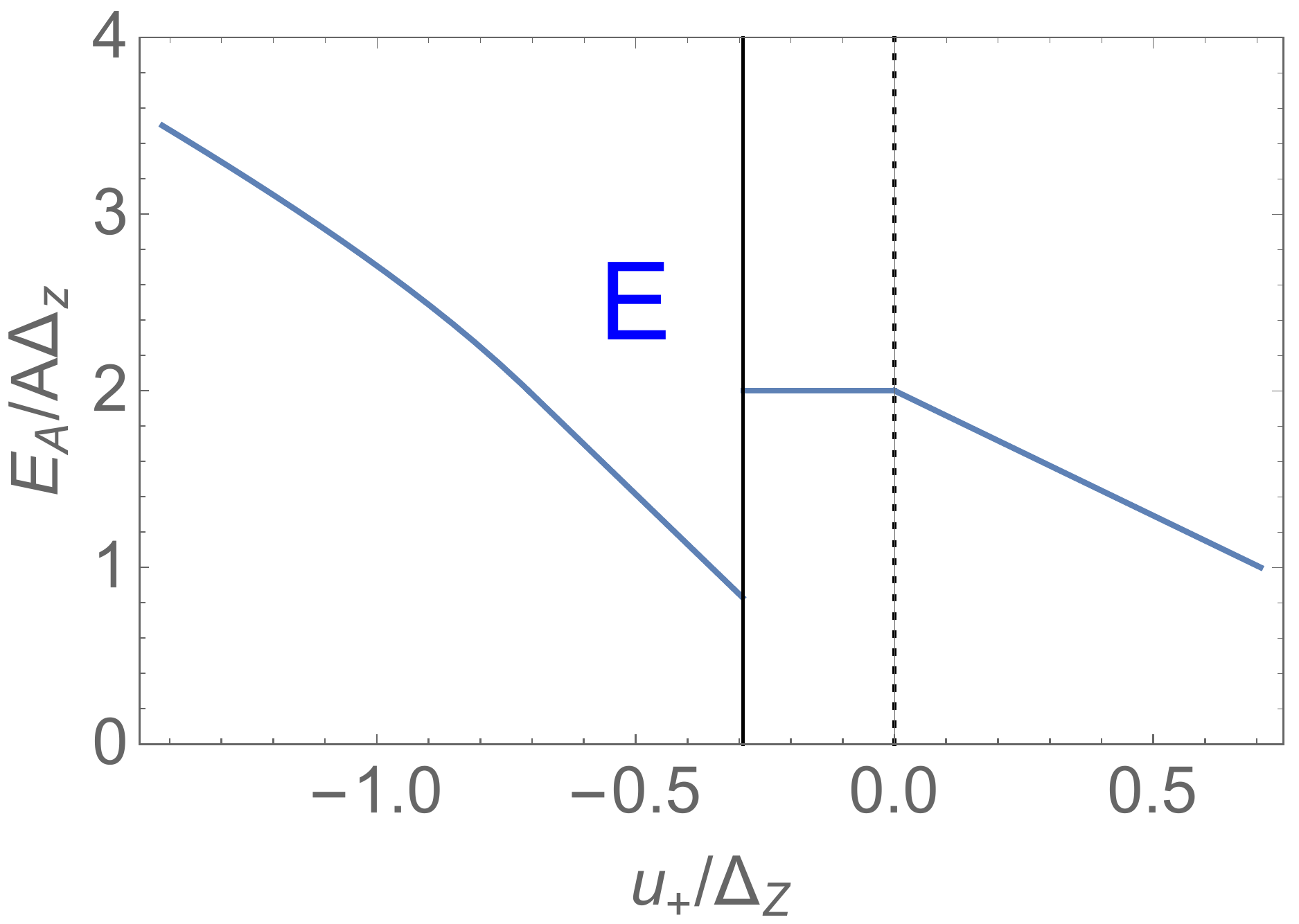}
	\caption{(a) Anisotropic energy of the skyrmion relative to the QHFM background. (b), (c), (d), (e) and (f) Anisotropic energy along the lines A, B, C, D and E in Fig. (a) with the new directions $u_+=(u_z+u_\perp)$ and $u_-=(u_z-u_\perp)$. The energy is continuous everywhere except at the F-CAF transition. The continuity of the energy at each transition can be explained in terms of symmetry or continuity of the observables.}
	\label{fig:Energy}
\end{figure}

As mentioned previously, the energy of a skyrmion is the sum of three terms
\begin{align}
    E_\text{sk}[Z]=E_\text{NLSM}[Z]+E_A[Z]+E_C[Z].
\end{align}
The first term is the nonlinear sigma model energy $E_{NLSM}=4\pi\rho_s$ which is independent of the size and of the angles of the skyrmion. The second term is the anisotropic energy 
\begin{align}
    E_A\propto A=\left(\frac{\lambda}{l_B}\right)^2\ln{\frac{\Lambda}{l_B}}
\end{align}
which is proportional to the area of the skyrmion. In an SU(2) spin skyrmion, this term is simply the Zeeman term, but due to the additional terms in Eq. (\ref{eq:aniso}) acting on the valley pseudo-spin, it can be seen as a \textit{generalized} or \textit{effective} Zeeman term here. Most saliently, it acts similarly on the skyrmion size as the true Zeeman effect. 
Lastly, the Coulomb interaction is proportional to $1/\lambda$ and favors large skyrmions. 

Since the NLSM energy is identical for all skyrmion, we will disregard it in the following discussion of the skyrmion energy. The Coulomb energy is inversely proportional to the size $\lambda=l_B\left(\frac{E_C[\check{Z}]}{2E_A[\check{Z}]}\right)^{1/3}$, but we will focus, for the moment, only on the anisotropic energy $E_A$ in the discussion of the skyrmion energy and size. 

When we compare the skyrmion diagram (Fig. \ref{fig:PhaseDiagSk}) to the QHFM phase diagram (Fig. \ref{fig:PhaseDiagQHFM}), we clearly see that one needs to distinguish three different types of transitions: (1) those that are in common in both diagrams, in which case both the skyrmion type and the QHFM background change over the transition; (2) skyrmion transitions where only the skyrmion type changes (i.e. the pattern at the skyrmion center) while maintaining the same background phase; and (3) transitions where the skyrmion type remains the same while changing the background phase. Notice that the second type cannot be viewed as a phase transition in the thermodynamic sense, such as those of the QHFM background, because they concern single objects. These \textit{skyrmion transitions} might eventually become true phase transitions if a large number of such skyrmions are taken into account that could be arranged, e.g., in a skyrmion crystal\cite{Brey1995,Brey1996b,Cote2008}, but this discussion is beyond the scope of the present paper. In order to study these transitions in more detail, we plot in Fig. \ref{fig:Energy}(a) the anisotropic part $E_A$ of the skyrmion energy in units of $A\Delta_Z$ such that it corresponds to scale-invariant skyrmions, while the Figs. \ref{fig:Energy}.(b)-(f) show different line cuts along the lines indicated in Fig.\ref{fig:Energy}(a). The plain vertical lines in Figs. \ref{fig:Energy}.(b)-(f) correspond to phase transitions, while the dashed lines refer to skyrmion transitions. In the remainder, we use the following notation A[B] for a skyrmion of type A (pattern A at the center) embedded in the QHFM background B. As an example, the CDW skyrmion in the F background is labelled as CDW[F], while the F skyrmion in the CDW background is labelled as F[CDW].

Let us first concentrate on the skyrmion transition where the background QHFM remains the same as for example along the line cut A [Fig. \ref{fig:Energy}(b)]. Along this line, the AF[F] changes into the EP[F] skyrmion at $u_\perp=u_z$. The anisotropic energy of these skyrmions are $E_A^{AF[F]}=2A(u_\perp+2\Delta_Z)$ and $E_A^{EP[F]}=A(u_\perp+u_z+2\Delta_Z)$. At the transition, the SU(2) pseudo-spin symmetry is restored such that there is no cost in energy and we have $E_A^{AF[F]}=E_A^{EP[F]}$. Furthermore, since the anisotropic energy of the EP[F] skyrmion depends only on the sum of the energies $u_\perp+u_z$, it remains constant along the cut A in the direction $u_-=(u_z-u_\perp)$.

Figure \ref{fig:Energy}(c) corresponds to the phase transition between the KD and CDW backgrounds along line B which also happens at $u_\perp=u_z$, but where one finds the same skyrmion type (CAF) on both sides of the transition. Similarly to Fig. \ref{fig:Energy}(b), the pseudo-spin symmetry is restored at the transition and the transitions corresponds simply to rotate the pseudo-spins from $|\mathbf{n}_z\rangle$ to $|\mathbf{n}_\perp\rangle$ without a cost in energy so that the energy is continuous. Since the anisotropic energy is non-zero even at the transition, the skyrmions experience still an effective Zeeman effect that delimits the skyrmion size.

Figure \ref{fig:Energy}(d) shows two transitions along line cut C. First, there is the F[CDW]-CDW[F] phase transition at $u_\perp+u_z=-\Delta_Z$, where both the skyrmion type and the background QHFM are changed, or even switched in the present case. Second, one notices in the ferromagnetic background the CDW[F]-EP[F] skyrmion transition at $u_\perp+u_z=0$. The CDW[F]-EP[F] is similar to the AF[F]-EP[F] since there is there is an SU(2) symmetry restoration, as discussed in Sec. \ref{sec:symmetryF}, that ensures energy continuity, albeit with a change in the slope. The F[CDW]-CDW[F] is even more interesting. In contrast to the transition along line B, the change in the background phase is accompanied by a change of the skyrmion type. Furthermore, as we can already see from the names of the phases, there is a duality between the background and the skyrmion such that the skyrmion center on one side of the transition becomes the background on the other side and vice versa. The energies of the two phases are $E_A^\text{F[CDW]}=-2A(u_\perp+u_z+\Delta_Z)$ and $E_A^\text{CDW[F]}=2A(u_\perp+u_z+\Delta_Z)$ such that at the transition, the anisotropic energy vanishes,  $E_A^\text{F[CDW]}=E_A^\text{CDW[F]}=0$.
Because the anisotropic energy energy acts like an effective Zeeman energy, when the anisotropic energy vanishes the size of the skyrmion diverges because of the Coulomb repulsion. Hence, as $u_+=(u_\perp+u_z)/\sqrt{2}$ gets closer to the transition, the skyrmion size increases and becomes the ground state at the transition. To be more specific, the spinor $|F_1\rangle=|K\uparrow\rangle$ is identical on both sides of the transition while on the F[CDW] side the second QHFM spinor is $|F_2\rangle_\text{F[CDW]}=|K\downarrow\rangle$ while the center spinor is $|C_2\rangle_{F[CDW]}=|K'\uparrow\rangle$. Upon crossing transition, we have $|C_2\rangle_\text{F[CDW]}\rightarrow|F_2\rangle_\text{CDW[F]}$ and $|F_2\rangle_\text{F[CDW]}\rightarrow|C_2\rangle_\text{CDW[F]}$ such that the ground state corresponds indeed to the F state with a CDW skyrmion. Moreover, we have seen in Sec. \ref{sec:F-CDW} that there is a SU(2) symmetry with the operator given by Eq. (\ref{eq:R}) that exchanges the levels $|K\downarrow\rangle$ and $|K'\uparrow\rangle$. The fact that the ground state is invariant under this symmetry implies that both levels have the same energy such that a skyrmion has no cost in anisotropic energy. Thereby, in that special case, the SU(2) symmetry at the transition explains the vanishing of the anisotropic energy of a skyrmion that is accompanied by a vanishing  Coulomb energy due to its divergent size, $\lambda\rightarrow \infty$.

Figure \ref{fig:Energy}(e) shows the transition between the TAF[CAF] and the AF[F] skyrmions along line cut D which happens at $u_\perp=-\Delta_Z/2$. The energy is continuous because, as we mentionned ealier, the canting (tilting) angle $\alpha$ of the TAF[CAF] skyrmion defined by Eq. (\ref{eq:alphaCAF}) reaches 0 at the transition which fits with the AF[F] skyrmion. The transition is thus continuous. However, we see a scenario analogous to the CDW[F]-F[CDW] transition where the energy diminish as we approach the transition from both sides. The energy of the skyrmions are $E_A^{TAF[CAF]}=-2Au_\perp$ and $E_A^{AF[F]}=2A(u_\perp+\Delta_Z)$ such that at the transition $E_A=A\Delta_Z$ remains finite unless the Zeeman effect vanishes ($\Delta_Z=0)$, in which case the  transition is located at $u_\perp=0$. When $\Delta_Z=0$, the F ground state remains unaffected but, as mentionned earlier, the CAF ground state becomes anti-ferromagnetic. We can see from Eqs. (\ref{eq:Z1TAF}) and (\ref{eq:Z2TAF}) for $\alpha=0$, that upon an SU(2) spin rotation (allowed because $\Delta_Z=0$), we obtain the skyrmion spinors
\begin{align}
	|F_2\rangle&=e^{i\alpha_f}|\mathbf{n}_z\rangle|\mathbf{s}\rangle \\
	|C_2\rangle&=e^{i\alpha_c}|-\mathbf{n}_z\rangle|\mathbf{s}\rangle ,
\end{align}
which correspond to a ferromagnetic skyrmion. In that limit, this transition becomes thus an F[AF]-AF[F] transition analogous to the F[CDW]-CDW[F] transition, along with a vanishing anisotropic energy and thus a zero (effective) Zeeman effect. 
The spinors $|C_2\rangle$ and $|F_2\rangle$ that are interchanged at the transition are $|K'\uparrow\rangle$ and $|K'\downarrow\rangle$. At the transition there is thus an $\text{SU}(2)^K\otimes\text{SU}(2)^{K'}$ spin symmetry in each valley generated by the operators $(\sigma_x\tau_z,\sigma_y\tau_z,\sigma_z\tau_z)$ as mentioned by Wu \textit{et al.}\cite{Wu2014a}. Because $|C_2\rangle$ and $|F_2\rangle$ have the same energy, the energy vanishes. However, in the general case $\Delta_Z\neq 0$, this symmetry is broken and the energy is $E_A=A\Delta_Z$, the energy of a pure spin skyrmion.

Figure \ref{fig:Energy}(f) shows the energy along line cut E. One finds transitions between the F[KD] and EAE[CAF] skyrmions, and later (at $u_z=-u_\perp$) between EAE[CAF] and TAF[CAF] skyrmions. The latter transition involves only the skyrmion center but not the QHFM background, and one realizes a similar behavior to the transitions along line A and along line C (at $u_z=-u_\perp$). The constant anisotropic energy in EAE[CAF] is again due to the fact that it is insensitive to the combination $u_-=u_z-u_\perp$. Notice that, in the case of the F[KD] to EAE[CAF] transition, the background as well as the skyrmion center spinors are completely unrelated on the two different sides of the transition, i.e the skyrmions are completely different on both sides of the transition. This is the reason why the anisotropic energy is discontinuous there, and indeed now there is no symmetry restoration at the transition. The sum of the Coulomb and anisotropic energies is
\begin{align}
    E_A[Z]+E_C[Z]=(2^{-2/3}+2^{1/3})E_C[\check{Z}]^{2/3}E_A[\check{Z}]^{1/3}
\end{align}
where $\check{Z}=Z|_{\lambda=l_B}$ is the scale invariant skyrmion. Because $E_C[\check{Z}]$ and $E_A[\check{Z}]$ are scale invariant, a discontinuity in the anisotropic energy $E_A[\check{Z}]$ also implies a discontinuity in the sum of the Coulomb and anisotropic energy i.e. they don't compensate each other at the transition.

\subsection{Size}

\begin{figure}[h]
    (a)\includegraphics[width=8cm]{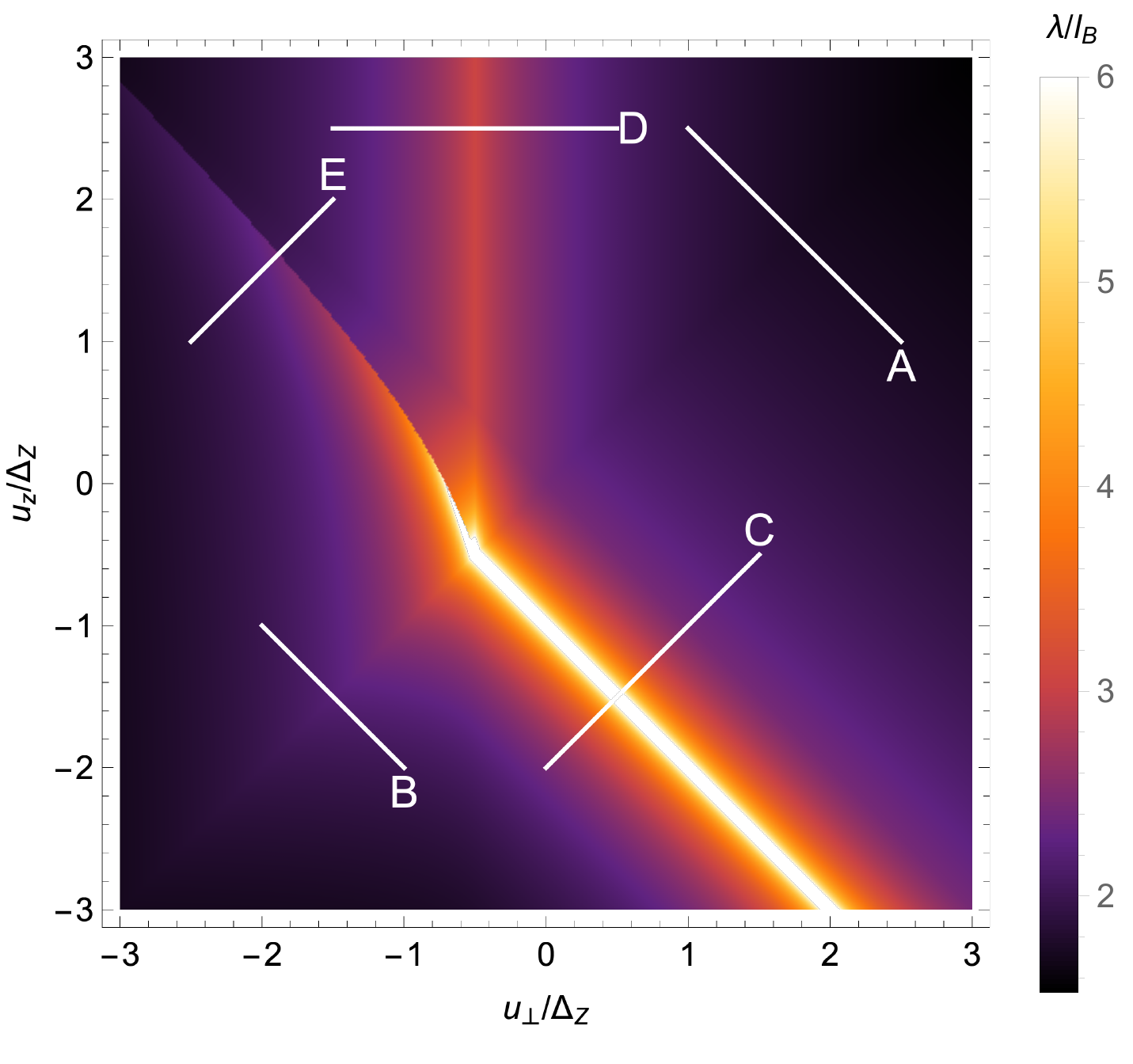}\\
    (b)\includegraphics[width=3.5cm]{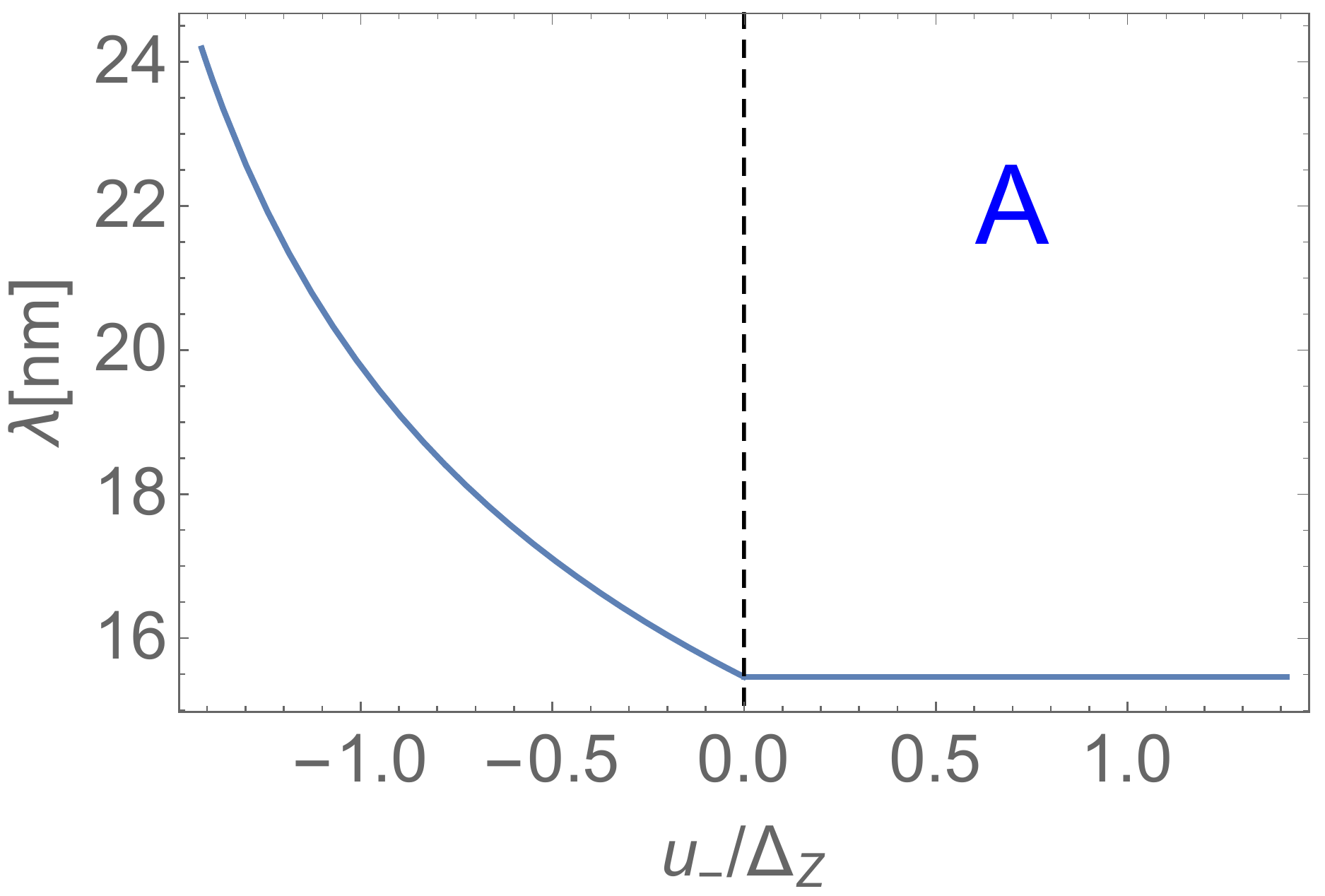}
    (c)\includegraphics[width=3.5cm]{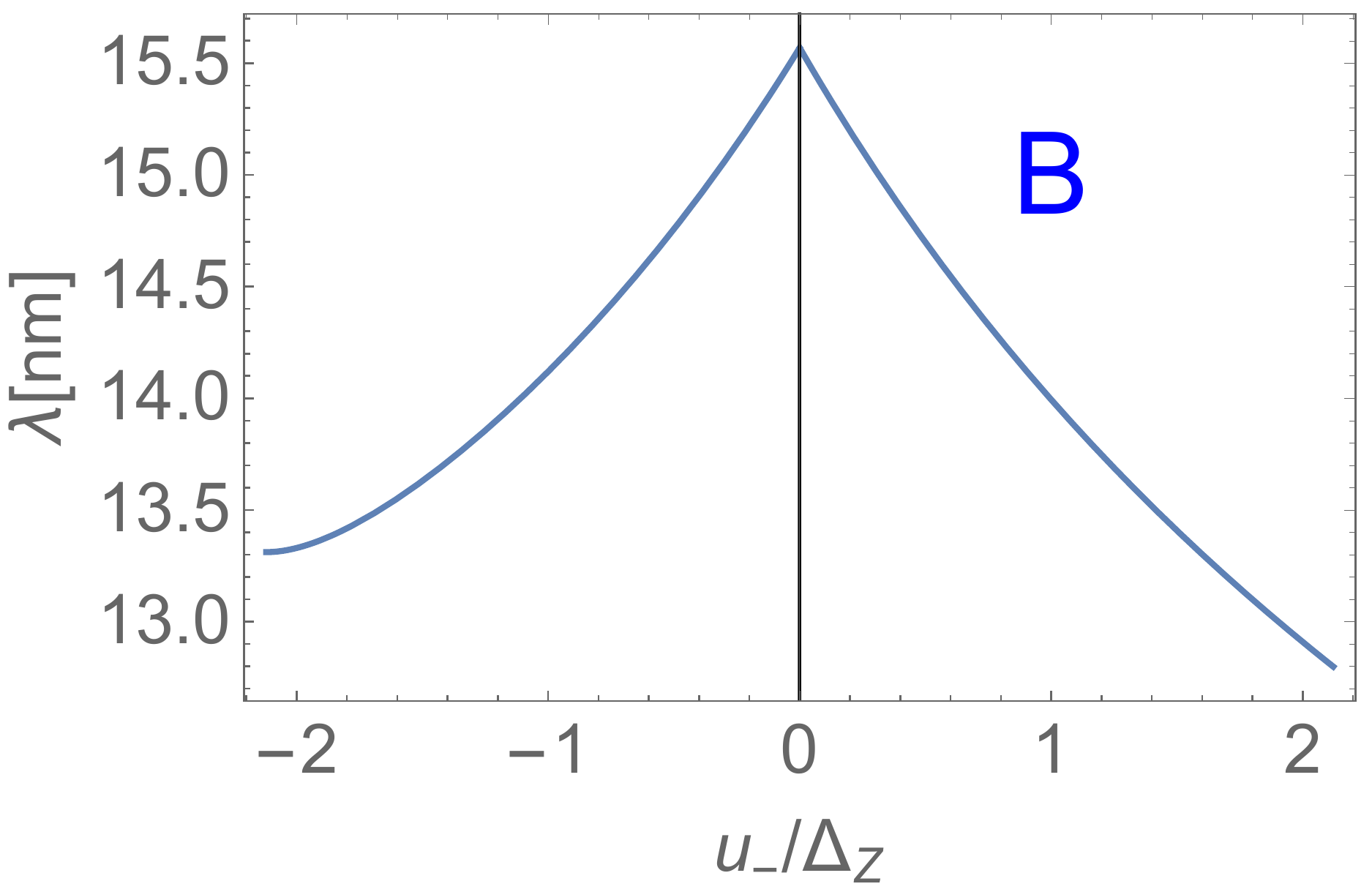}\\
    (d)\includegraphics[width=3.5cm]{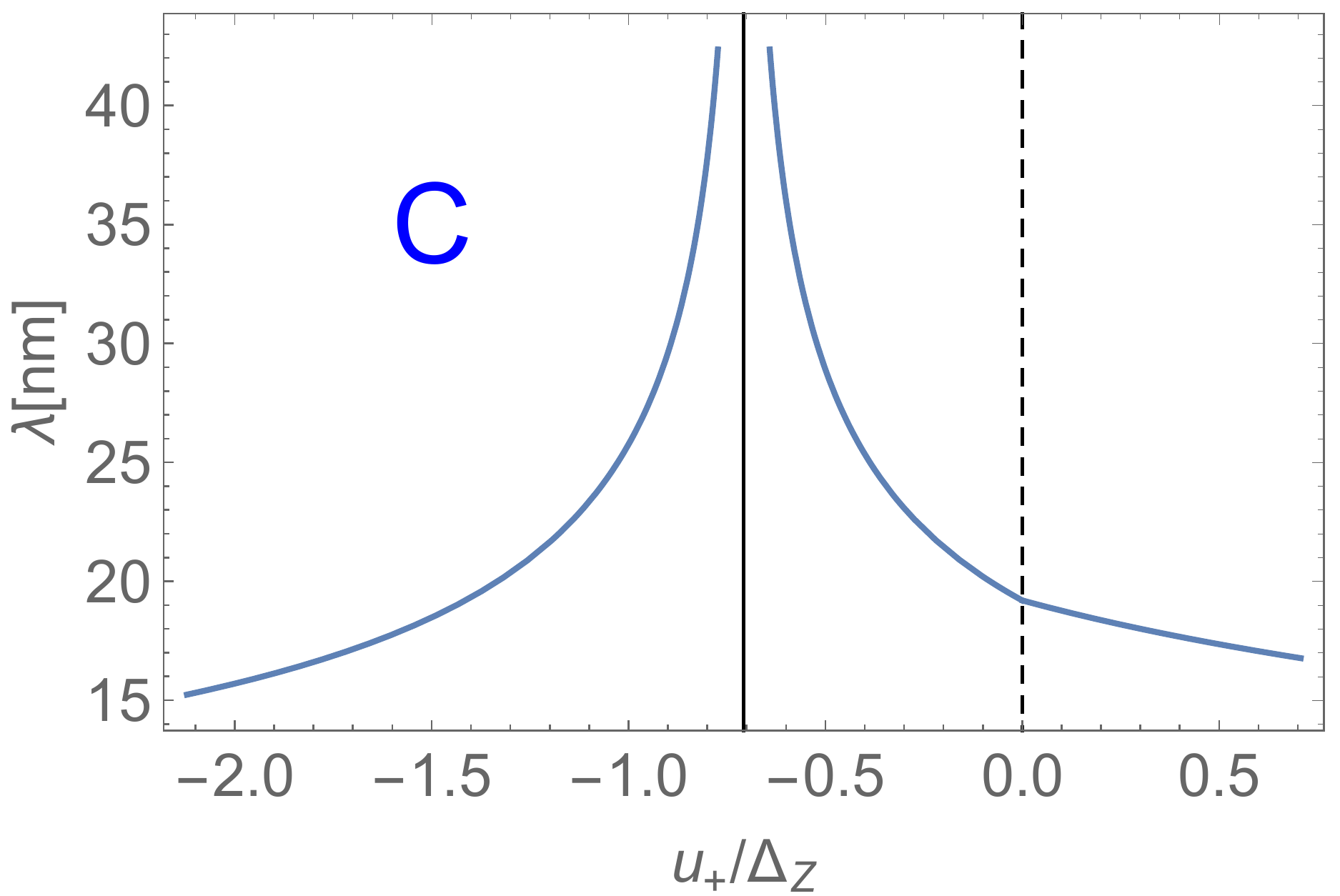}
    (e)\includegraphics[width=3.5cm]{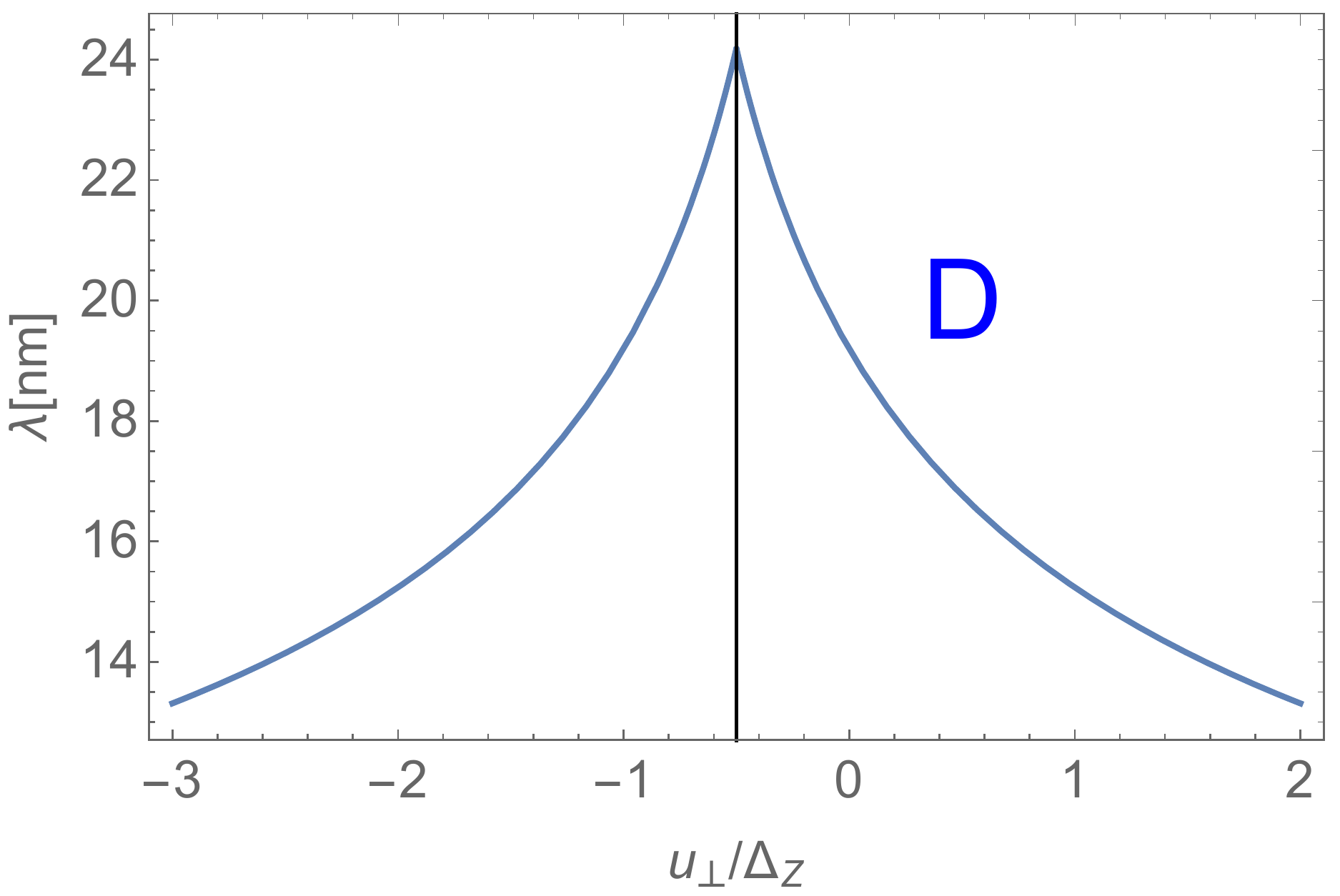}
    (f)\includegraphics[width=3.5cm]{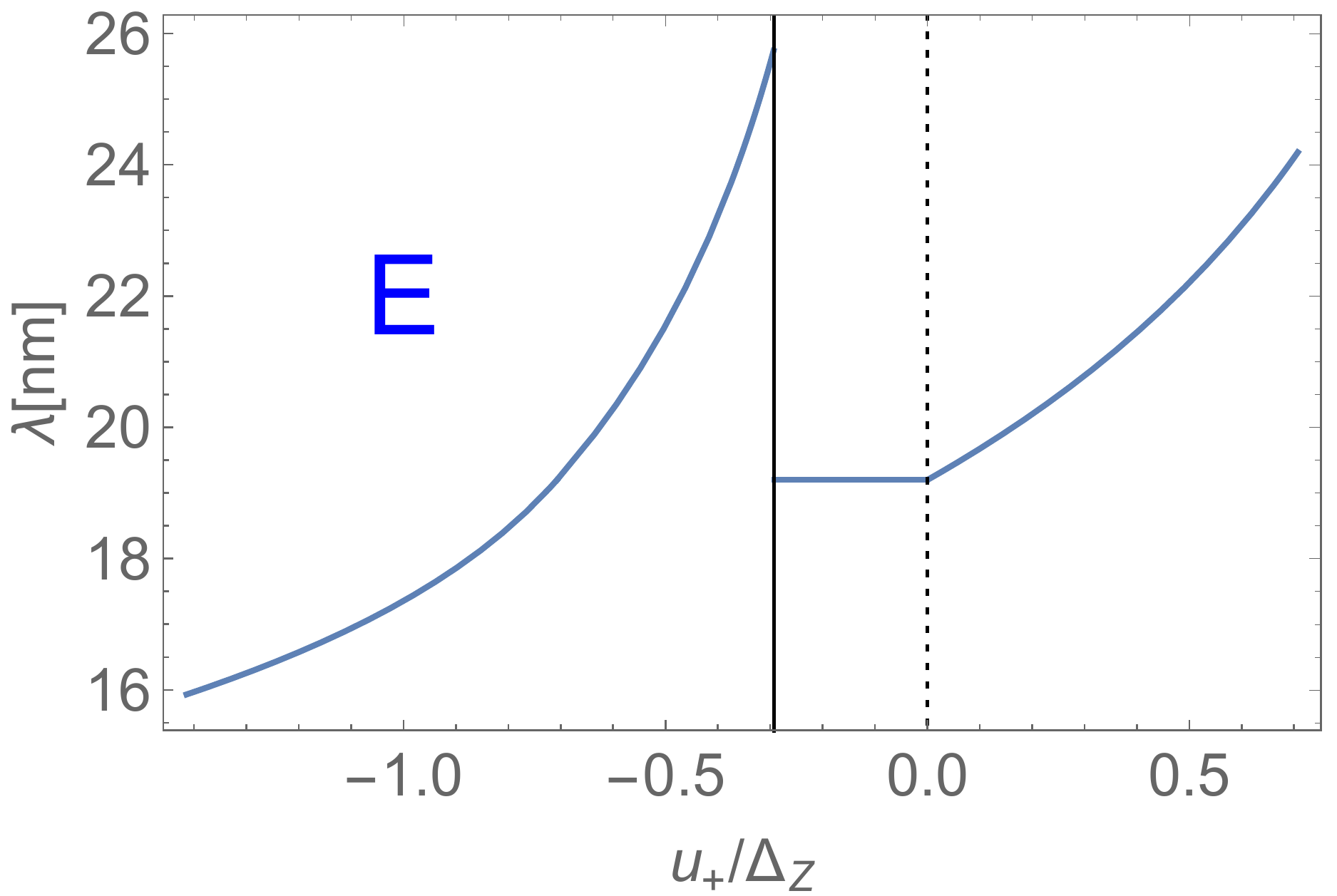}
	\caption{(a) Size of the skyrmions in units of the magnetic length for a magnetic field $B=10\text{T}$, a relative permittivity $\varepsilon_r=2.3$ corresponding to an hexagonal Boron-Nitride substrate and a cut-off $\Lambda=1000l_B$. The magnetic length is $l_B=7.9$nm. (b), (c), (d), (e) and (f) Size of the skyrmions in nanometers along the lines A, B, C, D and E and for the same parameters as Fig. (a). The plain lines indicate a phase transition while the dashed lines indicate a skyrmion transition.}
	\label{fig:Size}
\end{figure}

As we have already seen in Sec. \ref{sec:skyrmions} C, the size $\lambda$ of a skyrmion is given by 
\begin{align}\label{eq:SkyrmSize}
    \frac{\lambda}{l_B}=\left(\frac{E_C[\check{Z}]}{2E_A[\check{Z}]}\right)^{1/3}
\end{align}
where $\check{Z}=Z|_{\lambda=l_B}$ and is thus proportional to the ratio between the scale invariant Coulomb and anisotropic energy. Fig. \ref{fig:Size}.(a) shows the size of the skyrmions in units of the magnetic length for a magnetic field of 10\,T. We can see that in the black regions, the size of the skyrmion is close to the magnetic length such that it might be barely distinguishable from a quasiparticle excitation with a spin and/or valley pseudo-spin flip, while near the phase transitions its size increases. 

Figures \ref{fig:Size}.(b)-(f) present the size of the skyrmions in nanometers for graphene on a hexa-boron-nitride (hBN) substrate and a magnetic field of $10$ T. Along line A, which corresponds to the energy between the skyrmion phases AF[F] and EP[F] in the ferromagnetic phase, the energy increases in the AF[F] phase and then remains constant in EP[F] [see Fig. \ref{fig:Energy}(b)]. One can therefore already inspect from Eq. (\ref{eq:SkyrmSize}) that the skyrmion size decreases in AF[F] and then remains constant in EP[F], as one can clearly see in Fig. \ref{fig:Size}(b). The line B in Fig. \ref{fig:Size}(c) corresponds to the CAF[CDW]-CAF[KD] phase transition located at $u_\perp=u_z$. At the transition, the effective Zeeman energy is minimal, but non-zero [see Fig. \ref{fig:Energy}(c)], such that the skyrmion size reaches a maximum but does not diverge. Such divergence is seen in Fig. \ref{fig:Size}(d) at the F[CDW]-CDW[F] transition (along line C), as a consequence of a vanishing anisotropic energy and the restoration of the SU(2) symmetry associated with the transition as described in Sec. \ref{sec:energy}. Furthermore, one notices a kink in the skyrmion size at $u_z=-u_\perp$ ($u_+=0$) because of the change in the skyrmion type there, where the CDW[F] becomes an EP[F] skyrmion. The kink in the skyrmion size is therefore a consequence of the kink in the anisotropic energy at this transition. At the TAF[CAF]-AF[F] transition (Line D) in Fig.\ref{fig:Size}(e), the size of the skyrmion reaches a maximum but does not diverge because, as discussed, the $\text{SU(2)}^K\otimes \text{SU(2)}^{K'}$ is broken, and the two levels do not have the same energy. This symmetry is restored if the Zeeman effect, which constitutes alone the anisotropic energy, vanishes, as discussed above, in which case the skyrmion size diverges in the same manner as at the F[CDW] to CDW[F] transition. Notice finally the discontinuity in the skyrmion size along line E [Fig. \ref{fig:Size}(f)]. As already anticipated in the previous subsection, where we discussed the different energies, it is a direct consequence of the discontinuity in the anisotropic energy [see Fig. \ref{fig:Energy}(f)].
 
The Coulomb energy of a scale-invariant skyrmion $\check{Z}=Z|_{\lambda=l_B}$ is
\begin{align}
    E_C[\check{Z}]=\frac{3\pi^2}{64}\frac{e^2}{\varepsilon_0 \varepsilon_r l_B}\approx 335\frac{\sqrt{B[T]}}{\varepsilon_r}\text{meV},
\end{align}
where $\varepsilon_0$ is the vacuum permittivity and $\varepsilon_r$ is the relative permittivity of the medium surrounding the graphene sheet. 
As an example, we consider the skyrmion located at the F-CAF transition $u_\perp=-\Delta_Z/2$ in the phase diagram which is a pure spin skyrmion with energy $E_A[Z]=A\Delta_Z$. The scale-invariant anisotropic energy of this skyrmion is 
\begin{align}
    E_A[\check{Z}]=\ln\left(\frac{\Lambda}{l_B}\right)g\mu_B B\approx8.03 B[T]\text{meV}
\end{align}
where we have used a cut-off $\Lambda=1000l_B$ and  g-factor $g=2$. We note that the size of a skyrmion depends very weakly on the cut-off $\Lambda$ since it is proportional to $\ln(\Lambda/l_B)^{-1/3}$. The magnetic length equals
\begin{align}
    l_B\approx \frac{25}{\sqrt{B[T]}}\, \text{nm}
\end{align}
which is $l_B=7.9$ nm at a magnetic field of $B=10$ T. The size of this skyrmion is thus
\begin{align}
    \lambda=112B^{-2/3}[T]\,\text{nm}
\end{align}
for $\varepsilon_r=2.3$, which is the characteristic value for graphene on an hBN substrate and that gives a size $\lambda=24.2$nm at $B=10$ T. Such a skyrmion has thus a size which is three times the magnetic length, we thereby expect approximately 9 spins to be reversed in this pure spin skyrmion. However, in the phase diagram, generally, the skyrmions can have a larger anisotropic energy such that their size is smaller than the spin skyrmion we considered.

\subsection{Magnetic field tilting}

\begin{figure}[h]
    (a)\includegraphics[width=8cm]{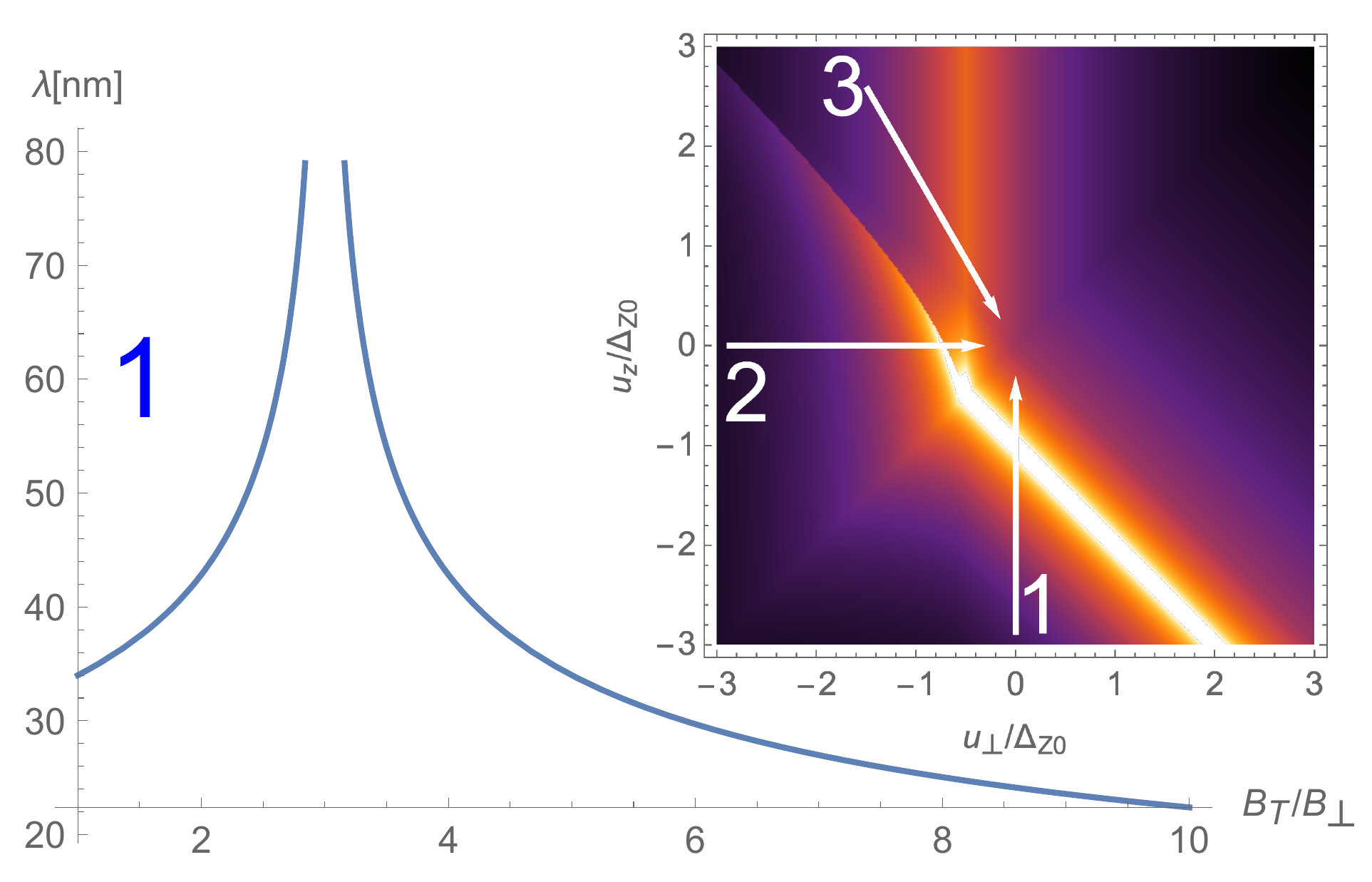}
    (b)\includegraphics[width=8cm]{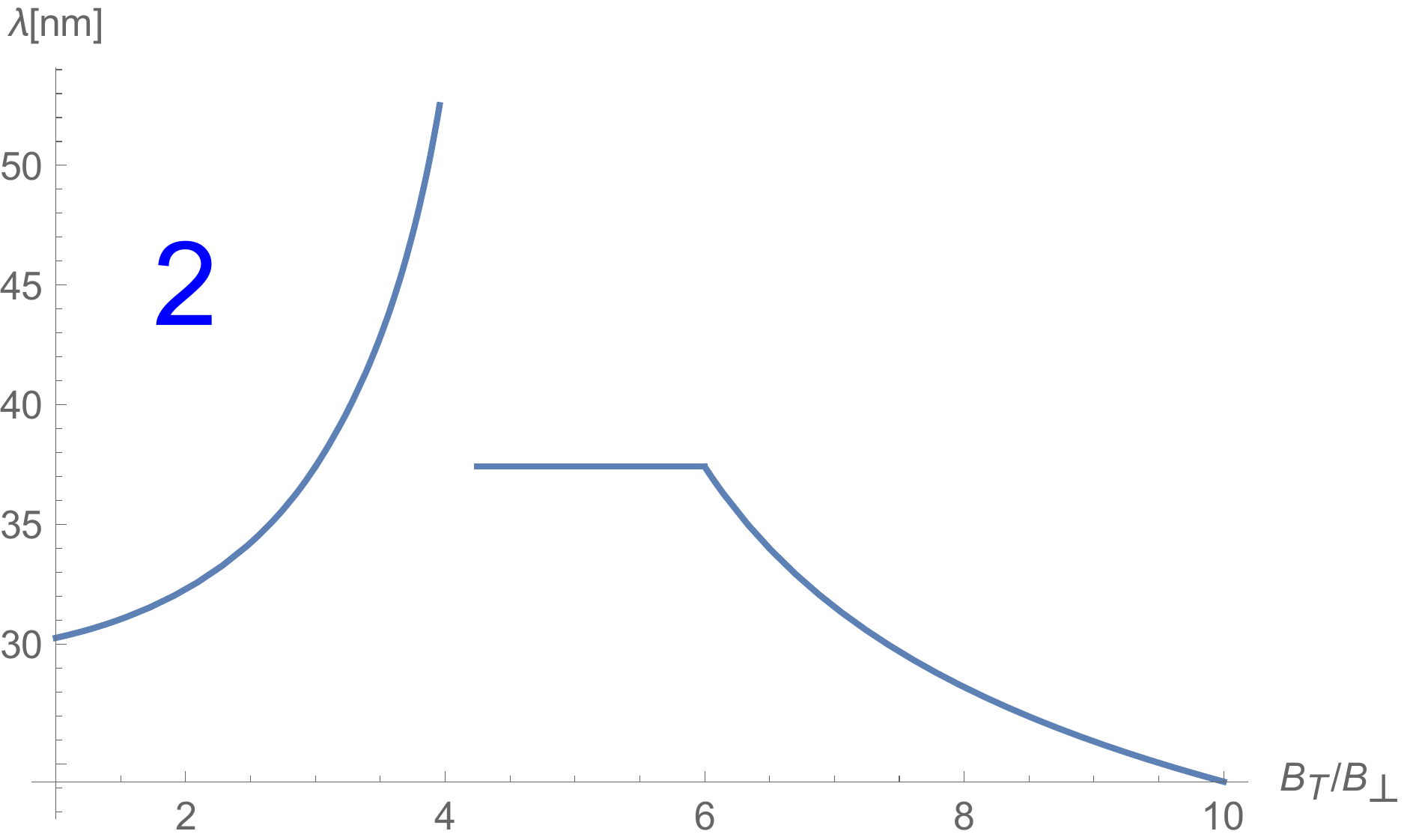}\\
    (c)\includegraphics[width=8cm]{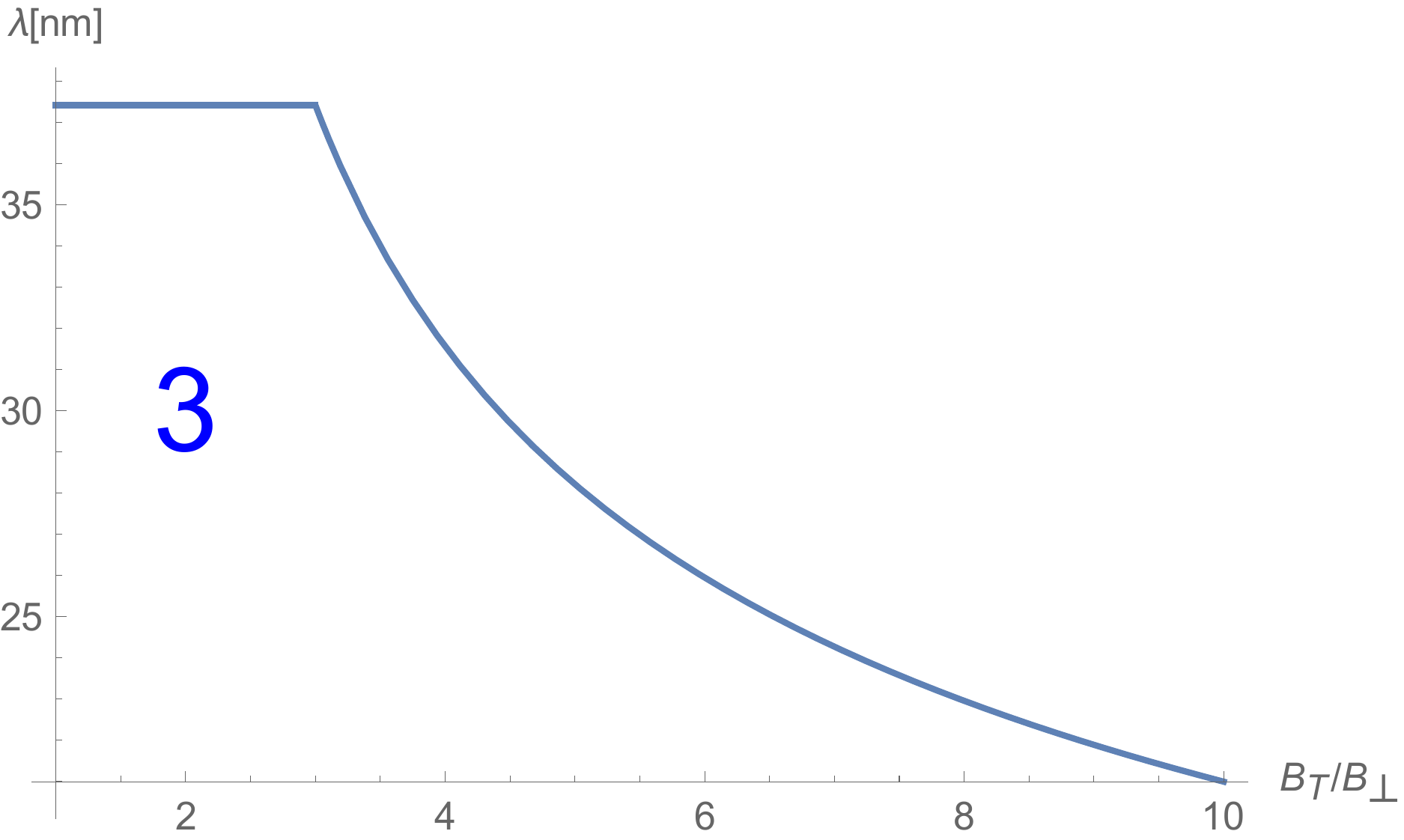}
	\caption{Effect of the magnetic field tilt. (a)-(c) Size of the skyrmions as a function of the total magnetic field $B_T$ for a fixed perpendicular magnetic field $B_\perp=3\text{T}$ along the lines 1-3 of the inset of Fig. (a). }
	\label{fig:Tilt}
\end{figure}

As we have noted earlier, tilting the magnetic field can modify the value of the anisotropic parameters relative to the Zeeman coupling. One can thus change the size of the skyrmions and eventually generate some phase and/or skyrmion transitions with the help of the tilt. Indeed, the anisotropic parameters $u_\perp$ and $u_z$ are related to the bare parameters $U_\perp$ and $U_z$ via 
\begin{align}
    u_{\perp,z}=\frac{U_{\perp,z}}{2\pi l_B^2},
\end{align}
where the magnetic length is proportional to the magnetic field \textit{perpendicular} to the sample $B_\perp$
\begin{align}
l_B=\sqrt{\frac{\hbar}{eB_\perp}}.
\end{align}
On the other side, the Zeeman energy is proportional to the \textit{total} magnetic field
\begin{align}
    \Delta_Z=g\mu_B B_T.
\end{align}
Therefore, let us consider an experiment similar to Ref. \onlinecite{Young2014} where the perpendicular component of the magnetic field is kept constant at $B_\perp=1.4$ T while the total magnetic field $B_T$ is increased ranging from $1.4$ to $34.5$ T. The angle $\theta$ between the direction of the magnetic field and that normal to the graphene plane is thus given by $\cos\theta=B_\perp/B_T$. In such an experiment, the anisotropic parameters are thus kept constant while the Zeeman energy is increased. The inset of \ref{fig:Tilt}(a) shows the skyrmion size in the absence of a tilt, $B_T=B_\perp$, such that $\Delta_{Z0}=g\mu_BB_\perp$. As the perpendicular magnetic field is increased, we have 
\begin{align}
    \frac{u_{\perp,z}}{\Delta_Z}=\frac{u_{\perp,z}}{\Delta_{Z0}}\frac{B_\perp}{B_T}
\end{align}
such that as $B_T/B_\perp$ increases, the values $(u_z/\Delta_Z,u_\perp/\Delta_Z)$ move to the center $(0,0)$ of the phase diagram. Figures \ref{fig:Tilt}(a)-(c) show the evolution of the size of a skyrmion as $B_T/B_\perp$ increases along the three arrows 1, 2, 3 indicated on the inset of Fig. \ref{fig:Tilt}(a) with $B_\perp=3$ T, corresponding to a magnetic length $l_B=14.4$ nm.  In the case of lines 1 and 2, as the magnetic field increases, the size of the skyrmion increases until the respective phase transitions are reached. This happens around around $B_T/B_\perp\sim 3-4$, which corresponds to an angle $\theta=70-75\si{\degree}$, where the skyrmion size reaches a maximum (for line 2) or even diverges (for line 1). For $4<B_T/B_\perp<6$ in the case of line 2 and for $B_T/B_\perp<3$ for line 3, the skyrmion size is not changed because the energies of the phases are independent of the Zeeman coupling. Therefore increasing the total magnetic field only modifies the location of the phase transitions but not the energies.

\section{Conclusion}

In conclusion, we have presented the phase diagram of skyrmions in graphene in the QH regime at filling $\nu=0$, corresponding to charge neutrality. We have considered the pseudo-spin anisotropic terms $u_\perp$ and $u_z$ (considered as control parameters in our study) and the Zeeman term that all break the approximate SU(4) symmetry of graphene and that had been used by Kharitonov in the elaboration of the phase diagram of the underlying QHFM phases \cite{Kharitonov2012}. After a brief discussion of this phase diagram, we have presented the different skyrmion types that are compatible with the QHFM background and identified those that are of lowest energy, within a variational study. Most saliently, the different skyrmion types have a unique signature in the spin magnetization on the two sublattices (A and B). This allows us to classify these skyrmion types by the spin-valley pattern at the skyrmion center and that of the QHFM background that is retrieved far away from the skyrmion center. Because the valley is identical to the sublattice in the $n=0$ LL, these pattern are expected to be visible in spin-resolved STM experiments.

Our findings are at first sight reminiscent of similar results obtained at a filling factor $\nu=\pm 1$ \cite{Lian2016,Lian2017}, but both the QHFM backgrounds and the different skyrmion types are more involved. Indeed, at $\nu=0$, two LL subbranches are completely filled and they are characterized, within a Grassmannian $\text{Gr}(2,4)$ approach, by two orthogonal spinors $f_1$ and $f_2$. In agreement with Kharitonov's calculations, we obtain four distinct background phases, two of which consist of a full pseudo-spin polarization (the KD and CDW phases), and two other phases with full or partial spin polarization (the FM and the CAF phases). The latter CAF phase shows entanglement which reduces the total spin magnetization. From an experimental point of view, the nature of the ground state is still debated. Reference \onlinecite{Young2014} shows signatures of the realization of a CAF phase, while Ref. \onlinecite{Li2019} shows the enlargement of the elementary unit cell characteristic of a KD phase. A possible origin of this discrepancy is the uncertainty in the precise values of the anisotropic parameters, which happen to be all on the same order of magnitude. The nature of the ground state might thus be sample-dependent and strongly sensitive to the dielectric environment. 

In the same manner as the QHFM background phases, the skyrmions, which are formed on these backgrounds, are classified by the Grassmannian $\text{Gr}(2,4)$. This approach gives rise to six variational parameters (angles) that discribe the different skyrmion types once the background spinors are fixed. We have then obtained the skyrmion diagram by minimizing the skyrmion energy, which contains a Coulomb repulsion and anisotropic energies in addition to the leading non-linear sigma model, with respect to these six parameters. 

Apart from the skyrmion zoo, the skyrmion diagram shows different types of transitions. The thermodynamic transitions between the underlying QHFM phases naturally have also an impact on the skyrmions since they determine a change in the spin-valley pattern far away from the skyrmion center. In most cases, the skyrmion type thus also changes across these transitions, with the notable difference of the CAF skyrmion that is the lowest-energy skyrmion at both sides of the CDW-KD background transition. Moreover, the background phase transitions are visible in the energy diagram and manifest themselves in the skyrmion size, which has a tendency to increase in the vicinity of these transitions. This is most prominent in the CDW[F] to F[CDW] at $u_z=-u_\perp-\Delta_Z<-\Delta_Z/2$, at which the skyrmion-center and background spinors are simply switched and where the skyrmion size even diverges. We have identified an SU(2) symmetry at this line that connects precisely the CDW and F spinors such that the anisotropic energy, which can be seen as an effective Zeeman term, vanishes. We are thus left with the Coulomb repulsion, which is minimized when the skyrmion size diverges. 

In addition to the signature of the underlying QHFM phase transitions in the skyrmion diagram, we obtain transitions between the different skyrmion types while maintaining the same QHFM background. Usually they are accompanied by a symmetry restoration that relates the different skyrmion-center spinors across the transition. This is particularly the case at $u_z=u_\perp$, where the SU(2) pseudospin symmetry is clearly restored, but also at $u_z=-u_\perp$. However, in these cases the effective Zeeman term, in the form of the anisotropic energy, does not vanish so that the skyrmion size does not diverge but just reveals a kink. Our findings can to great extent be tested with the help of an inplane magnetic field, which affects only the true Zeeman term while the terms $u_z$ and $u_\perp$, which are due to orbital effects, are sensitive solely to the perpendicular component of the magnetic field. Generically, the system follows lines from some point in the phase diagram towards the center, and one may therefore hope to cross different transition lines. 

Finally we emphasize the particular CAF phase, which is of great interest for two reasons. From an experimental point of view, there is evidence that it may be the ground state at least in some graphene samples \cite{Young2014}. From a theoretical point of view, the CAF phase is the only QHFM background that shows spin-valley entanglement, while the other phases (F,KD, and CDW) are formed either in the spin or in the valley pseudo-spin sector. It therefore shows a continuous transition to the neighboring ferromagnetic phase at $u_\perp=0$, where the spin magnetization is identical on the two sublattices and is then smoothly canted away from the $z$ axis with an inplane component depending on the sublattice. It is therefore natural to expect that the skyrmions, which are formed on top of this background, equally display spin-valley entanglement. This is indeed the case in both the EAE and TAF skyrmions, which we have identified as the low-energy skyrmions in this phase. Most saliently, we find a discontinuity in the skyrmion energy across the EAE[CAF] to F[KD] transition, which is most likely due to the fact that there is no symmetry that connects the two phases (nor the two skyrmion types) across the transition.

\label{sec:conclusion}

\begin{acknowledgements}
We acknowledge financial support from Agence Nationale de la Recherche (ANR project ``GraphSkyrm'') under Grant No. ANR-17-CE30-0029.
\end{acknowledgements}

\bibliographystyle{apsrev4-1}
\bibliography{library}

\end{document}